\documentclass[twocolumn,apj,numberedappendix]{openjournal}
\usepackage{graphicx}
\usepackage{amsmath,amssymb,mathtools}
\usepackage{booktabs}
\usepackage{microtype}
\usepackage{bm}
\usepackage{enumitem}
\usepackage{ifthen}
\usepackage{placeins}
\providecommand{\texorpdfstring}[2]{#1}

\shorttitle{From Quantum Fluctuations to High-Redshift AGN}
\shortauthors{Yogeswaren, Najmi \& Schaefer}

\newcommand{\insertorplaceholder}[3]{%
  \IfFileExists{#1}{%
    \includegraphics[width=#2]{#1}%
  }{%
    \fbox{\parbox[c][0.28\textheight][c]{\dimexpr#2-2\fboxsep-2\fboxrule\relax}{\centering
      Placeholder for figure file \texttt{\detokenize{#1}}.\\[0.5em]
      Place the image in the same directory as this \texttt{.tex} file to render the figure.\\[0.5em]
      #3}}%
  }%
}

\begin{document}

\makeatletter
\@firstsectionfalse
\makeatother

\onecolumngrid
{\scriptsize Preprint typeset using \LaTeX\ style openjournal v. 09/06/15}\par
\vspace{0.35in}

\begin{center}
{\normalsize\bfseries\MakeUppercase{From Primordial Quantum Fluctuations to Cosmic-Dawn Structure: A Review of the Cosmological Connections to Early Structure Statistics and Astrophysical Uncertainties}\par}
\vspace{0.20in}
{\small\textsc{Elan Yogeswaren}\\
\footnotesize Terra Nova Sciences LLC, Houston, TX, 77054, USA\par}
\vspace{0.10in}
{\small\textsc{Amir-Homayoon Najmi}\\
\footnotesize Johns Hopkins University Whiting School EP Applied Physics Program and the Applied Physics Laboratory, Laurel, MD, 20723 USA\par}
\vspace{0.10in}
{\small\textsc{Robert K. Schaefer}\\
\footnotesize Johns Hopkins University Whiting School EP Applied Physics Program and the Applied Physics Laboratory, Laurel, MD, 20723 USA\par}
\vspace{0.20in}
{\itshape\footnotesize Preprint -- 2026\par}
\vspace{0.18in}
{\footnotesize ABSTRACT\par}
\end{center}

\begin{quote}
\small
We develop a continuous framework connecting quantum field fluctuations in the early Universe to the abundance and observable properties of galaxies and active galactic nuclei (AGN) at cosmic dawn. Beginning with quantum fields in curved Friedmann--Lema\^itre--Robertson--Walker spacetime, we follow inflationary vacuum fluctuations into the primordial curvature spectrum, linear matter power spectrum, mass variance $\sigma(M,z)$, peak height $\nu(M,z)$, and the statistics of nonlinear halo formation. The Press--Schechter formalism establishes the historical foundation for gravitational collapse, while the Sheth--Tormen mass function provides an analytic description of halo abundance and its dependence on peak rarity. A complementary high-redshift halo mass function calibrated with the GUREFT simulations of Yung et al.\ provides a numerical description specifically constructed for the cosmic-dawn regime probed by JWST. The comparison between these analytic and simulation-calibrated prescriptions provides a direct measure of the sensitivity of predicted halo populations to the adopted halo mass function. We separate this cosmological supply of halos from the astrophysical and observational mappings that produce stellar masses, ultraviolet luminosities, and active nuclei. Controlled numerical experiments show that a percent-level perturbation in the primordial fluctuation amplitude produces only a modest change in $\sigma$ and $\nu$, but can be strongly amplified in the exponentially suppressed high-$\nu$ tail. This sensitivity increases toward high halo mass and high redshift, identifying rare cosmic-dawn populations as potentially powerful probes of primordial statistics while simultaneously exposing the importance of uncertainties in halo abundance. High-redshift AGN provide an additional constraint through the requirement that massive black holes satisfy both a finite growth-time condition and a sufficiently abundant population of host halos. Propagating the halo population into the galaxy ultraviolet luminosity function further demonstrates how information originating in primordial fluctuations can survive into observable populations while remaining coupled to halo statistics, halo-to-light physics, and observational selection. JWST observations are therefore interpreted as quantitative stress tests of the complete quantum-to-structure-to-observation chain rather than as direct evidence for a failure of $\Lambda$CDM or the standard primordial spectrum. This framework provides a systematic means of identifying whether any future robust discrepancy arises from primordial physics, nonlinear halo statistics, astrophysical mapping, or observational inference.
\end{quote}
\vspace{0.12in}
\twocolumngrid

%
%

\section{Introduction}
\label{sec:introduction}

The origin of cosmic structure remains one of the central questions of modern cosmology. Observations indicate that the Universe is remarkably homogeneous and isotropic on the largest scales while simultaneously exhibiting a rich hierarchy of structures ranging from galaxies and galaxy clusters to the filamentary cosmic web. Understanding how these structures emerged from the nearly uniform conditions of the early Universe requires a synthesis of quantum field theory, general relativity, statistical physics, and nonlinear gravitational dynamics.

Within the standard cosmological paradigm, structure formation begins with primordial fluctuations generated in the very early Universe. These perturbations subsequently evolve under gravitational instability, eventually producing the dark matter halos that host galaxies, stars, and active galactic nuclei (AGN). While the later evolution of cosmological perturbations can be described through linear and nonlinear gravitational dynamics, the physical origin of the primordial fluctuations lies within the domain of quantum field theory in curved spacetime.

Quantum field theory in curved spacetime occupies an intermediate position between quantum field theory and a complete theory of quantum gravity. Matter fields are quantized while the spacetime geometry remains classical and is described by general relativity. This framework provides the theoretical basis for particle creation in expanding universes and for the quantum origin of inflationary perturbations. Its mathematical formulation, including the construction of physically admissible quantum states and renormalized stress--energy tensors, provides the conceptual setting in which microscopic quantum fluctuations can be amplified by cosmic expansion and ultimately seed macroscopic structure \citep{Wald1994,DeutschNajmi1983,NajmiOttewill1984,NajmiOttewill1985}.

Inflation provides a mechanism for connecting these quantum fluctuations to cosmological initial conditions. Vacuum fluctuations of quantum fields are stretched to wavelengths larger than the Hubble radius during accelerated expansion and subsequently behave as effectively classical perturbations. Their statistical properties are encoded in the primordial curvature power spectrum, whose nearly scale-invariant form is strongly supported by measurements of the cosmic microwave background (CMB) \citep{Mukhanov1992,Planck2018}. After inflation, the perturbations evolve through the radiation- and matter-dominated eras. Transfer functions, horizon entry, and gravitational growth transform the primordial spectrum into the linear matter power spectrum from which the statistical abundance of collapsed structures can be calculated.

The connection between the linear density field and nonlinear structure is expressed through the mass variance $\sigma(M,z)$ and peak height $\nu(M,z)$. These quantities measure the amplitude and rarity of fluctuations associated with a given mass scale and redshift. Halo mass functions then translate the statistics of the density field into the abundance of gravitationally collapsed dark matter halos. The Press--Schechter formalism provides the historical foundation for this connection, while the Sheth--Tormen prescription incorporates the effects of nonspherical collapse and provides a useful analytic description of the rare-halo population \citep{PressSchechter1974,Bond1991,ShethTormen1999,ShethMoTormen2001}.

The high-redshift halo population is particularly important because rare objects occupy the exponentially suppressed tail of the fluctuation distribution. Small changes in the primordial fluctuation spectrum can produce relatively modest changes in $\sigma(M,z)$ and $\nu(M,z)$ while generating substantially larger fractional changes in the abundance of sufficiently rare halos. The high-mass, high-redshift halo population therefore provides a natural regime in which to investigate how information originating in primordial quantum fluctuations can propagate into observable cosmic structure.

The James Webb Space Telescope has greatly expanded access to this regime. Massive galaxies with substantial inferred stellar masses have been identified deep into the cosmic-dawn era, together with growing evidence for accreting black holes and AGN activity at similarly early epochs \citep{Labbe2023,Harikane2023,Kocevski2023,Maiolino2024,Atek2024}. These observations have stimulated renewed investigation of early star-formation efficiency, halo assembly, black-hole seeding and growth, and the abundance of sufficiently massive dark matter hosts. They also provide an opportunity to test the complete sequence connecting primordial fluctuations to nonlinear structure rather than any single component of that sequence in isolation.

Supermassive black holes provide an especially informative complementary probe. In the local Universe they exhibit strong empirical correlations with host-galaxy properties such as bulge mass and stellar velocity dispersion \citep{KormendyHo2013}. At high redshift, the appearance of massive black holes within the limited cosmic time available introduces an additional constraint beyond halo abundance alone. Models must simultaneously provide sufficiently massive and sufficiently numerous host halos while allowing black-hole seeds to grow to the inferred masses within the age of the Universe at observation \citep{Volonteri2010,Inayoshi2020}. High-redshift AGN therefore combine a structure-formation constraint with an independent growth-clock constraint.

The purpose of this paper is to construct a continuous theoretical and numerical framework linking these physical regimes. The calculation follows the information flow from quantum fields in an expanding spacetime through inflationary perturbations, the primordial curvature spectrum, the linear matter power spectrum, mass variance, peak rarity, nonlinear collapse, halo abundance, and finally the astrophysical populations observed by JWST. This organization makes it possible to identify where sensitivity to primordial physics is amplified and where additional astrophysical or observational uncertainties enter.

A central feature of the analysis is the explicit separation of three physical layers: the primordial origin of the fluctuation field, the cosmological supply of dark matter halos, and the astrophysical mapping from halos to observable galaxies and AGN. A change in the primordial curvature spectrum modifies the initial statistical conditions. A change in the halo mass function modifies the abundance of available hosts for a given density field. Changes in star-formation efficiency, black-hole occupation, duty cycle, luminosity scatter, or selection modify the observable population associated with those halos. These effects can produce partially degenerate observational consequences, but they represent physically distinct stages of the quantum-to-structure-to-observation chain.

The halo-abundance calculation is developed using complementary analytic and simulation-calibrated descriptions. Press--Schechter establishes the spherical-collapse foundation, and the Sheth--Tormen mass function provides the principal analytic description of the dependence of halo abundance on peak height. The high-redshift calculation is also evaluated with the halo mass function calibrated from the Gadget at Ultrahigh Redshift with Extra-Fine Timesteps (GUREFT) simulations of \citet{Yung2024}. GUREFT was designed specifically to resolve halo demographics and assembly histories in the high- and ultra-high-redshift Universe and provides halo mass functions across the cosmic-dawn regime relevant to the present JWST analysis. Comparing Sheth--Tormen with the GUREFT simulation-calibrated prescription allows the sensitivity of the predicted rare-halo population to the halo mass function itself to be distinguished from sensitivity to the primordial spectrum and from uncertainties introduced by the subsequent astrophysical mapping.

This distinction is particularly important in the extreme rare-halo regime. The GUREFT calibration directly covers the redshift interval used in the present cosmic-dawn calculations, but the most massive halos at the highest redshifts remain intrinsically sparse. Predictions approaching the extreme high-mass, high-redshift corner are therefore interpreted as rare-peak stress tests rather than precision determinations of halo abundance. The comparison between an analytic rare-peak prescription and a high-redshift simulation-calibrated mass function provides a means of assessing how robust the inferred propagation of primordial information remains under changes in the nonlinear halo-abundance model.

The analysis is organized around three complementary observational tests. The first is a massive-galaxy and baryonic-reservoir test, in which the cosmic baryon fraction and halo abundance jointly determine whether an inferred stellar mass can be supported by a plausible population of host halos. The second is an AGN test, in which the available host-halo reservoir is combined with black-hole occupation, duty cycle, host-mass assumptions, and the finite time available for black-hole growth. The third is a galaxy ultraviolet-luminosity-function test, in which halo masses are mapped into rest-frame UV luminosities and the resulting population prediction is compared with high-redshift JWST measurements.

These tests share the same cosmological halo reservoir but probe different astrophysical mappings. The stellar halo-to-UV relation used for the galaxy luminosity function is distinct from an AGN luminosity model, while black-hole occupation fractions and duty cycles enter naturally in the AGN calculation but not generically in the stellar galaxy luminosity function. Maintaining this separation allows an apparent discrepancy to be localized more clearly among primordial physics, halo statistics, baryonic conversion, black-hole growth and activity, and observational selection.

The aim is not to interpret unusually massive or luminous high-redshift sources as direct evidence for a failure of standard $\Lambda$CDM. Instead, JWST observations are treated as quantitative stress tests of the complete chain connecting early-Universe quantum fluctuations to observable cosmic structure. A robust inference of non-standard primordial physics would require a discrepancy that persists after uncertainties in halo abundance, stellar masses, star-formation efficiencies, black-hole masses, halo occupation, Eddington ratios, duty cycles, luminosity scatter, survey completeness, source classification, and cosmic variance are treated consistently.

The paper is organized as follows. Section~2 develops the elements of quantum field theory in curved Friedmann--Lema\^itre--Robertson--Walker spacetime required for the cosmological calculation, including field quantization, vacuum-state selection, Hadamard behavior, and the semiclassical framework. Section~3 develops inflationary vacuum fluctuations and the primordial curvature spectrum. Section~4 describes the gauge-invariant evolution of curvature perturbations into classical density fluctuations. Sections~5--7 construct the matter power spectrum, transfer functions, linear growth, mass variance, peak height, and spherical-collapse threshold. Section~8 develops the Press--Schechter and Sheth--Tormen halo mass functions and introduces the GUREFT high-redshift simulation-calibrated halo mass function. Sections~9--12 connect the resulting halo populations to massive galaxies, AGN, and rest-frame ultraviolet luminosity functions. Section~13 propagates controlled changes in the primordial spectrum through the numerical framework and examines their dependence on the halo-abundance prescription. Sections~14 and 15 interpret the results within the full quantum-to-structure chain and summarize the principal conclusions and limitations.

Throughout this work we adopt a spatially flat $\Lambda$CDM cosmology specified in Section~13, with parameters consistent with Planck measurement and employ natural units with $c=\hbar=1$ unless otherwise stated.

%
%

\section{Quantum Fields in Curved Spacetime}
\label{sec:qft_curved_spacetime}

The primordial perturbations used in structure-formation calculations are
ultimately statistical properties of quantum fields evolving in an expanding
spacetime. The purpose of this section is to establish only the elements of
quantum field theory in curved spacetime that are required for the
inflationary calculation that follows: field quantization, time-dependent
mode evolution, vacuum-state selection, and the ultraviolet regularity of
physical states.

The conceptual sequence is
\begin{equation}
\begin{split}
\text{quantized field}
&\longrightarrow \text{mode functions in FLRW}\\
&\longrightarrow \text{vacuum state}\\
&\longrightarrow \text{inflationary perturbations}.
\end{split}
\label{eq:sec2_conceptual_chain}
\end{equation}

\subsection{Scalar-field quantization}
\label{subsec:scalar_quantization}

A real scalar field in Minkowski spacetime is described by
\begin{equation}
S_\phi
=
-\frac{1}{2}
\int {\rm d}^4x
\left[
\eta^{\mu\nu}
\partial_\mu\phi\,
\partial_\nu\phi
+
m^2\phi^2
\right],
\label{eq:sec2_flat_action}
\end{equation}
with metric signature $(-,+,+,+)$. Variation gives
\begin{equation}
\left(
\Box-m^2
\right)\phi=0,
\qquad
\Box=\eta^{\mu\nu}\partial_\mu\partial_\nu .
\label{eq:sec2_flat_kg}
\end{equation}

Canonical quantization promotes $\phi$ and its conjugate momentum to
operators satisfying
\begin{equation}
\left[
\hat\phi({\bf x},t),
\hat\pi({\bf y},t)
\right]
=
i\hbar\,
\delta^3({\bf x}-{\bf y}),
\label{eq:sec2_canonical_comm}
\end{equation}
and the field operator is expanded in positive- and negative-frequency modes,
\begin{equation}
\begin{aligned}
\hat\phi({\bf x},t)
=
\int
\frac{{\rm d}^3k}{(2\pi)^{3/2}}
\frac{1}{\sqrt{2\omega_k}}
\left[
\hat a_{\bf k}
e^{i{\bf k}\cdot{\bf x}-i\omega_k t}
+
\hat a_{\bf k}^{\dagger}
e^{-i{\bf k}\cdot{\bf x}+i\omega_k t}
\right],
\end{aligned}
\label{eq:sec2_flat_modes}
\end{equation}
where
\begin{equation}
\omega_k=\sqrt{k^2+m^2}.
\label{eq:sec2_flat_frequency}
\end{equation}
The Minkowski vacuum is defined by
\begin{equation}
\hat a_{\bf k}|0\rangle=0.
\label{eq:sec2_minkowski_vacuum}
\end{equation}

In flat spacetime the existence of a global timelike Killing vector provides
a preferred notion of positive frequency and therefore a preferred vacuum
and particle interpretation \citep{Weinberg1995,PeskinSchroeder1995}. The
essential complication in cosmology is that an expanding spacetime is not
stationary.

\subsection{Quantum fields on a curved background}
\label{subsec:curved_scalar}

For a scalar field propagating on a classical curved spacetime,
\begin{equation}
S_\phi
=
-\frac{1}{2}
\int
{\rm d}^4x\,
\sqrt{-g}
\left[
g^{ab}\nabla_a\phi\nabla_b\phi
+
m^2\phi^2
+
\xi R\phi^2
\right],
\label{eq:sec2_curved_action}
\end{equation}
where $R$ is the Ricci scalar and $\xi$ controls the coupling to curvature.
The field equation is
\begin{equation}
\left(
\Box-m^2-\xi R
\right)\phi=0,
\label{eq:sec2_curved_kg}
\end{equation}
with
\begin{equation}
\Box\phi
=
\frac{1}{\sqrt{-g}}
\partial_a
\left(
\sqrt{-g}\,
g^{ab}\partial_b\phi
\right).
\label{eq:sec2_curved_box}
\end{equation}

The choices $\xi=0$ and $\xi=1/6$ correspond respectively to minimal and
conformal coupling for a massless scalar in four spacetime dimensions.

The semiclassical regime treats the spacetime geometry classically while
quantizing matter fields upon it. This is not a complete theory of quantum
gravity, but it is the appropriate framework for particle creation in
expanding universes and for the quantum-field-theoretic foundations of
inflationary perturbations
\citep{BirrellDavies1982,Fulling1989,Wald1994,ParkerToms2009}.

\subsection{Mode evolution in an FLRW universe}
\label{subsec:flrw_section2}

For a spatially flat FLRW spacetime,
\begin{equation}
{\rm d}s^2
=
-{\rm d}t^2
+
a^2(t)\,{\rm d}{\bf x}^2
=
a^2(\tau)
\left[
-{\rm d}\tau^2
+
{\rm d}{\bf x}^2
\right],
\label{eq:sec2_flrw_metric}
\end{equation}
where
\begin{equation}
{\rm d}\tau=\frac{{\rm d}t}{a(t)}.
\label{eq:sec2_conformal_time}
\end{equation}
The ordinary and conformal Hubble parameters are
\begin{equation}
H=\frac{\dot a}{a},
\qquad
\mathcal H=\frac{a'}{a}=aH,
\label{eq:sec2_hubble_parameters}
\end{equation}
with dots and primes denoting derivatives with respect to $t$ and $\tau$,
respectively.
Throughout this work, $H$ denotes the physical Hubble parameter and
$\mathcal H$ the conformal Hubble parameter. Thus horizon crossing
is expressed equivalently as $k=aH$ or $k=\mathcal H$ in units where $c=1$.

The scalar-field equation becomes
\begin{equation}
\phi''
+
2\mathcal H\phi'
-
\nabla^2\phi
+
a^2
\left(
m^2+\xi R
\right)\phi
=
0.
\label{eq:sec2_flrw_kg}
\end{equation}
After defining
\begin{equation}
\chi=a\phi,
\label{eq:sec2_rescaled_field}
\end{equation}
the Fourier modes satisfy
\begin{equation}
\chi_k''
+
\omega_k^2(\tau)\chi_k
=
0,
\label{eq:sec2_mode_oscillator}
\end{equation}
with
\begin{equation}
\omega_k^2(\tau)
=
k^2
+
a^2m^2
+
\left(
\xi-\frac{1}{6}
\right)a^2R.
\label{eq:sec2_effective_frequency}
\end{equation}

Equation~(\ref{eq:sec2_mode_oscillator}) is the central structural result:
each Fourier mode behaves as a harmonic oscillator with a time-dependent
frequency. The evolving background therefore changes the mode basis and
removes the unique global particle interpretation available in Minkowski
spacetime.

\subsection{Vacuum ambiguity and Bogoliubov transformations}
\label{subsec:vacuum_ambiguity}

The field operator may be expanded in any complete set of normalized mode
functions,
\begin{equation}
\hat\phi({\bf x},\tau)
=
\int
\frac{{\rm d}^3k}{(2\pi)^{3/2}}
\left[
\hat a_{\bf k}u_k(\tau)e^{i{\bf k}\cdot{\bf x}}
+
\hat a_{\bf k}^{\dagger}
u_k^*(\tau)e^{-i{\bf k}\cdot{\bf x}}
\right].
\label{eq:sec2_curved_mode_expansion}
\end{equation}
Two admissible mode bases can be related by
\begin{equation}
u_k
=
\alpha_k v_k
+
\beta_k v_k^*,
\qquad
|\alpha_k|^2-|\beta_k|^2=1.
\label{eq:sec2_bogoliubov}
\end{equation}
A state that is vacuum with respect to one basis generally contains particles
with respect to another, with occupation number
\begin{equation}
N_k=|\beta_k|^2.
\label{eq:sec2_particle_number}
\end{equation}

The absence of a unique particle definition in a nonstationary geometry is a
physical feature of curved-spacetime QFT. It also makes the selection of an
acceptable cosmological quantum state nontrivial. Deutsch and Najmi
emphasized the distinction between state selection and particle
interpretation in nonstationary backgrounds \citep{DeutschNajmi1983}, while
Najmi and Ottewill showed that simple instantaneous energy-minimization
prescriptions need not produce states with acceptable ultraviolet behaviour
\citep{NajmiOttewill1984,NajmiOttewill1985}.

\subsection{Hadamard states and semiclassical consistency}
\label{subsec:hadamard_states}

Physical states in curved spacetime must possess the universal short-distance
singularity structure of the Minkowski vacuum. This requirement is encoded
by the Hadamard condition \citep{BirrellDavies1982,Wald1994,ParkerToms2009}.
Schematically,
\begin{equation}
G^{(1)}(x,x')
=
\frac{U(x,x')}{\sigma(x,x')}
+
V(x,x')
\ln|\sigma(x,x')|
+
W(x,x'),
\label{eq:sec2_hadamard_form}
\end{equation}
Here $\sigma(x,x')$ is Synge's world function (one-half the squared geodesic distance within a normal neighbourhood), not the cosmological mass variance. The biscalar $U(x,x')$ sets the leading geometric singularity, $V(x,x')$ controls the logarithmic term through the field equation, and the smooth remainder $W(x,x')$ contains state-dependent information. This expression suppresses conventional normalization constants and the associated length scale inside the logarithm; it is intended to display the local Hadamard structure rather than a fully normalized parametrix.

The Hadamard structure permits a finite renormalized expectation value
\begin{equation}
\left\langle
\hat T_{ab}
\right\rangle_{\rm ren},
\label{eq:sec2_renormalized_stress}
\end{equation}
which enters the semiclassical Einstein equation
\begin{equation}
G_{ab}
+
\Lambda g_{ab}
=
8\pi G
\left\langle
\hat T_{ab}
\right\rangle_{\rm ren}.
\label{eq:sec2_semiclassical_einstein}
\end{equation}

The Hadamard condition identifies the short-distance singularity structure
required of physically admissible quantum states in curved spacetime and
provides the framework for defining renormalized local observables. Its role
here is foundational rather than a direct ingredient of the halo-abundance
calculation: the later numerical analysis does not evaluate the Hadamard
coefficients $U$, $V$, and $W$ or the renormalized stress--energy tensor
$\langle\hat T_{ab}\rangle_{\rm ren}$. Once the inflationary quantum state
has specified the two-point function of the gauge-invariant curvature
perturbation and hence its primordial spectrum
$\mathcal{P}_{\mathcal R}(k)$, the subsequent structure-formation calculation
proceeds through transfer functions, linear growth, mass variance, and halo
statistics. The short-wavelength Bunch--Davies prescription used in the next
section is motivated by local Minkowski-like vacuum behavior; this motivation
should not be confused with an independent proof of the full Hadamard
property for an arbitrary state.

\subsection{Connection to inflationary perturbations}
\label{subsec:sec2_to_inflation}

The essential lessons of curved-spacetime QFT are therefore compact:

\begin{enumerate}
\item quantum fields propagate on an evolving FLRW background;
\item their Fourier modes behave as time-dependent oscillators;
\item the vacuum state must be selected physically rather than by a unique
global particle prescription;
\item acceptable states require the correct ultraviolet singularity structure.
\end{enumerate}

Inflation applies these principles not to an independent spectator field but
to the coupled inflaton--metric perturbation. Its canonically normalized
gauge-invariant degree of freedom is the Mukhanov--Sasaki variable, whose
quantization generates the primordial curvature spectrum.

Thus the role of this section is summarized by
\begin{equation}
\boxed{
\begin{aligned}
\text{curved-spacetime QFT}
&\longrightarrow
\text{physical quantum state}
\\
&\longrightarrow
\text{time-dependent modes}
\\
&\longrightarrow
\text{inflationary perturbations}.
\end{aligned}
}
\label{eq:sec2_summary_chain}
\end{equation}
Section~\ref{sec:inflation_perturbations} develops this final step explicitly.

%
%

\section{Inflationary Vacuum Fluctuations and the Primordial Curvature Spectrum}
\label{sec:inflation_perturbations}

Section~\ref{sec:qft_curved_spacetime} established the curved-spacetime
quantum-field framework needed to describe time-dependent mode evolution and
vacuum-state selection. Inflation applies that framework to the coupled
inflaton--metric system and provides the mechanism by which microscopic
quantum fluctuations generate the primordial curvature perturbations that
seed later structure.

The relevant sequence is
\begin{equation}
\hat{\delta\phi}
\longrightarrow
\mathcal R
\longrightarrow
\mathcal P_{\mathcal R}(k),
\label{eq:sec3_chain}
\end{equation}
where $\mathcal R$ is the comoving curvature perturbation and
$\mathcal P_{\mathcal R}(k)$ is its dimensionless power spectrum.

\subsection{Inflationary background}
\label{subsec:inflation_background}

Inflation is a period of accelerated expansion,
\begin{equation}
\ddot a>0,
\label{eq:sec3_acceleration}
\end{equation}
during which the comoving Hubble radius decreases,
\begin{equation}
\frac{{\rm d}}{{\rm d}t}
\left(
\frac{1}{aH}
\right)<0.
\label{eq:sec3_comoving_hubble}
\end{equation}
A mode that initially satisfies $k\gg aH$ can therefore be stretched across
the Hubble radius, $k\simeq aH$, during inflation and re-enter at a later
epoch. This provides the basic mechanism by which fluctuations generated on
microscopic scales become relevant to cosmological structure
\citep{Guth1981,LiddleLyth2000,Mukhanov2005,Weinberg2008}.

For a canonical inflaton $\phi$ with potential $V(\phi)$,
\begin{equation}
S_\phi
=
\int {\rm d}^4x\,\sqrt{-g}
\left[
-\frac{1}{2}
g^{ab}\nabla_a\phi\nabla_b\phi
-
V(\phi)
\right].
\label{eq:sec3_inflaton_action}
\end{equation}
For a homogeneous background $\phi_0(t)$,
\begin{equation}
H^2
=
\frac{1}{3M_{\rm Pl}^2}
\left[
\frac{1}{2}\dot\phi_0^{\,2}
+
V(\phi_0)
\right],
\label{eq:sec3_friedmann}
\end{equation}
and
\begin{equation}
\ddot\phi_0
+
3H\dot\phi_0
+
V_{,\phi}
=
0,
\label{eq:sec3_background_field}
\end{equation}
where $M_{\rm Pl}$ denotes the reduced Planck mass. We distinguish it
from the unreduced Planck mass $M_{\rm P}$ by restoring $\hbar$ and $c$:
\begin{equation}
\begin{gathered}
M_{\rm P}=\left(\frac{\hbar c}{G}\right)^{1/2},\\
M_{\rm Pl}=\left(\frac{\hbar c}{8\pi G}\right)^{1/2}
=\frac{M_{\rm P}}{\sqrt{8\pi}}.
\end{gathered}
\label{eq:sec3_reduced_planck}
\end{equation}
In the natural units $c=\hbar=1$ adopted throughout this work,
$M_{\rm Pl}=(8\pi G)^{-1/2}$. All subsequent inflationary
and slow-roll expressions use this reduced Planck-mass convention.

In the slow-roll regime,
\begin{equation}
H^2
\simeq
\frac{V}{3M_{\rm Pl}^2},
\qquad
3H\dot\phi_0
\simeq
-V_{,\phi},
\label{eq:sec3_slowroll_background}
\end{equation}
with potential slow-roll parameters
\begin{equation}
\epsilon_V
=
\frac{M_{\rm Pl}^2}{2}
\left(
\frac{V_{,\phi}}{V}
\right)^2,
\qquad
\eta_V
=
M_{\rm Pl}^2
\frac{V_{,\phi\phi}}{V}.
\label{eq:sec3_slowroll_parameters}
\end{equation}
Slow roll requires $\epsilon_V\ll1$ and $|\eta_V|\ll1$.

\subsection{Gauge-invariant scalar perturbations}
\label{subsec:gauge_invariant_inflation}

The inflaton is perturbed as
\begin{equation}
\phi({\bf x},t)
=
\phi_0(t)
+
\delta\phi({\bf x},t).
\label{eq:sec3_phi_split}
\end{equation}
Because $\delta\phi$ is coupled to scalar metric perturbations, the physical
degree of freedom must be expressed in a gauge-invariant form
\citep{Bardeen1980,Mukhanov1992,Weinberg2008}.

For a single canonical scalar field, the comoving curvature perturbation
$\mathcal R$ is related to the inflaton fluctuation in spatially flat gauge by
\begin{equation}
\mathcal R
=
-\frac{H}{\dot\phi_0}
\delta\phi_{\rm flat},
\label{eq:sec3_R_deltaphi}
\end{equation}
up to the overall sign convention adopted for scalar metric perturbations. Although $\delta\phi$ alone changes under a first-order time-slicing transformation, $\delta\phi_{\rm flat}$ denotes the inflaton perturbation evaluated on spatially flat slices; the corresponding combination with the scalar spatial-curvature perturbation is gauge invariant at linear order. Thus $\mathcal R$ is a gauge-invariant variable, not an assertion that the unqualified $\delta\phi$ is gauge invariant.

The canonically normalized Mukhanov--Sasaki variable is
\begin{equation}
v
=
z\mathcal R,
\label{eq:sec3_v_def}
\end{equation}
with
\begin{equation}
z
=
a\frac{\dot\phi_0}{H}
=
a\sqrt{2\epsilon}\,M_{\rm Pl},
\qquad
\epsilon
=
-\frac{\dot H}{H^2}.
\label{eq:sec3_z_def}
\end{equation}

Expanding the coupled inflaton--gravity system to second order gives
\begin{equation}
S^{(2)}
=
\frac{1}{2}
\int
{\rm d}\tau\,{\rm d}^3x
\left[
(v')^2
-
(\nabla v)^2
+
\frac{z''}{z}v^2
\right].
\label{eq:sec3_second_order_action}
\end{equation}
The Fourier modes therefore satisfy
\begin{equation}
v_k''
+
\left(
k^2
-
\frac{z''}{z}
\right)
v_k
=
0.
\label{eq:sec3_MS_equation}
\end{equation}
This is the Mukhanov--Sasaki equation and is the central equation governing
the quantum generation of scalar cosmological perturbations.

\subsection{Quantization and the Bunch--Davies state}
\label{subsec:BD_sec3}

The canonical field is quantized through
\begin{equation}
\hat v({\bf x},\tau)
=
\int
\frac{{\rm d}^3k}{(2\pi)^{3/2}}
\left[
v_k(\tau)
\hat a_{\bf k}
e^{i{\bf k}\cdot{\bf x}}
+
v_k^*(\tau)
\hat a_{\bf k}^{\dagger}
e^{-i{\bf k}\cdot{\bf x}}
\right],
\label{eq:sec3_operator_expansion}
\end{equation}
with
\begin{equation}
\left[
\hat a_{\bf k},
\hat a_{{\bf k}'}^\dagger
\right]
=
\delta^3({\bf k}-{\bf k}').
\label{eq:sec3_commutator}
\end{equation}

Deep inside the Hubble radius, $k\gg aH$, spacetime curvature is negligible
over one wavelength and the physical mode should approach the local
Minkowski positive-frequency solution. This defines the Bunch--Davies
initial condition,
\begin{equation}
v_k(\tau)
\longrightarrow
\frac{1}{\sqrt{2k}}
e^{-ik\tau},
\qquad
k\gg aH.
\label{eq:sec3_BD}
\end{equation}
The choice is therefore the inflationary realization of the short-distance
vacuum condition discussed in Section~\ref{sec:qft_curved_spacetime}
\citep{BirrellDavies1982,Wald1994,ParkerToms2009}.

\subsection{Horizon exit and freeze-out}
\label{subsec:freezeout_sec3}

For quasi-de Sitter expansion,
\begin{equation}
a(\tau)
\simeq
-\frac{1}{H\tau},
\qquad
\tau<0,
\label{eq:sec3_desitter}
\end{equation}
and, to leading order in slow roll,
\begin{equation}
\frac{z''}{z}
\simeq
\frac{2}{\tau^2}.
\label{eq:sec3_zpp}
\end{equation}
The Mukhanov--Sasaki equation then has the Bunch--Davies solution
\begin{equation}
v_k(\tau)
=
\frac{1}{\sqrt{2k}}
\left(
1-\frac{i}{k\tau}
\right)
e^{-ik\tau}.
\label{eq:sec3_BD_solution}
\end{equation}

For $|k\tau|\gg1$, the mode oscillates as a vacuum fluctuation. After Hubble
exit, $|k\tau|\ll1$, the comoving curvature perturbation
\begin{equation}
\mathcal R_k
=
\frac{v_k}{z}
\label{eq:sec3_R_from_v}
\end{equation}
approaches a constant for adiabatic single-field inflation. This conservation
of $\mathcal R$ allows the inflationary perturbation spectrum to be carried
forward as the initial condition for the later radiation- and matter-era
evolution.

The resulting super-Hubble state is highly squeezed, and its correlation
functions may be represented by an effectively classical stochastic field
for the subsequent linear evolution. Thus the classical primordial spectrum
used in structure formation retains the statistical imprint of a quantum
two-point function.

\subsection{Primordial curvature spectrum}
\label{subsec:primordial_spectrum_sec3}

The dimensionless curvature power spectrum is defined through
\begin{equation}
\left\langle
\hat{\mathcal R}_{\bf k}
\hat{\mathcal R}_{{\bf k}'}
\right\rangle
=
(2\pi)^3
\delta^3({\bf k}+{\bf k}')
\frac{2\pi^2}{k^3}
\mathcal P_{\mathcal R}(k).
\label{eq:sec3_two_point}
\end{equation}

Throughout this work, $\mathcal P_{\mathcal R}(k)$ denotes the
\emph{dimensionless} primordial curvature spectrum, while
$P_{\mathcal R}^{\mathrm{dim}}(k)$ denotes its dimensional counterpart
(with units of comoving volume). The dimensional matter power spectrum
$P_m(k,z)$ also has units of comoving volume. This distinction is maintained throughout the subsequent transfer-function and structure-formation calculations.

Equivalently,
\begin{equation}
\mathcal P_{\mathcal R}(k)
=
\frac{k^3}{2\pi^2}
|\mathcal R_k|^2.
\label{eq:sec3_PR_definition}
\end{equation}

Evaluating the slowly varying background quantities at the time $t_*(k)$ when each comoving mode exits the Hubble radius during inflation, $k=a(t_*)H(t_*)$, gives
\begin{equation}
\mathcal P_{\mathcal R}(k)
\simeq
\frac{1}{8\pi^2M_{\rm Pl}^2}
\frac{H_*^2}{\epsilon_*}.
\label{eq:sec3_PR_slowroll}
\end{equation}
Here a star denotes evaluation at that mode's horizon-exit time, rather than at a single universal cosmic time. Using the slow-roll background equations,
\begin{equation}
\mathcal P_{\mathcal R}(k)
\simeq
\frac{1}{24\pi^2M_{\rm Pl}^4}
\frac{V_*}{\epsilon_{V,*}}.
\label{eq:sec3_PR_potential}
\end{equation}

Around a pivot scale $k_*$, the spectrum is written
\begin{equation}
\mathcal P_{\mathcal R}(k)
=
A_s
\left(
\frac{k}{k_*}
\right)^{n_s-1},
\label{eq:sec3_powerlaw}
\end{equation}
where
\begin{equation}
n_s-1
=
\frac{{\rm d}\ln\mathcal P_{\mathcal R}}
{{\rm d}\ln k}
\simeq
-6\epsilon_V+2\eta_V
\label{eq:sec3_ns}
\end{equation}
to leading order in slow roll. CMB observations find a primordial amplitude
of order $A_s\sim10^{-9}$ and a slightly red spectrum with $n_s<1$
\citep{Planck2018}.

These quantities are the observationally constrained imprint of the quantum
inflaton--metric system and form the primordial input to the structure
calculation.

\subsection{Interface with structure formation}
\label{subsec:sec3_structure_interface}

The result of the inflationary calculation is therefore
\begin{equation}
\boxed{
\hat{\delta\phi}
\longrightarrow
\mathcal R
\longrightarrow
\mathcal P_{\mathcal R}(k)
}.
\label{eq:sec3_final_chain}
\end{equation}

The gauge-invariant curvature perturbation $\mathcal R$ supplies the inflationary initial condition. Section~\ref{sec:gauge_invariant_perturbations} describes its relativistic evolution using gauge-invariant Bardeen potentials, which coincide with the metric potentials in conformal Newtonian gauge. The later sub-Hubble Newtonian density evolution is a controlled approximation within this framework, not a replacement for gauge invariance.

The primordial curvature spectrum is not yet the late-time matter power
spectrum. Its conversion into the density field requires the classical
relativistic evolution of scalar perturbations and the subsequent
scale-dependent transfer through radiation domination and
matter--radiation equality. Section~\ref{sec:gauge_invariant_perturbations}
develops that next step.

%
%

\section{Gauge-Invariant Cosmological Perturbation Theory}
\label{sec:gauge_invariant_perturbations}

The primordial curvature spectrum derived in Section~\ref{sec:inflation_perturbations}
provides the statistical initial condition for structure formation. To connect
that spectrum to the matter density field, the perturbations must be evolved
through the relativistic equations governing the metric and cosmic fluids.
Because general relativity contains coordinate freedom, this evolution is most
cleanly formulated in terms of gauge-invariant perturbation variables
\citep{Bardeen1980,Mukhanov1992,MaBertschinger1995,Weinberg2008}.

The role of this section is therefore limited and specific: to establish the
classical relativistic bridge between the primordial curvature perturbation
$\mathcal R$ and the density contrast $\delta$ that later enters the matter
power spectrum. Detailed transfer through radiation domination,
matter--radiation equality, and subsequent linear growth is developed in the
following sections.

\subsection{Perturbations of an FLRW spacetime}
\label{subsec:perturbed_flrw}

The unperturbed spatially flat FLRW metric is
\begin{equation}
{\rm d}s^2
=
a^2(\tau)
\left[
-{\rm d}\tau^2
+
\delta_{ij}{\rm d}x^i{\rm d}x^j
\right],
\label{eq:sec4_background_metric}
\end{equation}
with conformal Hubble parameter
\begin{equation}
\mathcal H
=
\frac{a'}{a}
=
aH.
\label{eq:sec4_conformal_hubble}
\end{equation}
A prime denotes differentiation with respect to conformal time $\tau$.

Because the background is homogeneous and isotropic, linear perturbations
separate into scalar, vector, and tensor sectors. The formation of dark matter
halos is controlled primarily by scalar perturbations, and only that sector is
needed here. A general scalar perturbation of the metric contains several
functions whose individual values depend on the choice of spacetime
coordinates. The resulting coordinate ambiguity is the gauge problem of
cosmological perturbation theory.

\subsection{Gauge freedom and physical perturbations}
\label{subsec:gauge_freedom_sec4}

Consider an infinitesimal coordinate transformation
\begin{equation}
x^\mu
\rightarrow
\widetilde{x}^{\mu}
=
x^\mu+\xi^\mu.
\label{eq:sec4_gauge_transform}
\end{equation}
For scalar perturbations one may write
\begin{equation}
\xi^0=T,
\qquad
\xi^i=\partial^iL,
\label{eq:sec4_scalar_generator}
\end{equation}
where $T$ and $L$ are scalar functions.

A perturbation of any background scalar quantity changes under the time shift.
For example,
\begin{equation}
\delta\rho
\rightarrow
\delta\rho-\bar\rho' T,
\label{eq:sec4_density_gauge_transform}
\end{equation}
and similarly the inflaton and metric perturbations change under the same
coordinate transformation. A nonzero value of $\delta\rho$ or
$\delta\phi$ can therefore contain both physical information and a change in
the choice of time slicing.

The observable content of the perturbation must consequently be expressed
either in a fully specified gauge or in combinations that are invariant under
Eq.~(\ref{eq:sec4_gauge_transform}). This is particularly important for the
present calculation because the curvature perturbation generated during
inflation must be matched consistently onto the post-inflationary density
field.

\subsection{Bardeen potentials and Newtonian gauge}
\label{subsec:bardeen_sec4}

A convenient gauge-invariant description of scalar metric perturbations is
provided by the Bardeen potentials $\Phi$ and $\Psi$ \citep{Bardeen1980}.
In conformal Newtonian gauge the perturbed line element is
\begin{equation}
{\rm d}s^2
=
a^2(\tau)
\left[
-(1+2\Psi){\rm d}\tau^2
+
(1-2\Phi)\delta_{ij}{\rm d}x^i{\rm d}x^j
\right].
\label{eq:sec4_newtonian_metric}
\end{equation}
In this gauge the metric perturbations coincide directly with the
corresponding gauge-invariant Bardeen potentials.

When anisotropic stress is negligible,
\begin{equation}
\Phi=\Psi.
\label{eq:sec4_equal_potentials}
\end{equation}
This approximation is appropriate for pressureless matter on many of the
scales and epochs relevant to the later halo calculation, although radiation
and free-streaming species can generate anisotropic stress at earlier times.

On sub-Hubble scales $\Psi$ plays the role of the Newtonian gravitational
potential governing nonrelativistic motion, while $\Phi+\Psi$ controls
relativistic effects such as gravitational lensing. The important point for
this paper is that these potentials inherit their initial statistical
distribution from the primordial curvature perturbation rather than being
introduced independently.

\subsection{Density and velocity perturbations}
\label{subsec:density_velocity_sec4}

The matter density is decomposed as
\begin{equation}
\rho({\bf x},\tau)
=
\bar\rho(\tau)
+
\delta\rho({\bf x},\tau),
\label{eq:sec4_density_decomposition}
\end{equation}
and the density contrast is
\begin{equation}
\delta({\bf x},\tau)
=
\frac{\delta\rho({\bf x},\tau)}
{\bar\rho(\tau)}.
\label{eq:sec4_density_contrast}
\end{equation}
Linear perturbation theory applies while
\begin{equation}
|\delta|\ll1.
\label{eq:sec4_linear_condition}
\end{equation}

For a pressureless fluid, the Fourier-space continuity and Euler equations in
Newtonian gauge take the form
\begin{equation}
\delta'
=
-\theta
+
3\Phi',
\label{eq:sec4_continuity}
\end{equation}
and
\begin{equation}
\theta'
+
\mathcal H\theta
=
k^2\Psi,
\label{eq:sec4_euler}
\end{equation}
where $\theta$ is the velocity-divergence variable. These equations describe
the response of matter to the metric perturbations.

At linear order each Fourier mode evolves independently. This property is
essential because it permits the primordial two-point function derived during
inflation to be propagated mode by mode through the later cosmological
evolution.

\subsection{The comoving curvature perturbation}
\label{subsec:comoving_curvature_sec4}

The comoving curvature perturbation $\mathcal R$ is the natural quantity
linking inflation to the post-inflationary Universe. One convenient
gauge-invariant form is
\begin{equation}
\mathcal R
=
\Phi
-
\frac{\mathcal H}
{\bar\rho+\bar p}
\,\delta q,
\label{eq:sec4_R_general}
\end{equation}
where $\delta q$ is the scalar momentum potential. Sign conventions for
$\mathcal R$ and the metric potentials vary in the literature; the convention
used here is chosen consistently with Section~\ref{sec:inflation_perturbations}.

For adiabatic perturbations on scales much larger than the Hubble radius,
\begin{equation}
\mathcal R'
\simeq0.
\label{eq:sec4_R_conservation}
\end{equation}
This conservation law is one of the most important simplifications in the
quantum-to-structure connection. It allows the primordial curvature spectrum
generated during inflation to survive through reheating and to provide the
initial condition for radiation- and matter-era evolution.

In simple single-field inflation the perturbations are predominantly
adiabatic, so a single primordial field $\mathcal R({\bf k})$ specifies the
initial scalar perturbations of the coupled cosmological components. Possible
isocurvature modes are not included in the baseline calculation.

\subsection{Linearized Einstein equations}
\label{subsec:linear_einstein_sec4}

The perturbed Einstein equations relate the metric potentials to the matter
perturbations. In Newtonian gauge, the time--time equation may be written
\begin{equation}
k^2\Phi
+
3\mathcal H
\left(
\Phi'
+
\mathcal H\Psi
\right)
=
-4\pi G a^2\delta\rho,
\label{eq:sec4_einstein_00}
\end{equation}
for the Fourier convention adopted here. The overall sign of the Fourier
Poisson term depends on the convention used for the spatial transform; all
subsequent equations are used consistently within the same convention.

On scales well inside the Hubble radius,
\begin{equation}
k\gg\mathcal H,
\label{eq:sec4_subhubble}
\end{equation}
the time-derivative terms become subdominant and the relativistic constraint
reduces to the familiar Poisson relation,
\begin{equation}
k^2\Phi
\simeq
-4\pi G a^2\bar\rho\,\delta.
\label{eq:sec4_poisson_fourier}
\end{equation}
Thus the Newtonian description of gravitational instability emerges as the
sub-Hubble limit of the relativistic perturbation equations.

Combining the continuity, Euler, and Poisson equations for pressureless matter
gives the familiar linear growth equation in cosmic time,
\begin{equation}
\ddot\delta
+
2H\dot\delta
-
4\pi G\bar\rho_m\delta
=
0.
\label{eq:sec4_growth_equation}
\end{equation}
The detailed solution and growth factor are developed in the section devoted
to linear structure growth.

\subsection{From primordial curvature to the density field}
\label{subsec:R_to_density_sec4}

The essential output of relativistic perturbation theory is a linear relation
between the primordial curvature perturbation and the later density contrast.
For adiabatic initial conditions one may write
\begin{equation}
\delta({\bf k},z)
=
\mathcal M(k,z)\,
\mathcal R({\bf k}),
\label{eq:sec4_delta_from_R}
\end{equation}
where $\mathcal M(k,z)$ contains the post-inflationary transfer through the
radiation era, matter--radiation equality, and subsequent linear growth.

The corresponding dimensional matter power spectrum therefore satisfies
\begin{equation}
P_m(k,z)
=
\mathcal M^2(k,z)\,
P_{\mathcal R}^{\mathrm{dim}}(k),
\label{eq:sec4_Pm_from_PR}
\end{equation}
where the dimensional curvature spectrum is defined in terms of the dimensionless primordial spectrum by
\begin{equation}
P_{\mathcal R}^{\mathrm{dim}}(k)
=
\frac{2\pi^2}{k^3}
\mathcal P_{\mathcal R}(k).
\label{eq:sec4_dimensional_PR}
\end{equation}

Equations~(\ref{eq:sec4_delta_from_R})--(\ref{eq:sec4_dimensional_PR})
provide the statistical bridge between the inflationary calculation and the
matter field used in structure formation. They also make clear why the
primordial amplitude and spectral tilt alone do not determine halo
abundances: the curvature perturbations must first be processed by the
post-inflationary expansion history.

The next section develops that processing explicitly through Hubble-radius
entry, matter--radiation equality, transfer functions, and linear growth. The
result will be the matter power spectrum needed to compute the mass variance
$\sigma(M,z)$ and, ultimately, the halo mass function.

%
%

\section{Transfer Functions, Horizon Entry, and the Matter Power Spectrum}
\label{sec:transfer_growth}

The primordial curvature spectrum $\mathcal P_{\mathcal R}(k)$ specifies the
initial statistics of scalar perturbations, but it is not itself the matter
power spectrum used in the halo calculation. After inflation, individual
Fourier modes evolve differently according to when they enter the Hubble
radius and which component dominates the cosmic energy density. Radiation
domination, matter--radiation equality, baryonic physics, and the subsequent
growth of matter perturbations therefore reshape the primordial spectrum
before nonlinear collapse begins
\citep{MaBertschinger1995,Peacock1999,EisensteinHu1998,Dodelson2003,Weinberg2008}.

The purpose of this section is to construct that bridge:
\begin{equation}
\mathcal P_{\mathcal R}(k)
\longrightarrow
T(k)
\longrightarrow
P_m(k,z),
\label{eq:sec5_bridge}
\end{equation}
where $T(k)$ contains the scale-dependent processing of perturbations and
$D(z)$ describes their subsequent linear growth.

\subsection{Hubble-radius entry and scale-dependent evolution}
\label{subsec:horizon_entry_sec5}

A Fourier mode of comoving wavenumber $k$ has physical wavelength
\begin{equation}
\lambda_{\rm phys}
=
\frac{2\pi a}{k}.
\label{eq:sec5_physical_wavelength}
\end{equation}
The mode enters the Hubble radius when
\begin{equation}
k
\simeq
aH
=
\mathcal H.
\label{eq:sec5_horizon_entry}
\end{equation}

Modes with sufficiently small $k$ remain outside the Hubble radius until
matter domination. Their primordial amplitudes are therefore preserved to a
good approximation until they begin ordinary gravitational growth. Modes with
larger $k$ enter during radiation domination. Because the dominant radiation
component does not cluster efficiently on sub-Hubble scales in the same way
as pressureless matter, the growth of cold-dark-matter perturbations is
suppressed relative to modes entering later.

The transition between these two behaviours is set primarily by
matter--radiation equality. Defining
\begin{equation}
1+z_{\rm eq}
=
\frac{\Omega_m}{\Omega_r},
\label{eq:sec5_zeq}
\end{equation}
the characteristic equality wavenumber is
\begin{equation}
k_{\rm eq}
=
a_{\rm eq}H_{\rm eq}.
\label{eq:sec5_keq}
\end{equation}
For standard cosmological parameters, $k_{\rm eq}$ is of order
$10^{-2}\ {\rm Mpc}^{-1}$. Halo-scale modes relevant to galaxies and AGN
typically satisfy $k\gg k_{\rm eq}$ and therefore entered the Hubble radius
during radiation domination. The suppression of these modes is a central
reason that the late-time matter spectrum differs strongly from the nearly
scale-invariant primordial curvature spectrum.

\subsection{The matter transfer function}
\label{subsec:transfer_function_sec5}

The scale-dependent processing of the perturbations is described by the
matter transfer function $T(k)$. It is conventionally normalized so that
\begin{equation}
T(k)\rightarrow1
\qquad
\text{as}
\qquad
k\rightarrow0.
\label{eq:sec5_T_large_scales}
\end{equation}
Thus large-scale modes retain the primordial spectral shape, whereas modes
that entered during radiation domination are increasingly suppressed.

For cold dark matter the qualitative asymptotic behaviour is
\begin{equation}
T(k)
\simeq
\begin{cases}
1, & k\ll k_{\rm eq},\\[0.4em]
\dfrac{\ln(k/k_{\rm eq})}{(k/k_{\rm eq})^2},
& k\gg k_{\rm eq},
\end{cases}
\label{eq:sec5_T_asymptotic}
\end{equation}
up to factors of order unity and the details of the baryonic component.

The numerical calculations in this paper use the Eisenstein--Hu transfer
function \citep{EisensteinHu1998}. This prescription provides an analytic
representation of the linear transfer of matter perturbations and permits the
same primordial spectrum to be propagated consistently to the halo-relevant
scales. Where a smooth broadband spectrum is required, the corresponding
no-wiggle form is sufficient; baryon-acoustic structure can be retained when
needed without changing the basic halo-abundance framework.

\subsection{From primordial curvature to matter density}
\label{subsec:curvature_to_matter_sec5}

Section~\ref{sec:gauge_invariant_perturbations} introduced the linear relation
\begin{equation}
\delta_m({\bf k},z)
=
\mathcal M(k,z)\,
\mathcal R({\bf k}).
\label{eq:sec5_delta_MR}
\end{equation}
For adiabatic perturbations in a spatially flat $\Lambda$CDM cosmology,
using the growth factor normalized by $D(0)=1$, the curvature-to-density
transfer function is
\begin{equation}
\mathcal M(k,z)
=
\frac{2}{5}
\frac{k^2T(k)g_0D(z)}
{\Omega_m H_0^2},
\label{eq:sec5_M_transfer}
\end{equation}
where $g_0=D_+(1)$ is the present-day growth-suppression factor relative
to the matter-dominated normalization, with $g_0\simeq 0.78827$ for the
adopted flat cosmology. The factor $2/5$ reflects the conversion between
the primordial comoving curvature perturbation and the matter-era
gravitational potential. Here $H_0$ is expressed in units with $c=1$.

The dimensional curvature power spectrum is related to the dimensionless
spectrum by
\begin{equation}
P_{\mathcal R}^{\mathrm{dim}}(k)
=
\frac{2\pi^2}{k^3}
\mathcal P_{\mathcal R}(k).
\label{eq:sec5_PR_dimensional}
\end{equation}
The dimensional matter power spectrum then follows as
\begin{equation}
P_m(k,z)
=
\mathcal M^2(k,z)
P_{\mathcal R}^{\mathrm{dim}}(k).
\label{eq:sec5_Pm_exact}
\end{equation}

Substituting the inflationary parameterization
\begin{equation}
\mathcal P_{\mathcal R}(k)
=
A_s
\left(
\frac{k}{k_*}
\right)^{n_s-1}
\label{eq:sec5_PR_parameterized}
\end{equation}
shows explicitly how the quantum-generated amplitude and tilt are carried
into the matter spectrum.

For the numerical halo calculation it is convenient to write the same result
in the equivalent shape form
\begin{equation}
P_m(k,z)
=
A\,k^{n_s}
T^2(k)
D^2(z),
\label{eq:sec5_Pm_shape}
\end{equation}
where the constant $A$ contains the dimensional normalization implied by the
primordial spectrum, the background cosmology, and the adopted growth
normalization. In the numerical implementation $A$ is fixed so that the
present-day matter spectrum reproduces the adopted value of $\sigma_8$.
Equation~(\ref{eq:sec5_Pm_shape}) should therefore be understood as the
computational form of Eq.~(\ref{eq:sec5_Pm_exact}), not as an independent
primordial-spectrum definition.

\subsection{Linear growth after matter domination}
\label{subsec:growth_sec5}

Once pressureless matter dominates and a mode is well inside the Hubble
radius, its linear evolution is governed by
\begin{equation}
\ddot\delta_m
+
2H\dot\delta_m
-
4\pi G\bar\rho_m\delta_m
=
0.
\label{eq:sec5_growth_equation}
\end{equation}
The growing solution defines the linear growth factor,
\begin{equation}
\delta_m({\bf k},z)
=
D(z)\,
\delta_m({\bf k},z_{\rm ref}),
\label{eq:sec5_growth_definition}
\end{equation}
with the normalization chosen so that
\begin{equation}
D(0)=1
\label{eq:sec5_D0}
\end{equation}
for the numerical calculations used later in the paper.

For a spatially flat matter--$\Lambda$ cosmology, define the
dimensionless expansion rate
\begin{equation}
E(a)
\equiv
\frac{H(a)}{H_0}
=
\left[
\Omega_m a^{-3}
+
\Omega_\Lambda
\right]^{1/2},
\label{eq:sec5_E_flatlcdm}
\end{equation}
when radiation is negligible. The unnormalized growing mode may then
be written as
\begin{equation}
D_+(a)
=
\frac{5\Omega_m}{2}
E(a)
\int_0^a
\frac{{\rm d}a'}
{a'^3 E^3(a')}.
\label{eq:sec5_growth_integral_raw}
\end{equation}
For the numerical calculations, the linear growth factor is normalized
to unity at the present epoch,
\begin{equation}
D(a)
=
\frac{D_+(a)}
{D_+(1)},
\qquad
D(1)=1.
\label{eq:sec5_growth_normalized}
\end{equation}

The unnormalized growing-mode solution satisfies $D_+(a)\rightarrow a$
in the early matter-dominated limit. Its present-day value
$g_0=D_+(1)$ accounts for the suppression of growth by dark energy;
the numerical calculations use $D(a)=D_+(a)/g_0$.

At high redshift in the matter-dominated limit,
\begin{equation}
D(a)\propto a
\qquad\Longrightarrow\qquad
D(z)\propto\frac{1}{1+z}.
\label{eq:sec5_matter_growth}
\end{equation}
The full $\Lambda$CDM growth factor, rather than the pure
$1/(1+z)$ approximation, is used when evaluating the matter spectrum. 

\subsection{Normalization by \texorpdfstring{$\sigma_8$}{sigma8}}
\label{subsec:sigma8_norm_sec5}

The amplitude of the linear matter spectrum is conventionally characterized
by the present-day rms density fluctuation in spheres of radius
$8\,h^{-1}{\rm Mpc}$,
\begin{equation}
\sigma_8^2
=
\frac{1}{2\pi^2}
\int_0^\infty
k^2P_m(k,0)
W^2(kR_8)\,
{\rm d}k,
\label{eq:sec5_sigma8}
\end{equation}
where
\begin{equation}
R_8
=
8\,h^{-1}{\rm Mpc},
\label{eq:sec5_R8}
\end{equation}
and
\begin{equation}
W(x)
=
3\,
\frac{\sin x-x\cos x}{x^3}
\label{eq:sec5_tophat}
\end{equation}
is the Fourier transform of a real-space spherical top-hat window.

The calculations adopt a Planck-consistent normalization with
$\sigma_8\simeq0.812$. Fixing the amplitude in this manner ensures that the
matter spectrum used in the halo calculation is consistent with the adopted
late-time cosmology while retaining the shape inherited from
$\mathcal P_{\mathcal R}(k)$ and $T(k)$.

\subsection{The quantum-to-matter interface}
\label{subsec:qft_matter_interface_sec5}

The physical content of this section can be summarized by separating the
primordial and post-inflationary contributions:
\begin{equation}
\begin{aligned}
\mathcal P_{\mathcal R}(k)
&\quad\text{sets the primordial amplitude and tilt},\\
T(k)
&\quad\text{encodes scale-dependent early-time processing},\\
D(z)
&\quad\text{encodes subsequent linear growth}.
\end{aligned}
\label{eq:sec5_three_components}
\end{equation}
Together they determine $P_m(k,z)$.

This separation is important for the interpretation of high-redshift halo
abundances. A change in the primordial spectrum, a change in the transfer
physics, and a change in the late-time growth history alter the matter
spectrum in physically different ways, even though all three ultimately
affect the abundance of rare halos.

The next step is to convert $P_m(k,z)$ into the fluctuation amplitude on a
specified mass scale. This is accomplished by smoothing the matter density
field with a spherical window to obtain $\sigma(M,z)$ and the peak-height
variable $\nu(M,z)$, which provide the immediate statistical inputs to the
halo mass function.

%
%

\section{Mass Variance, Peak Height, and the Statistics of Collapse}
\label{sec:mass_variance_peak_height}

The matter power spectrum developed in Section~\ref{sec:transfer_growth}
describes the variance of the density field as a function of Fourier scale.
Halo formation, however, is naturally described in terms of mass. The next
step is therefore to smooth the linear density field on a physical scale
corresponding to a mass $M$ and to determine how rare a fluctuation of that
mass is at redshift $z$.

The statistical bridge needed for the halo calculation is
\begin{equation}
P_m(k,z)
\longrightarrow
\sigma(M,z)
\longrightarrow
\nu(M,z),
\label{eq:sec6_statistical_bridge}
\end{equation}
where $\sigma(M,z)$ is the rms linear density fluctuation on mass scale $M$
and $\nu(M,z)$ is the corresponding peak-height parameter.

\subsection{Smoothing the density field}
\label{subsec:smoothing_density_sec6}

Let the linear matter density contrast be $\delta({\bf x},z)$. The density
field smoothed on comoving scale $R$ is
\begin{equation}
\delta_R({\bf x},z)
=
\int
{\rm d}^3x'\,
W_R(|{\bf x}-{\bf x}'|)
\delta({\bf x}',z),
\label{eq:sec6_smoothed_density}
\end{equation}
where $W_R$ is a normalized window function.

For halo calculations the standard choice is a spherical real-space top-hat.
Its Fourier transform is
\begin{equation}
W(kR)
=
3\,
\frac{\sin(kR)-kR\cos(kR)}
{(kR)^3}.
\label{eq:sec6_tophat}
\end{equation}

The smoothing radius is related to the enclosed comoving mass by
\begin{equation}
M
=
\frac{4\pi}{3}
\bar\rho_{m,0}
R^3,
\label{eq:sec6_mass_radius}
\end{equation}
where $\bar\rho_{m,0}$ is the present-day mean comoving matter density,
\begin{equation}
\bar\rho_{m,0}
=
\Omega_m\rho_{\rm crit,0},
\qquad
\rho_{\rm crit,0}
=
\frac{3H_0^2}{8\pi G}.
\label{eq:sec6_mean_density}
\end{equation}

Equation~(\ref{eq:sec6_mass_radius}) provides the conversion between Fourier
scale and the mass variable used by the halo mass function.

\subsection{Mass variance}
\label{subsec:mass_variance_sec6}

The variance of the smoothed linear density field is
\begin{equation}
\sigma^2(M,z)
=
\frac{1}{2\pi^2}
\int_0^\infty
k^2
P_m(k,z)
W^2(kR)\,
{\rm d}k.
\label{eq:sec6_sigma}
\end{equation}
Equivalently, if the growth factor is normalized by $D(0)=1$,
\begin{equation}
\sigma(M,z)
=
D(z)\,
\sigma(M,0).
\label{eq:sec6_sigma_growth}
\end{equation}

The quantity $\sigma(M,z)$ is the principal statistical input to analytic
halo-abundance models. It measures the typical linear fluctuation amplitude
associated with mass scale $M$. Because the $\Lambda$CDM matter spectrum
contains less power on progressively larger halo masses, $\sigma(M)$ generally
decreases with increasing $M$ over the mass range relevant to galaxies,
groups, and clusters.

This immediately explains why massive objects are rare. A fixed collapse
threshold represents a progressively larger excursion relative to the rms
fluctuation as $M$ increases.

\subsection{Gaussian initial conditions}
\label{subsec:gaussian_initial_sec6}

For the baseline calculation the primordial perturbations are taken to be
Gaussian to a very good approximation, consistent with the standard
single-field inflationary framework. After linear evolution and smoothing,
the one-point distribution of $\delta_R$ is therefore
\begin{equation}
p(\delta_R\,|\,M,z)
=
\frac{1}
{\sqrt{2\pi}\,\sigma(M,z)}
\exp\left[
-\frac{\delta_R^2}
{2\sigma^2(M,z)}
\right].
\label{eq:sec6_gaussian_pdf}
\end{equation}

The probability of obtaining a sufficiently large positive fluctuation is
therefore exponentially sensitive to the ratio between the collapse
threshold and $\sigma(M,z)$. This property underlies the strong sensitivity
of rare high-redshift halos to the amplitude and shape of the primordial
spectrum.

The present calculation assumes Gaussian initial conditions. Primordial
non-Gaussianity can alter the abundance of rare peaks, particularly in the
high-mass tail, but is not introduced as an additional degree of freedom in
the baseline model.

\subsection{Peak height}
\label{subsec:peak_height_sec6}

The rarity of a fluctuation is conventionally expressed through the
dimensionless peak-height parameter \citep{PressSchechter1974,Bond1991,ShethTormen1999}
\begin{equation}
\nu(M,z)
=
\frac{\delta_c}
{\sigma(M,z)},
\label{eq:sec6_nu}
\end{equation}
where $\delta_c$ is the linearly extrapolated density threshold for collapse.
For the standard spherical-collapse model,
\begin{equation}
\delta_c
\simeq
1.686
\label{eq:sec6_delta_c_reference}
\end{equation}
in an Einstein--de Sitter cosmology, with only weak cosmology dependence in
the applications considered here. The dynamical origin of this threshold is
developed in the following section.

The interpretation of $\nu$ is direct. Values $\nu\sim1$ correspond to
typical fluctuations on the chosen mass scale, whereas
\begin{equation}
\nu\gg1
\label{eq:sec6_high_nu}
\end{equation}
identifies rare peaks in the Gaussian density field. Massive halos at high
redshift are therefore high-$\nu$ objects because both increasing mass and
increasing redshift reduce $\sigma(M,z)$ relative to the fixed collapse
threshold.

\subsection{Sensitivity of rare peaks}
\label{subsec:rare_peak_sensitivity_sec6}

For a Gaussian field, the probability contained in the high-density tail
scales approximately as
\begin{equation}
p_{\rm tail}
\propto
\exp\left(
-\frac{\nu^2}{2}
\right)
\qquad
(\nu\gg1).
\label{eq:sec6_gaussian_tail}
\end{equation}
This exponential dependence is the key reason that apparently modest changes
in the fluctuation amplitude can lead to large changes in the predicted
abundance of rare halos.

Because
\begin{equation}
\sigma(M,z)
\propto
P_m^{1/2},
\label{eq:sec6_sigma_power_scaling}
\end{equation}
and the normalization of $P_m$ ultimately traces back to the primordial
curvature spectrum, high-$\nu$ halo abundances retain an indirect sensitivity
to the early-Universe fluctuation amplitude. The connection is not one-to-one,
however, because the transfer function, growth history, halo-abundance
prescription, and astrophysical mapping all enter at subsequent stages.

This distinction is particularly important in interpreting JWST observations:
an unexpectedly large number of massive high-redshift objects cannot by
itself identify the origin of a discrepancy. The statistical rarity encoded
by $\nu$ must first be separated from uncertainties in halo calibration and
from the mapping between halos and observable galaxies or AGN.

\subsection{Statistical input to the halo mass function}
\label{subsec:hmf_input_sec6}

The quantities developed in this section provide the immediate statistical
inputs to the halo-abundance calculation:
\begin{equation}
\begin{aligned}
P_m(k,z)
&\longrightarrow
\sigma(M,z)
\\
&\longrightarrow
\nu(M,z)
\longrightarrow
\text{halo abundance}.
\end{aligned}
\label{eq:sec6_to_hmf}
\end{equation}

At this stage the calculation describes the statistical rarity of linear
density fluctuations but not yet the nonlinear dynamics by which an
overdensity becomes a bound halo. The next section introduces the spherical
collapse model, which supplies the physical interpretation of the collapse
threshold $\delta_c$ and establishes the link between the linear density field
and the collapsed objects counted by analytic halo mass functions.

%
%

\section{Spherical Collapse and the Nonlinear Threshold}
\label{sec:spherical_collapse}

Section~\ref{sec:mass_variance_peak_height} characterized the linear density
field through $\sigma(M,z)$ and the peak-height variable
$\nu(M,z)=\delta_c/\sigma(M,z)$. The remaining ingredient is the physical
meaning of the collapse threshold $\delta_c$. The spherical-collapse model
provides the simplest analytic connection between a linearly extrapolated
overdensity and the formation of a gravitationally bound halo
\citep{GunnGott1972,Peebles1980,LaceyCole1993,Mo2010}.

The purpose of this section is therefore narrow:
\begin{equation}
\text{linear overdensity}
\longrightarrow
\text{nonlinear collapse}
\longrightarrow
\delta_c.
\label{eq:sec7_bridge}
\end{equation}
This threshold will be combined with $\sigma(M,z)$ in the halo mass functions
developed in the following section.

\subsection{Spherical top-hat perturbation}
\label{subsec:sec7_tophat}

Consider a spherical region of physical radius $R(t)$ containing a fixed mass
$M$ and having uniform density $\rho_p(t)$. Its mass is
\begin{equation}
M
=
\frac{4\pi}{3}
R^3\rho_p.
\label{eq:sec7_mass}
\end{equation}
The density contrast relative to the homogeneous background is
\begin{equation}
\delta
=
\frac{\rho_p-\bar\rho_m}
{\bar\rho_m}.
\label{eq:sec7_delta}
\end{equation}

For a pressureless isolated spherical perturbation, the shell enclosing $M$
obeys
\begin{equation}
\ddot R
=
-\frac{GM}{R^2},
\label{eq:sec7_eom}
\end{equation}
or, after one integration,
\begin{equation}
\frac{\dot R^2}{2}
-
\frac{GM}{R}
=
E,
\label{eq:sec7_energy}
\end{equation}
where $E$ is the specific energy. A bound overdensity has $E<0$: it initially
expands with the Hubble flow, reaches a maximum radius, and subsequently
collapses.

\subsection{Parametric collapse solution}
\label{subsec:sec7_parametric}

For an Einstein--de Sitter background, the bound solution may be written
parametrically as
\begin{equation}
R(\theta)
=
A(1-\cos\theta),
\label{eq:sec7_parametric_R}
\end{equation}
and
\begin{equation}
t(\theta)
=
B(\theta-\sin\theta),
\label{eq:sec7_parametric_t}
\end{equation}
where $A$ and $B$ are constants fixed by the perturbation.

The principal stages are
\begin{equation}
\theta=\pi
\quad\text{at turnaround},
\qquad
\theta=2\pi
\quad\text{at formal collapse}.
\label{eq:sec7_stages}
\end{equation}
Thus
\begin{equation}
R_{\rm ta}=2A,
\qquad
t_{\rm coll}=2\pi B.
\label{eq:sec7_turnaround_collapse}
\end{equation}

Although the exact nonlinear solution formally collapses to zero radius, the
same initial perturbation can be evolved simultaneously with linear theory.
It is this linearly extrapolated solution that supplies the threshold used in
halo statistics.

\subsection{The linear collapse threshold}
\label{subsec:sec7_delta_c}

In an Einstein--de Sitter universe, the growing linear mode satisfies
\begin{equation}
\delta_{\rm lin}
\propto
a
\propto
t^{2/3}.
\label{eq:sec7_linear_growth}
\end{equation}
Expanding the exact spherical solution at early times and matching it to the
linear growing mode gives
\begin{equation}
\delta_{\rm lin}(t)
=
\frac{3}{20}
(12\pi)^{2/3}
\left(
\frac{t}{t_{\rm coll}}
\right)^{2/3}.
\label{eq:sec7_linear_extrapolation}
\end{equation}
Evaluating the linear solution at the time when the nonlinear model formally
collapses yields
\begin{equation}
\delta_c
=
\frac{3}{20}
(12\pi)^{2/3}
\simeq
1.686.
\label{eq:sec7_delta_c}
\end{equation}

This is the quantity introduced in
Section~\ref{sec:mass_variance_peak_height}. It is not the nonlinear density
contrast of the collapsing object. Rather, it is the value that the
corresponding \emph{linear} perturbation would have reached at the collapse
time.

For the high-redshift applications considered here, adopting
$\delta_c\simeq1.686$ is an adequate approximation. In a general
$\Lambda$CDM cosmology the collapse threshold has a weak redshift and
cosmology dependence, but this correction is small compared with the
exponential sensitivity of rare-halo abundances to
$\nu=\delta_c/\sigma(M,z)$.

\subsection{Virialization}
\label{subsec:sec7_virialization}

Real collisionless systems do not collapse to zero radius. As the system
relaxes, gravitational potential energy is converted into random kinetic
motion and the object approaches virial equilibrium,
\begin{equation}
2K+U=0.
\label{eq:sec7_virial_theorem}
\end{equation}
For the idealized Einstein--de Sitter top-hat model, energy conservation gives
\begin{equation}
R_{\rm vir}
=
\frac{1}{2}
R_{\rm ta}.
\label{eq:sec7_rvir}
\end{equation}
The corresponding mean overdensity at virialization is
\begin{equation}
\Delta_{\rm vir}
=
18\pi^2
\simeq
178
\label{eq:sec7_delta_vir}
\end{equation}
relative to the background matter density. Modern halo catalogs often use
related operational definitions such as $M_{200}$ or other spherical
overdensity masses. These definitions affect the detailed calibration of a
halo mass function but do not change the role of $\delta_c$ as the analytic
collapse threshold.

\subsection{Role in halo statistics}
\label{subsec:sec7_role_hmf}

Spherical collapse supplies the dynamical criterion that complements the
statistical quantities developed earlier. The relevant sequence is
\begin{equation}
P_m(k,z)
\longrightarrow
\sigma(M,z)
\longrightarrow
\frac{\delta_c}{\sigma(M,z)}
\longrightarrow
\text{halo abundance}.
\label{eq:sec7_to_hmf}
\end{equation}
The first two steps characterize the linear density field; spherical collapse
provides the threshold that determines how exceptional a fluctuation must be
to form a halo.

The spherical model is deliberately idealized. Real collapse is affected by
tidal fields, nonsphericity, angular momentum, mergers, and environmental
effects. These departures motivate ellipsoidal-collapse corrections and
simulation-calibrated halo mass functions
\citep{Bond1991,ShethTormen1999,ShethMoTormen2001,Tinker2008}. They are treated
at the abundance level rather than by adding further nonlinear dynamics here.

The next section combines $\delta_c$ with $\sigma(M,z)$ to construct the
Press--Schechter and Sheth--Tormen mass functions and then introduces the
simulation-calibrated high-redshift halo mass function derived from the
GUREFT simulations of \citet{Yung2024}. This provides complementary analytic
and numerical descriptions of halo abundance for assessing the rare
high-redshift populations relevant to the JWST regime.

%
%

\section{Halo Mass Functions: Press--Schechter, Sheth--Tormen, and GUREFT}
\label{sec:halo_mass_functions}

Sections~\ref{sec:mass_variance_peak_height} and \ref{sec:spherical_collapse}
established the quantities required to predict halo abundances. The variance
$\sigma(M,z)$ measures the amplitude of the linearly evolved density field,
while the collapse threshold $\delta_c$ defines the linear overdensity
associated with collapse. Their ratio,
\begin{equation}
\nu(M,z)=\frac{\delta_c}{\sigma(M,z)},
\label{eq:sec8_peak_height}
\end{equation}
measures the rarity of a fluctuation of mass $M$ at redshift $z$.

The halo mass function converts these fluctuation statistics into the comoving
abundance of collapsed dark-matter halos. Press--Schechter theory provides the
historical spherical-collapse foundation; the Sheth--Tormen mass function
provides a transparent analytic description of the rare-peak tail; and the
GUREFT simulation calibration of \citet{Yung2024} provides a complementary
high-redshift description constructed for the cosmic-dawn regime. The
comparison between Sheth--Tormen and GUREFT is used to assess how the adopted
halo-abundance prescription propagates into predictions for high-redshift
galaxies and AGN.

The common upstream calculation is
\begin{equation}
\mathcal P_{\mathcal R}(k)\longrightarrow P_m(k,z)
\longrightarrow\sigma(M,z),
\label{eq:sec8_common_upstream}
\end{equation}
after which the analytic and simulation-calibrated prescriptions provide
alternative mappings from the same linear fluctuation field to halo abundance.

\subsection{Press--Schechter baseline}
\label{subsec:sec8_press_schechter}

For Gaussian initial conditions, the Press--Schechter collapsed fraction is
\begin{equation}
F_{\rm PS}(>M,z)
={\rm erfc}\left[
\frac{\delta_c}{\sqrt{2}\,\sigma(M,z)}
\right].
\label{eq:sec8_ps_fraction}
\end{equation}
The corresponding multiplicity function is
\begin{equation}
f_{\rm PS}(\nu)=
\sqrt{\frac{2}{\pi}}\,\nu
\exp\left(-\frac{\nu^2}{2}\right),
\label{eq:sec8_ps_multiplicity}
\end{equation}
and
\begin{equation}
\frac{{\rm d}n_{\rm PS}}{{\rm d}M}
=
\frac{\bar\rho_{m,0}}{M^2}f_{\rm PS}(\nu)
\left|\frac{{\rm d}\ln\sigma^{-1}}{{\rm d}\ln M}\right|.
\label{eq:sec8_ps_hmf}
\end{equation}
Press--Schechter theory establishes the connection between Gaussian initial
conditions, gravitational collapse, and bound-object abundance
\citep{PressSchechter1974,Bond1991}. Its spherical constant-barrier
description provides the physical foundation but is not used as the sole
quantitative description of the rare high-redshift halo population.

\subsection{Sheth--Tormen analytic mass function}
\label{subsec:sec8_sheth_tormen}

The Sheth--Tormen prescription incorporates nonspherical collapse through
\citep{ShethTormen1999,ShethMoTormen2001}
\begin{equation}
\begin{aligned}
f_{\rm ST}(\nu)
&= A_{\rm ST}\sqrt{\frac{2a_{\rm ST}}{\pi}}
\left[1+(a_{\rm ST}\nu^2)^{-p_{\rm ST}}\right]\\
&\quad\times\nu\exp\left(-\frac{a_{\rm ST}\nu^2}{2}\right),
\end{aligned}
\label{eq:sec8_st_multiplicity}
\end{equation}
with
\begin{equation}
A_{\rm ST}=0.3222,\qquad a_{\rm ST}=0.707,\qquad p_{\rm ST}=0.3.
\label{eq:sec8_st_parameters}
\end{equation}
The differential abundance is
\begin{equation}
\frac{{\rm d}n_{\rm ST}}{{\rm d}M}
=
\frac{\bar\rho_{m,0}}{M^2}f_{\rm ST}(\nu)
\left|\frac{{\rm d}\ln\sigma^{-1}}{{\rm d}\ln M}\right|.
\label{eq:sec8_st_hmf}
\end{equation}
In the rare-peak limit,
\begin{equation}
f_{\rm ST}(\nu)\propto
\exp\left(-\frac{a_{\rm ST}\nu^2}{2}\right),\qquad \nu\gg1,
\label{eq:sec8_st_rare_tail}
\end{equation}
which makes transparent the amplification by which modest changes in the
fluctuation amplitude can produce much larger fractional changes in rare-halo
abundance.

\subsection{The GUREFT high-redshift halo mass function}
\label{subsec:sec8_gureft}

The GADGET at Ultrahigh Redshift with Extra-Fine Timesteps (GUREFT) suite was
designed to characterize dark-matter halo populations at cosmic dawn
\citep{Yung2024}. It consists of four dark-matter-only $1024^3$-particle
simulations with box sizes of $5$, $15$, $35$, and $90\,h^{-1}{\rm Mpc}$.
The halo mass functions span approximately $6\lesssim z\lesssim20$ and are
supplemented at the massive end by MultiDark simulations.

The GUREFT halo mass function is written
\begin{equation}
\frac{{\rm d}n_{\rm G}}{{\rm d}M_{\rm vir}}
=
f_{\rm G}(\sigma)\,
\frac{\bar\rho_{m,0}}{M_{\rm vir}^{2}}
\left|\frac{{\rm d}\ln\sigma^{-1}}{{\rm d}\ln M_{\rm vir}}\right|,
\label{eq:sec8_gureft_hmf}
\end{equation}
with
\begin{equation}
f_{\rm G}(\sigma,z)=
A_{\rm G}(z)
\left[
\left(\frac{\sigma}{b_{\rm G}(z)}\right)^{-a_{\rm G}(z)}+1
\right]
\exp\left[-\frac{c_{\rm G}(z)}{\sigma^2}\right].
\label{eq:sec8_gureft_multiplicity}
\end{equation}
The fitting parameters evolve as
\begin{equation}
\chi_i(z)=\chi_{0,i}+\chi_{1,i}z+\chi_{2,i}z^2,
\qquad
\chi_i\in\{A_{\rm G},a_{\rm G},b_{\rm G},c_{\rm G}\}.
\label{eq:sec8_gureft_redshift}
\end{equation}
For the physical-unit fit of \citet{Yung2024},
\begin{equation}
\begin{array}{c|rrr}
\chi_i & \chi_{0,i} & \chi_{1,i} & \chi_{2,i} \\ \hline
A_{\rm G} & 0.13765772 & -0.01003821 & 0.00102964 \\
a_{\rm G} & 1.06641384 &  0.02475576 & -0.00283342 \\
b_{\rm G} & 4.86693806 &  0.09212356 & -0.01426283 \\
c_{\rm G} & 1.19837952 & -0.00142967 & -0.00033074
\end{array}
\label{eq:sec8_gureft_coefficients}
\end{equation}
with the combined GUREFT+MultiDark fit calibrated over $6\leq z\leq19$.

\citet{Yung2024} also provide an approximation for $\sigma(M_{\rm vir})$.
Here $\sigma(M,z)$ is instead calculated directly from the matter power
spectrum through Eq.~(\ref{eq:sec6_sigma}) and supplied to the GUREFT
multiplicity function. This choice preserves the continuous information flow
from the adopted primordial curvature spectrum through $P_m(k,z)$ and into
the halo calculation.

\subsection{Halo mass definition and calibration domain}
\label{subsec:sec8_mass_definition}

GUREFT identifies halos with the ROCKSTAR phase-space halo finder and adopts
the Bryan--Norman virial spherical-overdensity definition
\citep{BryanNorman1998,Yung2024}. We therefore denote the mass entering the
GUREFT fit by $M_{\rm vir}$. Masses defined as $M_{200{\rm c}}$,
$M_{200{\rm m}}$, or by another overdensity convention are not assumed to be
numerically identical to $M_{\rm vir}$.

For the controlled ST–GUREFT comparison, both mass functions are evaluated on a common physical-mass grid identified with the GUREFT virial-mass convention; no additional conversion between alternative spherical-overdensity mass definitions is performed, and any resulting calibration dependence is treated as a systematic limitation.

The GUREFT simulations adopt $\Omega_m=0.307$, $\Omega_\Lambda=0.693$,
$H_0=67.8\,{\rm km\,s^{-1}\,Mpc^{-1}}$, $\sigma_8=0.829$, and $n_s=0.960$
\citep{Yung2024}. These parameters are close to, but not identical with, the
Planck-consistent cosmology adopted here. GUREFT is therefore used as a
simulation-calibrated comparison rather than as an exact universal mapping
independent of cosmology.

The fitted redshift range directly covers the principal cosmic-dawn epochs
used here, including $z=6$, $8$, $10$, and $13$. Calibration is nevertheless
not equally constraining at every mass and redshift. The most massive halos
become intrinsically rare, and finite simulation volume limits direct sampling
of the extreme high-$\nu$ tail. Predictions involving the most massive halos
at the highest redshifts are therefore interpreted as rare-peak stress tests
rather than precision abundance determinations.

\subsection{Cumulative halo abundance and collapsed mass}
\label{subsec:sec8_cumulative}

For any prescription $X$,
\begin{equation}
n_X(>M,z)=
\int_M^\infty\frac{{\rm d}n_X}{{\rm d}M'}\,{\rm d}M',
\label{eq:sec8_cumulative_number}
\end{equation}
and
\begin{equation}
\rho_{h,X}(>M,z)=
\int_M^\infty M'\frac{{\rm d}n_X}{{\rm d}M'}\,{\rm d}M'.
\label{eq:sec8_cumulative_mass}
\end{equation}
The cumulative number density measures the supply of potential hosts, while
$\rho_{h,X}$ measures the matter reservoir contained in those halos.

For the quantitative comparisons,
\begin{equation}
X\in\{{\rm ST,GUREFT}\}.
\label{eq:sec8_hmf_branches}
\end{equation}
The diagnostic ratio
\begin{equation}
\mathcal R_{\rm HMF}(M,z)=
\frac{({\rm d}n/{\rm d}M)_{\rm GUREFT}}
{({\rm d}n/{\rm d}M)_{\rm ST}}
\label{eq:sec8_hmf_ratio}
\end{equation}
is generally both mass- and redshift-dependent. The ST--GUREFT comparison
therefore probes changes in the shape as well as the normalization of the
predicted halo population.

\subsection{Separating primordial information from halo calibration}
\label{subsec:sec8_interpretation}

The primordial and linear calculations determine
\begin{equation}
\mathcal P_{\mathcal R}(k)\longrightarrow P_m(k,z)
\longrightarrow\sigma(M,z),
\label{eq:sec8_primordial_to_sigma}
\end{equation}
while the halo mass function specifies how these fluctuation statistics are
mapped into collapsed objects. For the controlled comparison,
\begin{equation}
\begin{aligned}
{\rm ST}&\longrightarrow{\rm GUREFT},\\
\mathcal P_{\mathcal R},\;P_m,\;\sigma&\quad\text{fixed}.
\end{aligned}
\label{eq:sec8_controlled_hmf_test}
\end{equation}
The resulting change in halo abundance can therefore be distinguished from a
change in the primordial fluctuation field itself.

A difference between an observed abundance and a Sheth--Tormen prediction
does not by itself identify new primordial physics. The result may also depend
on nonlinear halo calibration, halo-mass definition, the mapping between
halos and luminous objects, and observational selection. The GUREFT
comparison provides a high-redshift numerical benchmark with which to expose
this part of the uncertainty chain.

\subsection{From halo abundance to observable populations}
\label{subsec:sec8_to_observables}

The halo mass function determines the abundance of potential gravitational
hosts; it does not by itself determine the abundance of observable galaxies
or active galactic nuclei. The next stage requires a mapping from halo mass
to baryonic mass, stellar mass, black-hole occupation, luminosity, duty cycle,
and observational selection.

\begin{equation}
\boxed{
\text{halo supply}\quad\neq\quad\text{halo-to-observable mapping}
}
\label{eq:sec8_supply_mapping}
\end{equation}

The following sections carry the Sheth--Tormen and GUREFT prescriptions
through the same astrophysical forward models. This makes it possible to
distinguish sensitivity to the primordial fluctuation field from sensitivity
to nonlinear halo statistics and from the subsequent mapping between
dark-matter halos and the high-redshift populations observed by JWST.

%
%

\section{From Halo Supply to Observable Populations: Occupation, Selection, and Inference}
\label{sec:halo_to_observable_interface}

Section~\ref{sec:halo_mass_functions} determines the cosmological supply of
collapsed halos. Observations, however, do not measure halo mass functions
directly. They measure stellar masses, ultraviolet luminosities, emission
lines, black-hole masses, X-ray fluxes, and other properties of baryonic
systems embedded within halos. The transition from halo abundance to an
observed population is therefore not another stage of gravitational collapse;
it is an inference problem.

This distinction is especially important in the present paper because the
calculation begins with quantum initial conditions. The chain
\begin{equation}
\mathcal P_{\mathcal R}(k)
\longrightarrow
P_m(k,z)
\longrightarrow
\sigma(M,z)
\longrightarrow
\frac{{\rm d}n_X}{{\rm d}M_h}
\label{eq:sec9_cosmological_chain}
\end{equation}
is primarily cosmological, whereas the subsequent conversion
\begin{equation}
M_h
\longrightarrow
\text{baryonic system}
\longrightarrow
\text{observable}
\label{eq:sec9_astrophysical_chain}
\end{equation}
depends on galaxy formation, black-hole growth, radiative physics, and
selection effects.

The purpose of this section is to keep these two layers explicit and to
formulate a common forward model that can be used for both galaxies and AGN.

\subsection{A general forward model}
\label{subsec:sec9_forward_model}

Let $O$ denote an observable quantity, such as stellar mass, UV magnitude,
or AGN luminosity. For the quantitative halo mass-function prescriptions
\begin{equation}
X\in\{{\rm ST,GUREFT}\},
\label{eq:sec9_hmf_branches}
\end{equation}
the differential abundance of observable systems can be written schematically
as
\begin{equation}
\frac{{\rm d}n_{{\rm obj},X}}{{\rm d}O}
=
\int_0^\infty
\frac{{\rm d}n_X}{{\rm d}M_h}
P(O\mid M_h,z)\,
{\rm d}M_h,
\label{eq:sec9_general_forward}
\end{equation}
where $P(O\mid M_h,z)$ contains the astrophysical mapping between halo mass
and the observable.

Here $M_h$ denotes the halo mass appropriate to the adopted HMF prescription.
For the GUREFT branch this is the virial mass $M_{\rm vir}$ defined in
Section~\ref{sec:halo_mass_functions}. Comparisons with observational or
empirical relations expressed using a different halo-mass convention require
a consistent mass conversion or an explicit allowance for the associated
systematic uncertainty.

If the survey selection function is $S(O,z)$, the predicted number density
entering an observational comparison is
\begin{equation}
n_{{\rm det},X}(z)
=
\int
S(O,z)
\frac{{\rm d}n_{{\rm obj},X}}{{\rm d}O}
\,{\rm d}O.
\label{eq:sec9_selection}
\end{equation}

Equations~(\ref{eq:sec9_general_forward}) and
(\ref{eq:sec9_selection}) define the conceptual boundary between cosmology
and astrophysics in this work. The halo mass function specifies the available
population of gravitational hosts; the conditional probability and selection
function determine how that population appears in a telescope.

\subsection{Deterministic mappings and scatter}
\label{subsec:sec9_deterministic_scatter}

A deterministic halo-to-observable relation,
\begin{equation}
O=\bar O(M_h,z),
\label{eq:sec9_deterministic_mapping}
\end{equation}
corresponds formally to
\begin{equation}
P(O\mid M_h,z)
=
\delta_{\rm D}
\left[
O-\bar O(M_h,z)
\right].
\label{eq:sec9_delta_mapping}
\end{equation}
In this limit, the observable abundance is obtained by the usual change of
variables,
\begin{equation}
\frac{{\rm d}n_{{\rm obj},X}}{{\rm d}O}
=
\left.
\frac{{\rm d}n_X}{{\rm d}M_h}
\left|
\frac{{\rm d}M_h}{{\rm d}O}
\right|
\right|_{M_h=M_h(O,z)}.
\label{eq:sec9_change_variables}
\end{equation}

This form separates two distinct effects. Changing the halo mass function
alters the number of available halos and, for ST and GUREFT, can also alter
the mass dependence of that supply. Changing the mapping
$\bar O(M_h,z)$ instead alters which halo mass is associated with a given
observable. These operations are not equivalent.

Real systems possess scatter. A finite-width $P(O\mid M_h,z)$ allows
lower-mass, more numerous halos to scatter into an observationally extreme
bin and higher-mass halos to scatter out of it. Because the halo abundance
falls steeply at the massive end, this asymmetry can significantly affect
rare-object predictions. The deterministic relations used later should
therefore be interpreted as controlled baseline mappings rather than complete
descriptions of galaxy or AGN populations
\citep{Moster2013,SomervilleDave2015}.

\subsection{Galaxies: the baryonic branch}
\label{subsec:sec9_galaxy_branch}

For galaxies, the first constraint is the cosmic baryon reservoir. A halo of
mass $M_h$ contains at most
\begin{equation}
M_b=f_bM_h,
\qquad
f_b=\frac{\Omega_b}{\Omega_m}.
\label{eq:sec9_baryon_mass}
\end{equation}
If a fraction $\epsilon_\star$ of those baryons is converted into stars,
\begin{equation}
M_\star
=
\epsilon_\star f_b M_h,
\qquad
0\leq\epsilon_\star\leq1.
\label{eq:sec9_stellar_mapping}
\end{equation}
The corresponding individual-object ceiling is
\begin{equation}
M_\star\leq f_bM_h.
\label{eq:sec9_baryon_ceiling}
\end{equation}

This relation does not predict a galaxy abundance by itself. It asks whether
a halo of a given mass contains enough baryonic material to host a galaxy of
inferred stellar mass. Population consistency additionally requires that such
halos exist in sufficient numbers.

The two requirements are
\begin{equation}
\boxed{
\begin{aligned}
\text{mass feasibility:}\quad&
M_\star\leq\epsilon_\star f_bM_h,\\
\text{population feasibility:}\quad&
n_{\rm gal}\leq n_X(>M_h,z).
\end{aligned}}
\label{eq:sec9_two_galaxy_conditions}
\end{equation}
They constrain different parts of the inference chain.

\subsection{Active nuclei: the black-hole branch}
\label{subsec:sec9_agn_branch}

AGN require a different mapping. Their observed luminosity is tied first to
an actively accreting black hole and only indirectly to the host halo. For a
black hole of mass $M_{\rm BH}$,
\begin{equation}
L_{\rm Edd}
=
1.26\times10^{38}
\left(\frac{M_{\rm BH}}{M_\odot}\right)
{\rm erg\,s^{-1}},
\label{eq:sec9_eddington_luminosity}
\end{equation}
and
\begin{equation}
L_{\rm bol}
=
\lambda_{\rm Edd}L_{\rm Edd},
\label{eq:sec9_bolometric_luminosity}
\end{equation}
where $\lambda_{\rm Edd}$ is the Eddington ratio.

A relation between black-hole mass and halo mass may be summarized by
\begin{equation}
\epsilon_{\rm BH}
=
\frac{M_{\rm BH}}{M_h},
\label{eq:sec9_bh_halo_ratio}
\end{equation}
so that an observed luminosity can be mapped, conditionally, through
\begin{equation}
L_{\rm bol}
\longrightarrow
M_{\rm BH}
\longrightarrow
M_h.
\label{eq:sec9_agn_mapping_chain}
\end{equation}

Unlike the stellar-mass ceiling, this mapping is not a conservation bound.
It depends on the accretion state and on the black-hole--host relation and
therefore carries substantial astrophysical uncertainty
\citep{Volonteri2010}.

\subsection{Occupation and duty cycle}
\label{subsec:sec9_occupation_duty}

Not every eligible halo contains the type of black hole under consideration,
and not every occupied halo is observed during an active phase. We therefore
write
\begin{equation}
P_{\rm AGN}(M_h,z)
=
f_{\rm occ}(M_h,z)\,
f_{\rm duty}(M_h,z),
\label{eq:sec9_agn_probability}
\end{equation}
where $f_{\rm occ}$ is the black-hole occupation fraction and
$f_{\rm duty}$ is the active duty cycle.

For a minimum host mass $M_{\min}$,
\begin{equation}
n_{{\rm AGN},X}(z)
=
\int_{M_{\min}}^\infty
f_{\rm occ}(M_h,z)
f_{\rm duty}(M_h,z)
\frac{{\rm d}n_X}{{\rm d}M_h}
\,{\rm d}M_h.
\label{eq:sec9_agn_general}
\end{equation}
If the two factors are treated as constants above the threshold,
\begin{equation}
n_{{\rm AGN},X}(z)
=
f_{\rm occ}f_{\rm duty}
n_X(>M_{\min},z).
\label{eq:sec9_agn_simple}
\end{equation}

The limiting choice
\begin{equation}
f_{\rm occ}=f_{\rm duty}=1
\label{eq:sec9_maximal_occupation}
\end{equation}
provides an upper envelope: the predicted active population cannot exceed the
supply of eligible halos under the adopted host threshold.

For identical occupation, duty-cycle, and host-mass assumptions, the effect
of changing the halo prescription is
\begin{equation}
\begin{aligned}
&\frac{n_{{\rm AGN},{\rm GUREFT}}}{n_{{\rm AGN},{\rm ST}}}\\
&\quad=\frac{
\displaystyle\int_{M_{\min}}^\infty
f_{\rm occ}f_{\rm duty}
({\rm d}n_{\rm GUREFT}/{\rm d}M_h)\,{\rm d}M_h
}{
\displaystyle\int_{M_{\min}}^\infty
f_{\rm occ}f_{\rm duty}
({\rm d}n_{\rm ST}/{\rm d}M_h)\,{\rm d}M_h
}.
\end{aligned}
\label{eq:sec9_agn_hmf_ratio}
\end{equation}
When $f_{\rm occ}$ and $f_{\rm duty}$ are common constants above the same
threshold, they cancel from this ratio. Unlike a mass-independent rescaling,
however, the ST--GUREFT difference cannot in general be represented by a
single redshift-only factor because the two HMFs differ as functions of both
mass and redshift. The halo prescription changes the cosmological host
reservoir; it does not prescribe a different black-hole lifetime, accretion
history, or Eddington-ratio distribution.

\subsection{Where information is preserved and where degeneracy enters}
\label{subsec:sec9_information_flow}

The full calculation can be viewed as an information-flow problem:
\begin{equation}
\begin{aligned}
\mathcal P_{\mathcal R}(k)
&\longrightarrow P_m(k,z)
\longrightarrow \sigma(M,z)\\
&\longrightarrow \frac{{\rm d}n_X}{{\rm d}M_h}
\longrightarrow P(O\mid M_h,z)\\
&\longrightarrow S(O,z)
\longrightarrow \text{observed sample}.
\end{aligned}
\label{eq:sec9_information_chain}
\end{equation}

The first line carries information about the primordial spectrum, linear
evolution, collapse statistics, and halo calibration. The second introduces
galaxy formation, black-hole physics, variability, attenuation, measurement
scatter, and survey selection. Consequently, an apparent mismatch at the end
of the chain cannot be assigned uniquely to the beginning without controlling
the intermediate mappings.

This motivates the separation
\begin{equation}
\boxed{
\begin{array}{ll}
\text{primordial/QFT} & \mathcal P_{\mathcal R}(k)\\[2pt]
\text{halo abundance} & {\rm d}n_X/{\rm d}M_h\\[2pt]
\text{observable mapping} & P(O\mid M_h,z),\,S(O,z)
\end{array}}
\label{eq:sec9_uncertainty_layers}
\end{equation}

This three-layer structure gives the manuscript a different emphasis from a
conventional halo-mass-function analysis. The HMF is not the endpoint. It is
the intermediate statistical operator that transmits the imprint of quantum
initial conditions into a population of potential hosts, after which
astrophysical and observational kernels determine what is actually seen.

\subsection{Connection to the observational tests}
\label{subsec:sec9_next}

The framework above is applied in two deliberately distinct ways. Massive
galaxies are tested through their baryonic and stellar-mass requirements and,
separately, through their rest-frame UV luminosity distribution. Active
nuclei are tested through black-hole mass, Eddington ratio, host-halo
requirements, occupation, and duty cycle.

These mappings should not be conflated. In particular, the AGN duty cycle is
not inserted into the stellar galaxy UV luminosity function, and the stellar
halo-to-light prescription is not used as an AGN luminosity model.

The following sections use this separation to ask a more precise question
than whether ``JWST objects are too abundant'': at which stage of the chain
from primordial fluctuations to detected sources does the strongest
constraint arise?

%
%

\section{Massive JWST Galaxies as Rare-Peak Consistency Tests}
\label{sec:massive_galaxy_stress_test}

The framework developed in the preceding sections allows a high-redshift
galaxy to be interpreted in two complementary ways. Astrophysically, its
stellar mass constrains the baryonic reservoir required of its host halo.
Statistically, that host halo corresponds to a fluctuation of peak height
$\nu(M_h,z)$ drawn from the primordial density field.

Massive JWST galaxies therefore provide a consistency test of the full chain
\begin{equation}
\begin{aligned}
\mathcal P_{\mathcal R}(k)
&\longrightarrow
P_m(k,z)
\longrightarrow
\sigma(M_h,z)
\longrightarrow
\nu(M_h,z)
\\
&\longrightarrow
n_X(>M_h,z),
\qquad
X\in\{{\rm ST,GUREFT}\}.
\end{aligned}
\label{eq:sec10_quantum_to_rare_peak}
\end{equation}
The observable stellar mass enters after this cosmological chain has
established the available population of potential hosts.

\subsection{From an observed stellar mass to a host requirement}
\label{subsec:sec10_host_requirement}

For a galaxy with inferred stellar mass $M_\star$, the baryonic relation from
Section~\ref{sec:halo_to_observable_interface} gives
\begin{equation}
M_\star
\leq
\epsilon_\star f_b M_h,
\label{eq:sec10_stellar_bound}
\end{equation}
where
\begin{equation}
f_b=\frac{\Omega_b}{\Omega_m}
\label{eq:sec10_fb}
\end{equation}
is the cosmic baryon fraction and $\epsilon_\star$ is an effective integrated
stellar-conversion efficiency. The minimum halo mass capable of supporting
the inferred stellar mass is therefore
\begin{equation}
M_{h,\min}
=
\frac{M_\star}{\epsilon_\star f_b}.
\label{eq:sec10_min_halo_mass}
\end{equation}

The limiting case $\epsilon_\star=1$ is an absolute baryonic ceiling rather
than a realistic star-formation model: it assigns the entire cosmic-mean
baryonic allotment of the halo to stars. Any value $\epsilon_\star<1$ moves
the required host to a larger halo mass and therefore farther into the rare
high-mass tail of the halo distribution
\citep{Behroozi2013,Moster2013,BoylanKolchin2023}.

Equation~(\ref{eq:sec10_min_halo_mass}) thus translates an observed stellar
mass into a cosmological host-mass and rarity requirement. When the GUREFT
mass function is used, the halo mass entering the abundance calculation is
$M_{\rm vir}$, as defined in Section~\ref{sec:halo_mass_functions}. An
observational host relation expressed in another halo-mass convention must
therefore be converted consistently or treated as an additional systematic
uncertainty.

\subsection{The number-density consistency condition}
\label{subsec:sec10_number_density}

For a halo-abundance prescription $X$, the cumulative number density of
eligible hosts is
\begin{equation}
n_X(>M_{h,\min},z)
=
\int_{M_{h,\min}}^\infty
\frac{{\rm d}n_X}{{\rm d}M_h}\,
{\rm d}M_h.
\label{eq:sec10_halo_abundance}
\end{equation}

If the observed comoving number density of galaxies above a stellar-mass
threshold is $n_{\rm gal}(>M_\star,z)$, a necessary consistency condition is
\begin{equation}
\boxed{
n_{\rm gal}(>M_\star,z)
\leq
n_X\left(
>
\frac{M_\star}{\epsilon_\star f_b},
z
\right)
}.
\label{eq:sec10_number_density_bound}
\end{equation}

This inequality expresses the massive-galaxy problem in a direct form. A stellar-population synthesis fit to a galaxy's spectral energy distribution, for example one allowing an older stellar component or a different star-formation history, may assign a large $M_\star$ to an observed source. This inferred mass is model-dependent; independently, the cosmological model must supply enough halos massive enough to contain the required baryonic reservoir.

The constraint becomes rapidly stronger with increasing $M_\star$ because
the high-mass halo abundance is steeply sensitive to peak rarity. A modest
shift in the inferred stellar mass, stellar-conversion efficiency, or
redshift can therefore produce a much larger change in the predicted host
abundance.

\subsection{A population-level mass-budget test}
\label{subsec:sec10_mass_density}

The number-density test counts potential hosts but does not use the total mass
contained in them. A complementary constraint follows from the cumulative
halo mass density,
\begin{equation}
\rho_{h,X}(>M_h,z)
=
\int_{M_h}^\infty
M_h'
\frac{{\rm d}n_X}{{\rm d}M_h'}\,
{\rm d}M_h'.
\label{eq:sec10_halo_mass_density}
\end{equation}
The maximum stellar mass density that can be assembled in halos capable of
hosting galaxies above $M_\star$ is then
\begin{equation}
\boxed{
\rho_\star(>M_\star,z)
\leq
\epsilon_\star f_b\,
\rho_{h,X}\left(
>
\frac{M_\star}{\epsilon_\star f_b},
z
\right)
}.
\label{eq:sec10_stellar_mass_density_bound}
\end{equation}

This population-level bound is physically distinct from the individual
condition in Equation~(\ref{eq:sec10_stellar_bound}). An individual galaxy
may satisfy the baryonic ceiling while an ensemble of such galaxies may still
require more collapsed baryonic mass than the halo population can provide.
The two tests therefore probe
\begin{equation}
\begin{aligned}
\text{individual object:}\quad&
M_\star\leq\epsilon_\star f_b M_h,
\\
\text{population:}\quad&
\rho_\star\leq\epsilon_\star f_b\rho_{h,X}.
\end{aligned}
\label{eq:sec10_individual_population}
\end{equation}

\subsection{Peak height as the link back to primordial fluctuations}
\label{subsec:sec10_peak_height}

The connection to the primordial fluctuation field is made explicit through
\begin{equation}
\nu(M_h,z)
=
\frac{\delta_c}{\sigma(M_h,z)}.
\label{eq:sec10_peak_height}
\end{equation}
A massive halo at high redshift has large $\nu$ because the linearly evolved
variance on that mass scale is small relative to the collapse threshold.
Such systems therefore probe the extreme tail of the initial fluctuation
distribution.

The mass variance is inherited from the primordial curvature spectrum,
\begin{equation}
\sigma^2(M_h,z)
=
\frac{1}{2\pi^2}
\int_0^\infty
k^2P_m(k,z)
W^2(kR)\,
{\rm d}k,
\label{eq:sec10_sigma_recall}
\end{equation}
with
\begin{equation}
P_m(k,z)
=
\mathcal M^2(k,z)P_{\mathcal R}^{\mathrm{dim}}(k).
\label{eq:sec10_matter_from_curvature}
\end{equation}

The statistical rarity of a massive JWST host is therefore the late-time
expression of the primordial two-point function after transfer, growth,
smoothing, and collapse. This does not imply that a single extreme galaxy
directly measures inflation. It identifies why rare high-redshift hosts can
be especially sensitive to changes in the primordial fluctuation amplitude or
shape: such changes alter $\sigma(M_h,z)$, which changes $\nu$ and hence the
predicted abundance.

\subsection{Sheth--Tormen and GUREFT as complementary abundance prescriptions}
\label{subsec:sec10_st_gureft}

The Sheth--Tormen and GUREFT calculations use the same primordial curvature
spectrum, transfer function, growth history, matter power spectrum, and
self-consistently calculated mass variance. They differ in the mapping from
those linear fluctuation statistics to the abundance of collapsed halos.
Sheth--Tormen provides an analytic peak-height description, whereas GUREFT
provides a simulation-calibrated high-redshift multiplicity function
\citep{ShethTormen1999,Yung2024}.

For the differential abundance, define
\begin{equation}
\mathcal R_{\rm HMF}(M_h,z)
=
\frac{
({\rm d}n/{\rm d}M_h)_{\rm GUREFT}
}{
({\rm d}n/{\rm d}M_h)_{\rm ST}
}.
\label{eq:sec10_hmf_ratio}
\end{equation}
The corresponding cumulative-number comparison is
\begin{equation}
\mathcal R_n(>M_h,z)
=
\frac{
n_{\rm GUREFT}(>M_h,z)
}{
n_{\rm ST}(>M_h,z)
},
\label{eq:sec10_cumulative_number_ratio}
\end{equation}
and the cumulative-mass comparison is
\begin{equation}
\mathcal R_\rho(>M_h,z)
=
\frac{
\rho_{h,{\rm GUREFT}}(>M_h,z)
}{
\rho_{h,{\rm ST}}(>M_h,z)
}.
\label{eq:sec10_cumulative_mass_ratio}
\end{equation}

These ratios are not assumed to depend on redshift alone. Because the GUREFT
multiplicity function has its own mass- and redshift-dependent calibration,
the ST--GUREFT comparison can modify both the normalization and the shape of
the available halo population. Consequently, the difference between the two
prescriptions must be propagated through the cumulative integrals rather than
represented by a single multiplicative factor.

The comparison separates two logically distinct questions:
\begin{enumerate}
\item How rare is the underlying linear fluctuation, as characterized by
$\sigma(M_h,z)$ and $\nu(M_h,z)$?
\item How many collapsed halos are assigned to those fluctuation statistics
by the adopted halo-abundance prescription?
\end{enumerate}

This distinction is central to the quantum-to-structure argument. Peak rarity
traces the primordial and linear fluctuation field, while the ST--GUREFT
difference measures sensitivity to the nonlinear halo-abundance calibration.
A discrepancy in halo counts should therefore not be attributed to the
primordial spectrum until the sensitivity to the HMF prescription has been
quantified.

The GUREFT redshift calibration directly includes the principal epochs used
in this work, including $z=6$, $8$, $10$, and $13$. At the same time, the
most massive halos at the highest redshifts occupy an increasingly sparse
tail of the simulated population. Results in that extreme region are treated
as rare-peak stress tests rather than precision abundance measurements
\citep{Yung2024}.

\subsection{JWST as a stress test of the full inference chain}
\label{subsec:sec10_interpretation}

Early JWST studies identified candidate galaxies with large inferred stellar
masses at redshifts approaching and exceeding $z\sim7$--10
\citep{Labbe2023,Harikane2023,BoylanKolchin2023}. Such systems motivate
cumulative abundance tests because demanding combinations of stellar mass,
redshift, and conversion efficiency can imply rare host halos.

The interpretation is conditional. The strength of a massive-galaxy
constraint depends on the inferred stellar mass, redshift, survey volume,
completeness, cosmic variance, lensing where relevant, stellar-population
modeling, halo-mass convention, and the assumed $\epsilon_\star$. Because the
high-mass HMF is steep, changes in these quantities can propagate
nonlinearly into the inferred host abundance.

The calculation is therefore formulated as a stress test of the inference
chain rather than as a binary verdict on $\Lambda$CDM. The observational
requirement can be propagated in reverse through
\begin{equation}
\boxed{
\begin{aligned}
M_\star^{\rm obs}
&\longrightarrow
M_{h,\min}
\longrightarrow
\nu(M_{h,\min},z)
\\
&\longrightarrow
n_X(>M_{h,\min},z).
\end{aligned}}
\label{eq:sec10_observation_to_primordial}
\end{equation}

This reverse reading complements the forward quantum-to-structure chain. The
observation constrains a host requirement; the host requirement identifies a
peak rarity; the adopted HMF determines the corresponding halo supply; and
comparison of ST with GUREFT exposes the sensitivity of that inference to the
nonlinear abundance prescription. Only after the halo-calibration,
astrophysical, and observational uncertainties have been separated can a
persistent discrepancy be meaningfully propagated farther upstream toward
the primordial fluctuation field.

The next section applies the same forward-model philosophy to active galactic
nuclei, where the observable mapping introduces an additional hierarchy:
AGN luminosity $\rightarrow$ black-hole mass $\rightarrow$ host-halo mass
$\rightarrow$ occupation and duty cycle.

%
%

\section{High-Redshift AGN as Coupled Growth-Clock and Rare-Host Tests}
\label{sec:highz_agn_growth_clock}

Section~\ref{sec:massive_galaxy_stress_test} treats massive galaxies as
rare-peak consistency tests by translating an observed stellar mass into a
host-halo requirement and then into the rarity and abundance of the required
halo population. Active galactic nuclei introduce an additional layer of
physics. A luminous AGN must not only inhabit an allowed halo; it must also
contain a black hole that formed early enough and grew rapidly enough to
reach the mass implied by its luminosity.

High-redshift AGN therefore provide two simultaneous clocks:
\begin{equation}
\boxed{
\text{structure clock}
\quad+\quad
\text{black-hole growth clock}
}
\label{eq:sec11_two_clocks}
\end{equation}
The first asks whether a sufficiently massive host halo exists at the
observed redshift. The second asks whether a black-hole seed can reach the
inferred mass within the available cosmic time. The two clocks are coupled
but physically distinct.

\subsection{From AGN luminosity to black-hole mass}
\label{subsec:sec11_luminosity_to_bh}

The characteristic luminosity scale of an accreting black hole is the
Eddington luminosity,
\begin{equation}
L_{\rm Edd}
=
\frac{4\pi G M_{\rm BH}m_p c}{\sigma_{\rm T}},
\label{eq:sec11_eddington}
\end{equation}
or numerically,
\begin{equation}
L_{\rm Edd}
=
1.26\times10^{38}
\left(\frac{M_{\rm BH}}{M_\odot}\right)
{\rm erg\,s^{-1}}.
\label{eq:sec11_eddington_numeric}
\end{equation}
Writing the bolometric luminosity as
\begin{equation}
L_{\rm bol}
=
\lambda_{\rm Edd}L_{\rm Edd},
\label{eq:sec11_lambda_edd}
\end{equation}
the black-hole mass inferred from an observed luminosity is
\begin{equation}
M_{\rm BH}
=
\frac{L_{\rm bol}}
{1.26\times10^{38}\lambda_{\rm Edd}}
\,M_\odot,
\label{eq:sec11_bh_from_luminosity}
\end{equation}
with $L_{\rm bol}$ in ${\rm erg\,s^{-1}}$.

This conversion is conditional rather than unique. The inferred $M_{\rm BH}$
depends directly on the assumed Eddington ratio and, in practice, on
bolometric corrections, obscuration, variability, and the spectral model
used to estimate $L_{\rm bol}$. Equation~(\ref{eq:sec11_bh_from_luminosity})
nevertheless provides the first step in translating a luminous high-redshift
AGN into a physical growth requirement.

\subsection{Seed formation and the growth clock}
\label{subsec:sec11_seed_growth}

The origin of the first black holes remains uncertain. Commonly discussed
channels include light seeds from Population~III stellar remnants, more
massive seeds formed through direct collapse, and primordial black holes
\citep{Abel2002,BrommLarson2004,Begelman2006,LodatoNatarajan2006,
Inayoshi2020,Carr2020}. The distinction matters because the time required to
reach a given $M_{\rm BH}$ depends logarithmically on
$M_{\rm BH}/M_{\rm seed}$.

If a fraction $\epsilon_r$ of the accreted rest-mass energy is radiated, the
black-hole growth equation may be written
\begin{equation}
\frac{{\rm d}M_{\rm BH}}{{\rm d}t}
=
\frac{1-\epsilon_r}{\epsilon_r}
\frac{L_{\rm bol}}{c^2}.
\label{eq:sec11_growth_equation}
\end{equation}
For accretion at a constant Eddington ratio,
\begin{equation}
L_{\rm bol}
=
\lambda_{\rm Edd}L_{\rm Edd},
\end{equation}
the solution is exponential,
\begin{equation}
M_{\rm BH}(t)
=
M_{\rm seed}
\exp\left(
\frac{t-t_{\rm seed}}{t_{\rm ef}}
\right),
\label{eq:sec11_exponential_growth}
\end{equation}
with e-folding time
\begin{equation}
t_{\rm ef}
=
\frac{\epsilon_r}{1-\epsilon_r}
\frac{\sigma_{\rm T}c}{4\pi Gm_p}
\frac{1}{\lambda_{\rm Edd}}.
\label{eq:sec11_efolding_time}
\end{equation}

For $\epsilon_r\simeq0.1$ and $\lambda_{\rm Edd}\simeq1$, this is the
familiar Salpeter-like growth time of order a few times $10^7\,{\rm yr}$.
The elapsed growth time required to reach an observed mass is
\begin{equation}
\Delta t_{\rm grow}
=
t_{\rm ef}
\ln\left(\frac{M_{\rm BH}}{M_{\rm seed}}\right).
\label{eq:sec11_growth_time}
\end{equation}
A viable model must satisfy
\begin{equation}
\boxed{
\Delta t_{\rm grow}
\leq
t(z_{\rm obs})-t(z_{\rm seed})
}
\label{eq:sec11_time_constraint}
\end{equation}
for the adopted cosmology and seed epoch.

The growth requirement can be relaxed by increasing $M_{\rm seed}$,
increasing $\lambda_{\rm Edd}$, lowering the radiative efficiency, allowing
super-Eddington episodes, or invoking black-hole mergers. Conversely, light
seeds growing intermittently below the Eddington rate face the most
restrictive timing requirement.

\subsection{From black-hole mass to host-halo requirement}
\label{subsec:sec11_bh_to_halo}

The growth clock alone does not determine whether the surrounding large-scale
structure can support the observed AGN. A second mapping is required from
black-hole mass to host-halo mass. A compact parameterization is
\begin{equation}
\epsilon_{\rm BH}
=
\frac{M_{\rm BH}}{M_h},
\label{eq:sec11_epsilon_bh}
\end{equation}
so that
\begin{equation}
M_{h,\min}
=
\frac{M_{\rm BH}}{\epsilon_{\rm BH}}.
\label{eq:sec11_min_host_mass}
\end{equation}

Equation~(\ref{eq:sec11_min_host_mass}) is not assumed to be a fundamental
universal relation. It summarizes the uncertain mapping between black-hole
mass and halo mass at an epoch where local scaling relations may not apply.
Its purpose is to make the inference explicit:
\begin{equation}
L_{\rm bol}
\longrightarrow
M_{\rm BH}
\longrightarrow
M_{h,\min}.
\label{eq:sec11_observable_to_host}
\end{equation}

A lower assumed $\lambda_{\rm Edd}$ increases the inferred black-hole mass.
A lower assumed $\epsilon_{\rm BH}$ then increases the required host-halo
mass. Both changes push the system toward a rarer part of the halo
distribution.

For the GUREFT abundance calculation, $M_h$ is identified with the virial
mass $M_{\rm vir}$ defined in Section~\ref{sec:halo_mass_functions}. A
black-hole--halo relation expressed using another halo-mass convention must
therefore be converted consistently or treated as a systematic uncertainty.

\subsection{The rare-host clock}
\label{subsec:sec11_rare_host}

Once $M_{h,\min}$ is specified, the required host corresponds to the peak
height
\begin{equation}
\nu_{\min}(z)
=
\frac{\delta_c}{\sigma(M_{h,\min},z)}.
\label{eq:sec11_nu_min}
\end{equation}
The available host abundance is
\begin{equation}
n_X(>M_{h,\min},z)
=
\int_{M_{h,\min}}^\infty
\frac{{\rm d}n_X}{{\rm d}M_h}\,
{\rm d}M_h,
\qquad
X\in\{{\rm ST,GUREFT}\}.
\label{eq:sec11_host_abundance}
\end{equation}

The observed active population cannot exceed the supply of eligible hosts.
Including black-hole occupation and activity gives
\begin{equation}
n_{{\rm AGN},X}(z)
=
\int_{M_{h,\min}}^\infty
f_{\rm occ}(M_h,z)
f_{\rm duty}(M_h,z)
\frac{{\rm d}n_X}{{\rm d}M_h}\,
{\rm d}M_h.
\label{eq:sec11_agn_abundance_general}
\end{equation}
For constant occupation and duty factors above the threshold,
\begin{equation}
n_{{\rm AGN},X}(z)
=
f_{\rm occ}f_{\rm duty}
n_X(>M_{h,\min},z).
\label{eq:sec11_agn_abundance_simple}
\end{equation}

Because
\begin{equation}
0\leq f_{\rm occ},f_{\rm duty}\leq1,
\end{equation}
the cumulative halo abundance itself provides a strict upper envelope for a
given $M_{h,\min}$. If an observed active population is difficult to
accommodate even for $f_{\rm occ}=f_{\rm duty}=1$, reducing either factor
strengthens rather than weakens the host-supply constraint.

\subsection{A coupled time--abundance consistency condition}
\label{subsec:sec11_joint_constraint}

The black-hole growth and host-abundance requirements form a joint
consistency problem:
\begin{equation}
\boxed{
\begin{aligned}
\Delta t_{\rm grow}
&\leq
t(z_{\rm obs})-t(z_{\rm seed}),
\\[3pt]
n_{\rm AGN}^{\rm obs}(z)
&\leq
f_{\rm occ}f_{\rm duty}
n_X(>M_{h,\min},z).
\end{aligned}}
\label{eq:sec11_joint_condition}
\end{equation}

The first inequality is a time constraint; the second is an abundance
constraint. A viable interpretation must satisfy both. Four regimes can be
distinguished:
\begin{enumerate}
\item growth is readily accommodated and hosts are abundant;
\item growth is restrictive while hosts remain abundant;
\item growth is readily accommodated but suitable hosts are rare;
\item both growth and host supply are restrictive.
\end{enumerate}
The fourth regime couples the black-hole assembly problem most strongly to
the rare tail of cosmic structure formation.

\subsection{Sheth--Tormen and GUREFT in the rare-host test}
\label{subsec:sec11_st_gureft}

The black-hole growth calculation is independent of the choice between the
Sheth--Tormen and GUREFT halo mass functions. For fixed
$M_{\rm seed}$, $\epsilon_r$, $\lambda_{\rm Edd}$, and $z_{\rm seed}$, the
quantities $M_{\rm BH}$ and $\Delta t_{\rm grow}$ are unchanged when the HMF
prescription is changed. The HMF enters only when the inferred host threshold
is translated into an available population of halos.

For identical black-hole--halo, occupation, and duty-cycle assumptions,
define
\begin{equation}
\mathcal R_{\rm AGN}(M_{h,\min},z)
=
\frac{
n_{{\rm AGN},{\rm GUREFT}}(z)
}{
n_{{\rm AGN},{\rm ST}}(z)
}.
\label{eq:sec11_agn_hmf_ratio}
\end{equation}
Using the general occupation model,
\begin{equation}
\mathcal R_{\rm AGN}
=
\frac{
\displaystyle
\int_{M_{h,\min}}^\infty
f_{\rm occ}f_{\rm duty}
({\rm d}n_{\rm GUREFT}/{\rm d}M_h)\,{\rm d}M_h
}{
\displaystyle
\int_{M_{h,\min}}^\infty
f_{\rm occ}f_{\rm duty}
({\rm d}n_{\rm ST}/{\rm d}M_h)\,{\rm d}M_h
}.
\label{eq:sec11_agn_hmf_ratio_general}
\end{equation}
If the same constant occupation and duty factors are applied above the same
host threshold, these factors cancel and
\begin{equation}
\mathcal R_{\rm AGN}
=
\frac{
n_{\rm GUREFT}(>M_{h,\min},z)
}{
n_{\rm ST}(>M_{h,\min},z)
}.
\label{eq:sec11_agn_hmf_ratio_simple}
\end{equation}

Unlike a global redshift-dependent rescaling, this ratio depends on the host
threshold because the ST and GUREFT mass functions differ as functions of
both mass and redshift. The comparison therefore isolates the sensitivity of
the AGN host reservoir to the nonlinear halo-abundance prescription without
altering the black-hole growth law, seed model, Eddington ratio, occupation
model, or duty cycle.

The GUREFT calibration covers the principal redshifts considered here,
including $z=6$, $8$, $10$, and $13$ \citep{Yung2024}. The most massive host
requirements at the highest redshifts nevertheless probe an increasingly
sparse part of the simulation-calibrated population. Such cases are
interpreted as rare-host stress tests rather than precision abundance
measurements.

\subsection{The reverse path from an AGN to primordial fluctuations}
\label{subsec:sec11_reverse_chain}

The connection to the primordial fluctuation field becomes clearest when the
calculation is read backward from the observation:
\begin{equation}
\begin{aligned}
L_{\rm bol}^{\rm obs}
&\longrightarrow
M_{\rm BH}
\longrightarrow
M_{h,\min}
\longrightarrow
\nu(M_{h,\min},z)
\\
&\longrightarrow
n_X(>M_{h,\min},z)
\longrightarrow
\sigma(M_h,z)
\longrightarrow
\mathcal P_{\mathcal R}(k).
\end{aligned}
\label{eq:sec11_reverse_chain}
\end{equation}

The last arrow is not a direct inversion from one AGN to an inflationary
spectrum. It identifies where the cosmological sensitivity enters. The
abundance of a rare high-redshift host is strongly sensitive to peak rarity,
while the peak height depends on the variance inherited from the primordial
curvature spectrum. At the same time, the ST--GUREFT comparison measures how
the nonlinear abundance calibration affects the inferred host supply at
fixed primordial and linear inputs.

High-redshift AGN are therefore unusual probes: the same population can test
both the finite time available for black-hole assembly and the abundance of
rare collapsed structures seeded by primordial fluctuations.

\subsection{Interpretation}
\label{subsec:sec11_interpretation}

The existence of a luminous AGN at high redshift does not by itself identify
a discrepancy with the standard cosmological model. The inference contains
several astrophysical layers, including seed formation, Eddington-ratio
distributions, radiative efficiency, mergers, obscuration, variability,
host-halo mapping, occupation, duty cycle, and survey selection. These
uncertainties can be substantial.

The more informative question is whether a physically plausible set of seed,
growth, host, and occupation assumptions can satisfy the growth clock and the
rare-host clock simultaneously. A persistent abundance discrepancy must also
be tested against the choice of halo-abundance prescription before it is
propagated upstream toward the primordial fluctuation field.

In this formulation, early AGN provide a bridge between two otherwise
separate problems: the assembly of supermassive black holes and the
statistics of rare primordial fluctuations. The observable population lies
at the point where those two histories meet.

The next section turns from individual and threshold-based constraints to the
population-level luminosity description, where halo abundance and
halo-to-light mappings can be compared with observed high-redshift galaxy and
AGN number densities.

%
%

\section{From Halo Populations to Light: UV Luminosity Functions as an Information-Transfer Test}
\label{sec:uvlf_information_transfer}

Sections~\ref{sec:massive_galaxy_stress_test} and
\ref{sec:highz_agn_growth_clock} use individual objects and threshold
populations to probe the rare-halo regime. A luminosity function asks a
different question: whether the distribution of observed luminosities can be
generated from the cosmological halo population after applying a physically
specified halo-to-light mapping.

The ultraviolet luminosity function (UVLF) is therefore a population-level
test of information propagated through
\begin{equation}
\begin{aligned}
\mathcal P_{\mathcal R}(k)
&\longrightarrow
P_m(k,z)
\longrightarrow
\frac{{\rm d}n_X}{{\rm d}M_h}
\\
&\longrightarrow
P(M_{\rm UV}\mid M_h,z)
\longrightarrow
\Phi_X(M_{\rm UV},z),
\end{aligned}
\label{eq:sec12_full_uv_chain}
\end{equation}
where
\begin{equation}
X\in\{{\rm ST,GUREFT}\}.
\label{eq:sec12_hmf_branches}
\end{equation}
The UVLF does not test the halo mass function in isolation. It tests the
combination of halo supply and the baryonic mapping from halo mass to emitted
light.

\subsection{From halo mass to star formation}
\label{subsec:sec12_halo_to_sfr}

For a halo of mass $M_h$, the cosmic-mean baryonic mass is
\begin{equation}
M_b=f_bM_h,
\label{eq:sec12_baryonic_mass}
\end{equation}
with $f_b=\Omega_b/\Omega_m$. If an effective fraction $\epsilon_\star$ of
this reservoir is converted into stars over a characteristic timescale
$t_\star$, the star-formation rate is
\begin{equation}
{\rm SFR}(M_h,z)
=
\frac{\epsilon_\star f_bM_h}{t_\star(z)}.
\label{eq:sec12_sfr_general}
\end{equation}

For the transparent baseline model used here, the timescale is related to the
Hubble time,
\begin{equation}
t_\star(z)=f_{\rm dyn}H^{-1}(z),
\label{eq:sec12_tstar}
\end{equation}
so that
\begin{equation}
{\rm SFR}(M_h,z)
=
\frac{\epsilon_\star f_b}{f_{\rm dyn}}
M_hH(z).
\label{eq:sec12_sfr}
\end{equation}

At fixed $M_h$, a shorter characteristic timescale at higher redshift can
raise the instantaneous star-formation rate and hence the UV luminosity even
while the cosmological abundance of such halos declines rapidly. A bright
individual galaxy and an abundant galaxy population are therefore different
requirements.

For the GUREFT branch, the halo mass entering the abundance calculation is
the virial mass $M_{\rm vir}$ defined in
Section~\ref{sec:halo_mass_functions}. A halo-to-light relation expressed in
another halo-mass convention must be converted consistently or included as a
systematic uncertainty.

\subsection{From star formation to rest-frame UV magnitude}
\label{subsec:sec12_sfr_to_uv}

For a stellar population forming stars continuously, the rest-frame UV
specific luminosity may be written
\begin{equation}
L_{\nu,{\rm UV}}
=
\kappa_{\rm UV}\,{\rm SFR},
\label{eq:sec12_uv_conversion}
\end{equation}
where $\kappa_{\rm UV}$ depends on the stellar initial mass function,
metallicity, star-formation history, and population-synthesis assumptions.

The corresponding AB absolute magnitude is
\begin{equation}
M_{\rm UV}
=
51.60
-
2.5\log_{10}
\left[
\frac{L_{\nu,{\rm UV}}}
{{\rm erg\,s^{-1}\,Hz^{-1}}}
\right].
\label{eq:sec12_ab_magnitude}
\end{equation}
Combining Equations~(\ref{eq:sec12_sfr})--(\ref{eq:sec12_ab_magnitude})
defines the deterministic baseline mapping
\begin{equation}
M_h
\longrightarrow
M_{\rm UV}(M_h,z;\epsilon_\star,f_{\rm dyn},\kappa_{\rm UV}).
\label{eq:sec12_halo_to_uv}
\end{equation}
This mapping determines the light assigned to an individual halo; it does not
determine how many such halos exist.

\subsection{The UV luminosity function}
\label{subsec:sec12_uvlf}

For a deterministic and monotonic halo-to-light relation,
\begin{equation}
\Phi_X(M_{\rm UV},z)
=
\left.
\frac{{\rm d}n_X}{{\rm d}M_h}
\left|
\frac{{\rm d}M_h}{{\rm d}M_{\rm UV}}
\right|
\right|_{M_h=M_h(M_{\rm UV},z)}.
\label{eq:sec12_uvlf_deterministic}
\end{equation}
The more general formulation is
\begin{equation}
\Phi_X(M_{\rm UV},z)
=
\int_0^\infty
\frac{{\rm d}n_X}{{\rm d}M_h}
P(M_{\rm UV}\mid M_h,z)\,
{\rm d}M_h,
\label{eq:sec12_uvlf_general}
\end{equation}
where $P(M_{\rm UV}\mid M_h,z)$ contains astrophysical scatter and
stochasticity in the halo-to-light relation.

Equation~(\ref{eq:sec12_uvlf_general}) is the population-level counterpart of
the forward model in Section~\ref{sec:halo_to_observable_interface}. It makes
explicit that the predicted luminosity function combines two conceptually
distinct ingredients:
\begin{equation}
\boxed{
\begin{gathered}
\text{cosmological halo supply}\\
\otimes\\
\text{baryonic luminosity mapping}
\end{gathered}
}.
\label{eq:sec12_convolution_statement}
\end{equation}

\subsection{Why scatter matters most at the bright end}
\label{subsec:sec12_scatter}

The bright end of the UVLF is associated with the steep part of the halo mass
function. Scatter in $M_{\rm UV}$ at fixed $M_h$ can therefore have a
disproportionate effect on predicted bright-source abundances. More numerous,
moderately massive halos may scatter into a rare bright bin, while the much
smaller population of very massive halos provides fewer systems that can
scatter in the opposite direction.

This Eddington-bias-like asymmetry can enhance the bright end relative to a
deterministic mapping, with the effect increasing as the underlying halo
abundance becomes steeper. A bright-end mismatch can consequently arise from
several distinct layers:
\begin{equation}
\begin{aligned}
&\text{primordial fluctuation statistics},\\
&\text{halo-abundance calibration},\\
&\text{scatter in the halo-to-light mapping}.
\end{aligned}
\label{eq:sec12_three_bright_end_layers}
\end{equation}
A luminosity-function discrepancy therefore cannot be assigned uniquely to
cosmology without controlling the baryonic mapping.

\subsection{Propagation of ST and GUREFT into luminosity space}
\label{subsec:sec12_st_gureft_uvlf}

The Sheth--Tormen and GUREFT branches are propagated through the same
halo-to-light kernel. Their luminosity functions are
\begin{equation}
\Phi_{\rm ST}(M_{\rm UV},z)
=
\int
\frac{{\rm d}n_{\rm ST}}{{\rm d}M_h}
P(M_{\rm UV}\mid M_h,z)\,
{\rm d}M_h
\label{eq:sec12_st_uvlf}
\end{equation}
and
\begin{equation}
\begin{aligned}
&\Phi_{\rm GUREFT}(M_{\rm UV},z)\\
&\quad=\int
\frac{{\rm d}n_{\rm GUREFT}}{{\rm d}M_h}
P(M_{\rm UV}\mid M_h,z)\,{\rm d}M_h.
\end{aligned}
\label{eq:sec12_gureft_uvlf}
\end{equation}

The corresponding diagnostic ratio is
\begin{equation}
\mathcal R_{\Phi,{\rm HMF}}(M_{\rm UV},z)
=
\frac{
\Phi_{\rm GUREFT}(M_{\rm UV},z)
}{
\Phi_{\rm ST}(M_{\rm UV},z)
}.
\label{eq:sec12_uvlf_hmf_ratio}
\end{equation}
Because the GUREFT-to-ST halo-abundance ratio depends on both mass and
redshift, $\mathcal R_{\Phi,{\rm HMF}}$ is generally luminosity dependent as
well as redshift dependent. It cannot be represented by a single
redshift-only multiplicative factor.

For a deterministic one-to-one mapping, the common Jacobian cancels at fixed
$M_{\rm UV}$, giving
\begin{equation}
\mathcal R_{\Phi,{\rm HMF}}(M_{\rm UV},z)
=
\left.
\frac{
({\rm d}n/{\rm d}M_h)_{\rm GUREFT}
}{
({\rm d}n/{\rm d}M_h)_{\rm ST}
}
\right|_{M_h=M_h(M_{\rm UV},z)}.
\label{eq:sec12_uvlf_ratio_deterministic}
\end{equation}
With scatter, the ratio instead becomes a weighted integral over a range of
halo masses through Equation~(\ref{eq:sec12_uvlf_general}).

This comparison changes the cosmological halo supply while leaving the
halo-to-light kernel fixed. By contrast, changing $\epsilon_\star$,
$f_{\rm dyn}$, $\kappa_{\rm UV}$, dust attenuation, or the scatter
distribution changes the astrophysical mapping between $M_h$ and
$M_{\rm UV}$. Separating these operations allows the numerical results to
identify whether a change in the predicted UVLF entered through halo
statistics or through the baryonic light assignment.

\subsection{Primordial sensitivity in luminosity space}
\label{subsec:sec12_primordial_uv_response}

A perturbation of the primordial spectrum enters the UVLF at a different
stage from a change of HMF prescription. At fixed spectral shape and
background cosmology, a controlled change in $A_s$ propagates through
\begin{equation}
\delta A_s
\longrightarrow
\delta P_m
\longrightarrow
\delta\sigma
\longrightarrow
\delta n_X
\longrightarrow
\delta\Phi_X.
\label{eq:sec12_primordial_uv_chain}
\end{equation}
For a chosen HMF prescription, define
\begin{equation}
\mathcal R_{\Phi,A_s}(M_{\rm UV},z)
=
\frac{
\Phi_X(M_{\rm UV},z;1.02A_{s,0})
}{
\Phi_X(M_{\rm UV},z;A_{s,0})
}.
\label{eq:sec12_uvlf_As_ratio}
\end{equation}

This diagnostic should not be confused with
$\mathcal R_{\Phi,{\rm HMF}}$. The first changes the primordial fluctuation
amplitude and therefore the upstream mass variance; the second changes the
mapping from the same calculated fluctuation field into collapsed halo
abundance. Their different locations in the causal chain are central to the
numerical experiments developed in Section~\ref{sec:numerical_information_flow}.

\subsection{JWST comparison as a population-level closure test}
\label{subsec:sec12_jwst_closure}

Representative JWST UVLF measurements at $z\simeq6$--10 provide an
observational test of the complete forward chain
\citep{Donnan2023,Donnan2024,Franco2025}. The comparison moves beyond
individual extreme galaxies by testing source abundances over a range of
luminosities.

The observational uncertainties include photometric-redshift errors,
completeness, cosmic variance, selection functions, lensing where relevant,
AGN contamination, and finite survey volume. The theoretical uncertainties
include star-formation efficiency, star-formation timescale, stellar
populations, metallicity, dust, feedback, burstiness, scatter in the
halo-to-light relation, and the halo-abundance prescription. The comparison
is therefore treated as a closure test:
\begin{equation}
\boxed{
\begin{aligned}
&\text{Does one physically consistent chain connect}\\
&\mathcal{P}_{\mathcal R}(k)
\quad \text{to} \quad
\Phi_{\rm obs}(M_{\rm UV},z)\,?
\end{aligned}
}
\label{eq:sec12_closure_question}
\end{equation}

Agreement indicates that the primordial spectrum, halo statistics, and
baryonic mapping can be mutually consistent over the observed luminosity
range. A persistent discrepancy instead identifies a point at which the chain
must be examined more carefully; it does not by itself specify which layer is
responsible.

\subsection{Galaxies and AGN as separate luminosity channels}
\label{subsec:sec12_galaxy_agn_separation}

The galaxy UVLF developed above is a stellar halo-to-light calculation and is
not an AGN luminosity function. For galaxies,
\begin{equation}
M_h
\longrightarrow
M_b
\longrightarrow
{\rm SFR}
\longrightarrow
L_{\nu,{\rm UV}}
\longrightarrow
M_{\rm UV}.
\label{eq:sec12_galaxy_light_chain}
\end{equation}
For AGN, the corresponding luminosity chain is
\begin{equation}
M_h
\longrightarrow
M_{\rm BH}
\longrightarrow
\lambda_{\rm Edd}
\longrightarrow
L_{\rm bol},
\label{eq:sec12_agn_light_chain}
\end{equation}
with occupation, duty cycle, obscuration, and variability entering the
selection probability.

The two channels share the underlying halo supply but not the same
astrophysical kernel. In particular, the AGN duty cycle is not multiplied
into the stellar galaxy UVLF. A general AGN luminosity function would have
the form
\begin{equation}
\begin{aligned}
&\Phi_{{\rm AGN},X}(L,z)\\
&\quad=\int \frac{{\rm d}n_X}{{\rm d}M_h}
f_{\rm occ}(M_h,z)f_{\rm duty}(M_h,z)\\
&\qquad\times P(L\mid M_h,z,{\rm active})\,{\rm d}M_h,
\end{aligned}
\label{eq:sec12_agn_lf_general}
\end{equation}
but constructing such a model lies beyond the stellar UVLF calculation used
here.

\subsection{Reading the UVLF backward toward the primordial spectrum}
\label{subsec:sec12_reverse_uvlf}

The forward calculation begins with quantum initial conditions and ends with
an observable luminosity function. The interpretation can also be read in
the opposite direction:
\begin{equation}
\begin{aligned}
\Phi_{\rm obs}(M_{\rm UV},z)
&\longrightarrow
P(M_{\rm UV}\mid M_h,z)
\longrightarrow
\frac{{\rm d}n_X}{{\rm d}M_h}
\\
&\longrightarrow
\sigma(M_h,z)
\longrightarrow
\mathcal P_{\mathcal R}(k).
\end{aligned}
\label{eq:sec12_reverse_uvlf}
\end{equation}

This is not a unique inversion. The baryonic kernel, halo calibration, and
observational selection must be constrained independently. Nevertheless, a
luminosity function samples a range of halo masses and therefore carries
information about the shape as well as the normalization of the underlying
halo population.

The UVLF can thus be viewed as an information-compression observable: it
combines the effects of the primordial spectrum, transfer function, growth,
collapse statistics, HMF calibration, and baryonic emission physics into a
single measurable population distribution. The numerical calculations in
the following section separate these contributions by comparing a controlled
perturbation of the primordial amplitude with a controlled change from the
Sheth--Tormen to the GUREFT halo-abundance prescription.

%
%

\section{Numerical Results: Propagating Primordial Information to JWST Observables}
\label{sec:numerical_information_flow}

The preceding sections establish the physical sequence connecting primordial
quantum fluctuations to the populations observed at cosmic dawn. We now use
that sequence as a set of controlled numerical experiments designed to
identify where information about the primordial fluctuation field is
preserved, where it is amplified, and where independent halo and
astrophysical assumptions enter.

Two perturbations of the forward model are considered. In the first, the
primordial amplitude $A_s$ is changed slightly while the spectral shape,
transfer function, background cosmology, growth history, and halo-collapse
prescription are held fixed. In the second, the primordial and linear fields
are held fixed while the halo-abundance prescription is changed from
Sheth--Tormen to the GUREFT calibration of \citet{Yung2024}. The two
experiments enter the causal chain at different locations:
\begin{equation}
\begin{aligned}
\delta A_s
&\rightarrow
\delta\sigma
\rightarrow
\delta\nu
\rightarrow
\delta n_h
\rightarrow
\delta\Phi_{\rm UV},
\\
{\rm ST}\rightarrow{\rm GUREFT}
&:\quad
\mathcal P_{\mathcal R},\,P_m,\,\sigma
\ \text{fixed},
\quad
\delta n_h
\rightarrow
\delta\Phi_{\rm UV}.
\end{aligned}
\label{eq:sec13_two_experiments}
\end{equation}
The first experiment changes the upstream fluctuation field. The second
changes the nonlinear mapping from that same fluctuation field to collapsed
halo abundance. Their observable consequences can therefore be compared
without conflating primordial sensitivity with HMF calibration.

\subsection{Baseline cosmology and numerical sequence}
\label{subsec:sec13_baseline}

We adopt a spatially flat Planck-like $\Lambda$CDM cosmology, using parameters consistent with the Planck CMB constraints \citep{Planck2018}, with
\begin{equation}
\Omega_m=0.3158,\qquad
\Omega_b=0.0493,\qquad
h=0.6732,
\label{eq:sec13_background_cosmology}
\end{equation}
and
\begin{equation}
n_s=0.96605,\qquad
\sigma_{8,0}=0.812,\qquad
\delta_c=1.686.
\label{eq:sec13_perturbation_parameters}
\end{equation}
The corresponding baryon fraction is
\begin{equation}
f_b=\frac{\Omega_b}{\Omega_m}\simeq0.156.
\label{eq:sec13_baryon_fraction}
\end{equation}

The fiducial matter spectrum is normalized so that the baseline model
satisfies $\sigma_{8,0}=0.812$, and the Eisenstein--Hu transfer prescription
is used \citep{EisensteinHu1998}. In the controlled $A_s$ experiment below,
the perturbed spectra are \emph{not} renormalized back to the fiducial
$\sigma_8$. Instead, the imposed primordial-amplitude change is allowed to
propagate into $P_m$, $\sigma$, and the halo population. Numerical
implementation and convergence tests are described in
Appendix~\ref{app:numerical_pipeline}.

The numerical information flow is
\begin{equation}
\mathcal P_{\mathcal R}(k)
\rightarrow
P_m(k,z)
\rightarrow
\sigma(M,z)
\rightarrow
\nu(M,z)
\rightarrow
\frac{{\rm d}n_X}{{\rm d}M_h},
\qquad
X\in\{{\rm ST,GUREFT}\},
\label{eq:sec13_forward_core}
\end{equation}
followed by the galaxy or AGN observable mapping appropriate to the test.

For the GUREFT branch, the same self-consistently calculated
$\sigma(M,z)$ is supplied to the simulation-calibrated multiplicity
function. The GUREFT approximation for $\sigma(M)$ is therefore not used.
This preserves a common upstream calculation for ST and GUREFT and ensures
that their difference enters only at the HMF stage.

\subsection{Primordial input and transfer to the matter field}
\label{subsec:sec13_primordial_transfer}

\begin{figure}
\centering
\includegraphics[width=\columnwidth]
{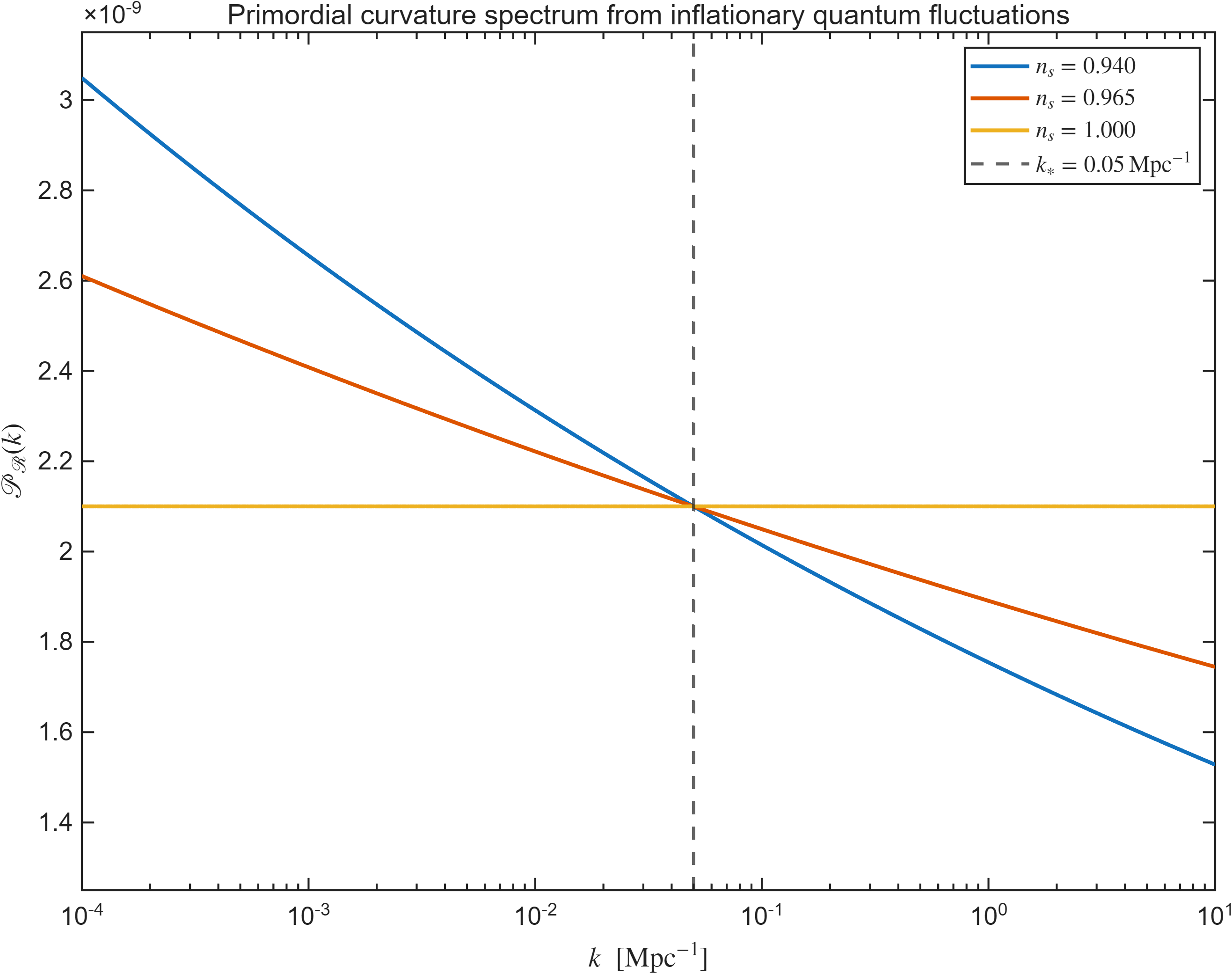}
\caption{Primordial curvature spectrum used as the numerical initial
condition. The spectrum is not itself a JWST observable; it is the upstream
statistical input subsequently processed by transfer, growth, smoothing,
collapse, and halo-to-observable mappings.}
\label{fig:sec13_primordial}
\end{figure}

The numerical initial condition is the primordial curvature spectrum
\begin{equation}
\mathcal P_{\mathcal R}(k)
=
A_s
\left(\frac{k}{k_*}\right)^{n_s-1}.
\label{eq:sec13_PR}
\end{equation}
Figure~\ref{fig:sec13_primordial} shows this upstream statistical input. Its
amplitude and tilt are inherited from the inflationary perturbation
calculation developed in
Sections~\ref{sec:inflation_perturbations} and
\ref{sec:gauge_invariant_perturbations}.

\begin{figure}
\centering
\includegraphics[width=\columnwidth]
{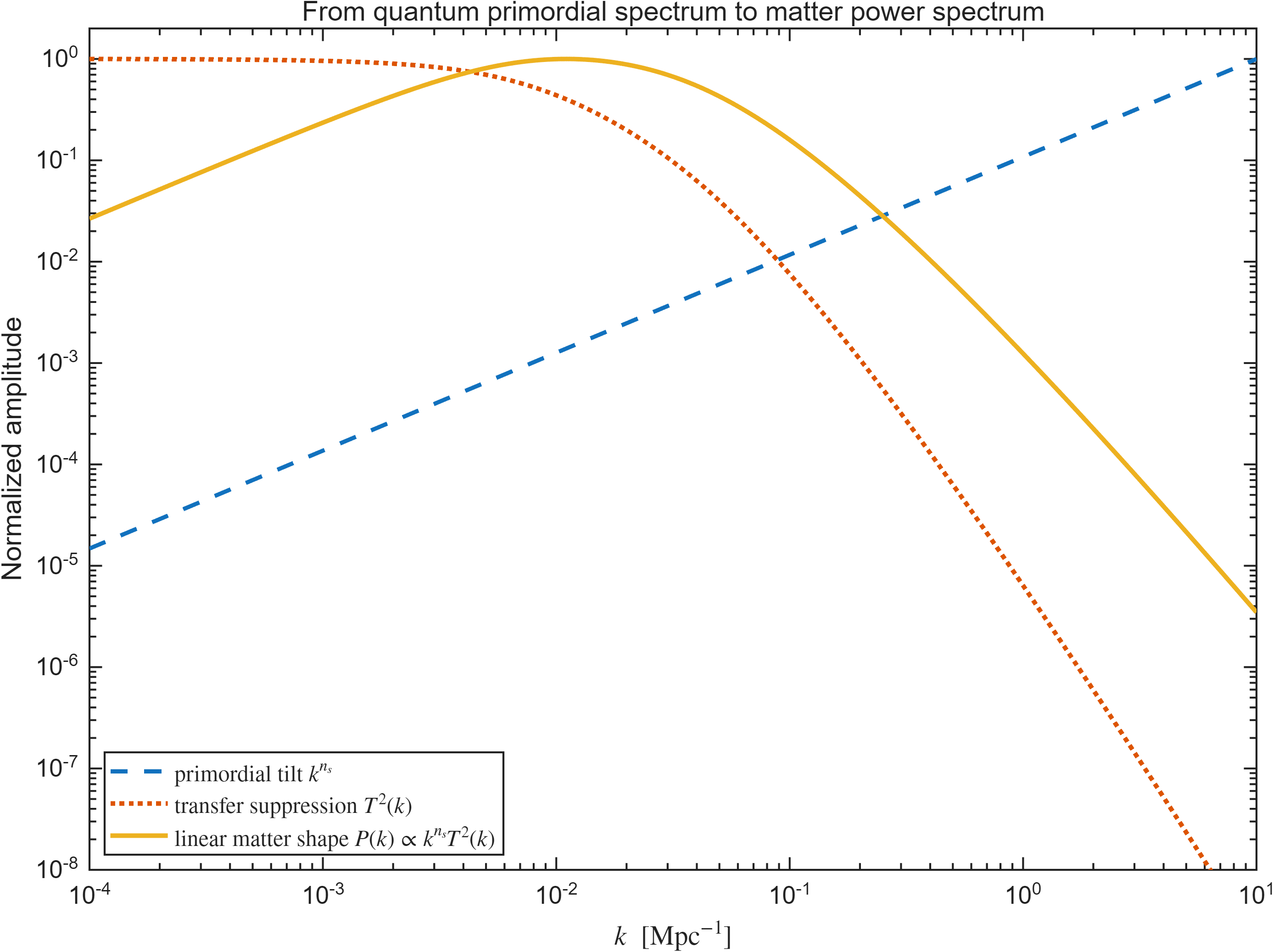}
\caption{Propagation of primordial spectral information into the linear
matter-power spectrum. The transfer function changes the scale dependence of
the fluctuations before they are smoothed on halo scales, providing the
numerical bridge between inflationary initial conditions and the statistics
of collapsed structure.}
\label{fig:sec13_matter}
\end{figure}

The corresponding linear matter spectrum is obtained after transfer-function
processing and growth,
\begin{equation}
P_m(k,z)\propto k^{n_s}T^2(k)D^2(z),
\label{eq:sec13_Pm}
\end{equation}
with the fiducial normalization fixed by $\sigma_{8,0}$.
Figure~\ref{fig:sec13_matter} illustrates this transformation. The
primordial tilt is preserved as part of the input, while the late-time
spectrum acquires the scale dependence associated with horizon entry and
matter--radiation equality.

\subsection{Controlled primordial-amplitude response}
\label{subsec:sec13_primordial_response}

The first sensitivity experiment changes the primordial amplitude by
$\pm2$ per cent while holding $n_s$, the transfer-function shape, the
background cosmology, and the growth prescription fixed. Under this
controlled rescaling,
\begin{equation}
P_m\propto A_s,\qquad
\sigma\propto A_s^{1/2},\qquad
\nu\propto A_s^{-1/2}.
\label{eq:sec13_scalings}
\end{equation}
Consequently,
\begin{equation}
\frac{\sigma(A_s)}{\sigma(A_{s,0})}
=
\left(\frac{A_s}{A_{s,0}}\right)^{1/2},
\qquad
\frac{\nu(A_s)}{\nu(A_{s,0})}
=
\left(\frac{A_s}{A_{s,0}}\right)^{-1/2}.
\label{eq:sec13_sigma_nu_response}
\end{equation}

\begin{figure}
\centering
\includegraphics[width=\columnwidth]
{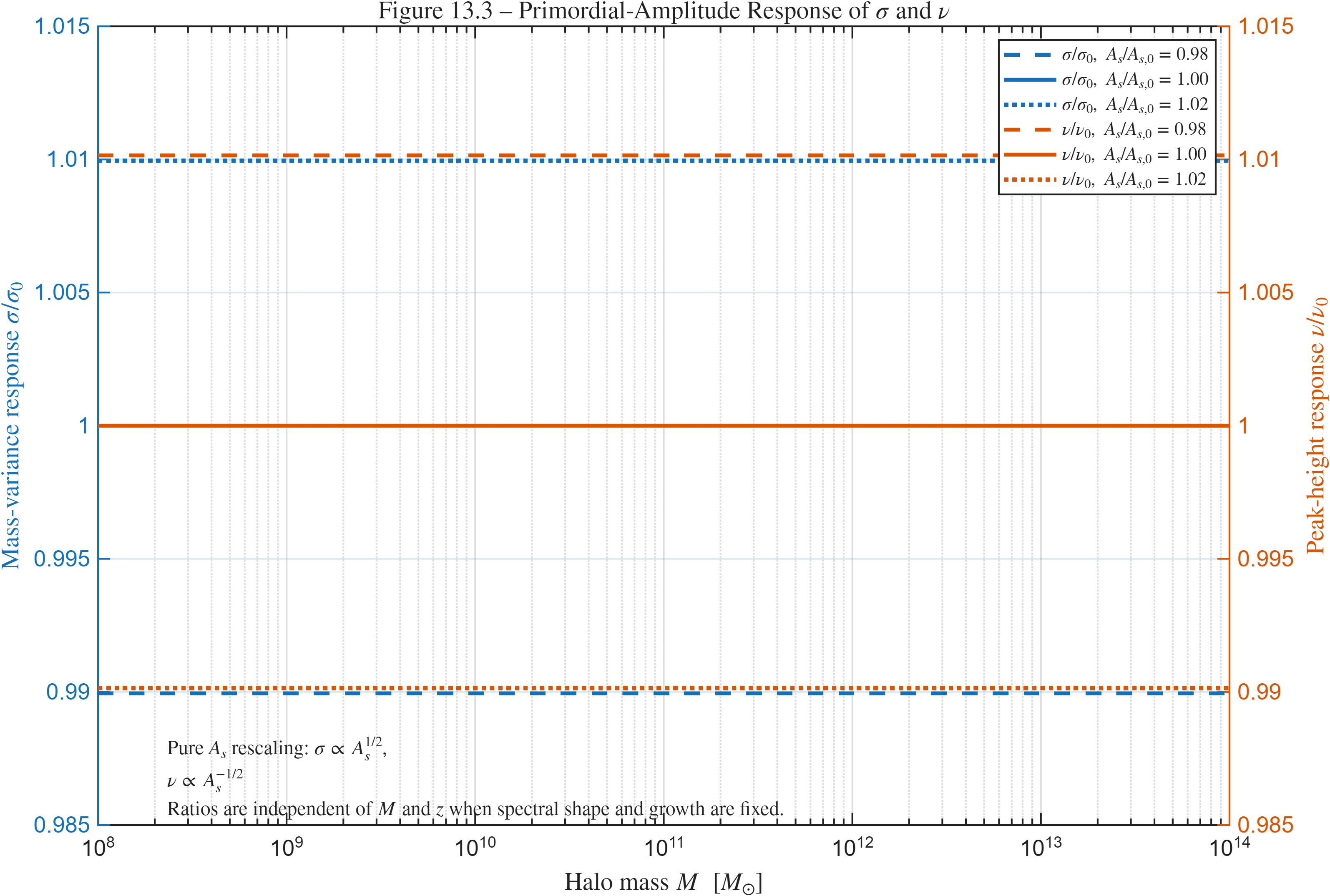}
\caption{Controlled response of the mass variance and peak height to a
$\pm2$ per cent perturbation of the primordial amplitude. With the spectral
shape and growth fixed, $\sigma/\sigma_0=(A_s/A_{s,0})^{1/2}$ and
$\nu/\nu_0=(A_s/A_{s,0})^{-1/2}$. The small response at this stage contrasts
with the larger abundance response produced by the rare-tail mapping.}
\label{fig:sec13_sigma_nu_response}
\end{figure}

Figure~\ref{fig:sec13_sigma_nu_response} makes an important distinction.
A two per cent change in $A_s$ produces only an approximately one per cent
change in $\sigma$ and an oppositely directed change of comparable magnitude
in $\nu$. The much larger response of rare halos arises when this small shift
in peak height is passed through the exponentially sensitive high-$\nu$
collapse tail.

\subsection{Exponential amplification in the Sheth--Tormen rare tail}
\label{subsec:sec13_rare_tail}

The analytic Sheth--Tormen branch makes the rare-peak amplification
particularly transparent. At high $\nu$ its dominant dependence is
\begin{equation}
n_{\rm rare}\propto
\exp\left(-\frac{a_{\rm ST}\nu^2}{2}\right),
\qquad
a_{\rm ST}=0.707.
\label{eq:sec13_ST_tail}
\end{equation}
Combining this with Equation~(\ref{eq:sec13_scalings}) gives, to leading
order,
\begin{equation}
{\rm d}\ln n_{\rm rare}
\simeq
\frac{a_{\rm ST}\nu^2}{2}\,
{\rm d}\ln A_s.
\label{eq:sec13_rare_derivative}
\end{equation}
The amplification therefore increases approximately as $\nu^2$.
Appendix~\ref{app:rare_peak_sensitivity} gives the corresponding derivation
and discusses the limits of the high-$\nu$ approximation.

\begin{figure}
\centering
\includegraphics[width=\columnwidth]
{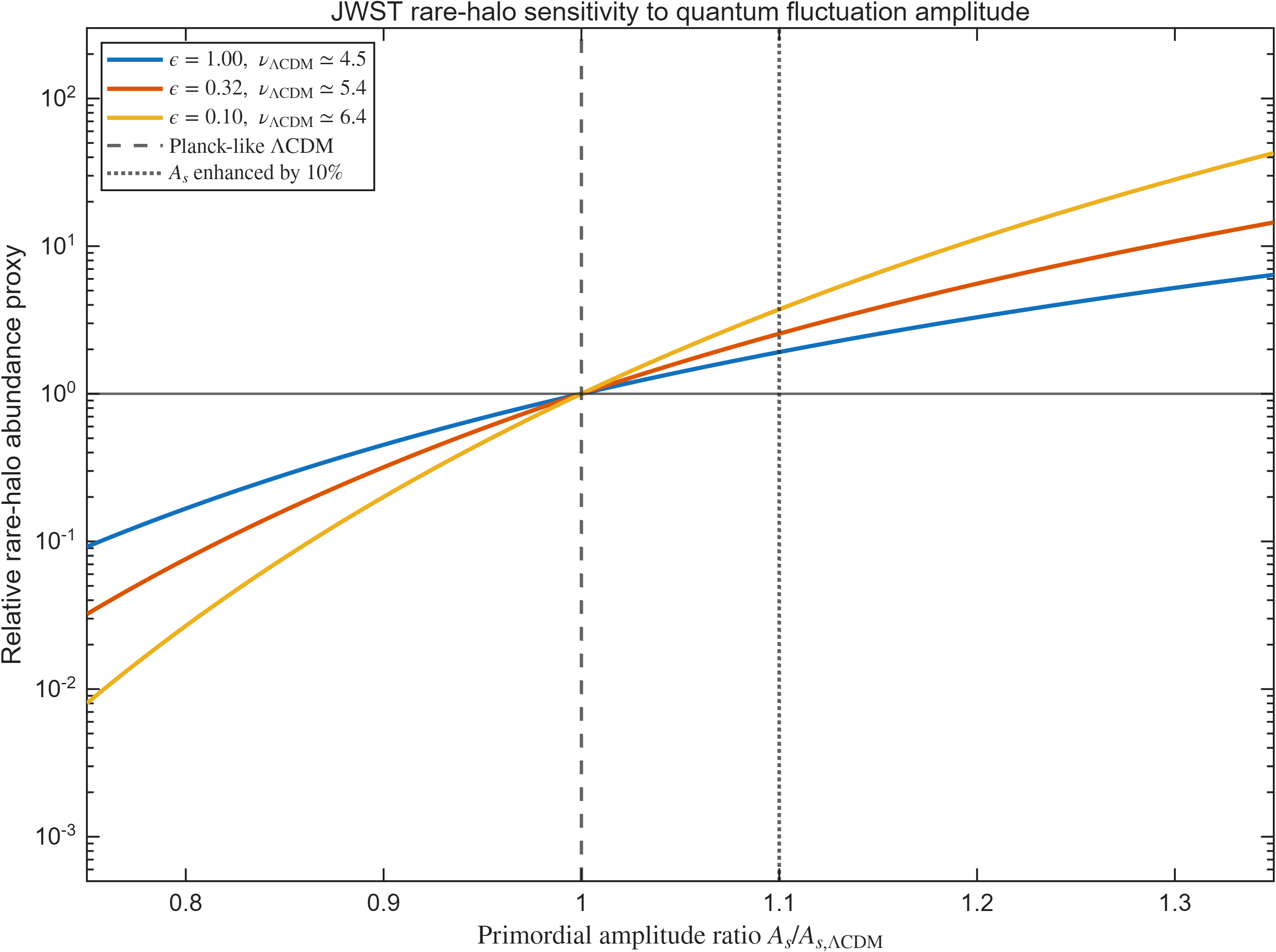}
\caption{Rare-tail amplification of a controlled change in the primordial
fluctuation amplitude. Increasing peak rarity increases the fractional
response of the predicted abundance. The calculation is a sensitivity
diagnostic rather than an alternative determination of $A_s$.}
\label{fig:sec13_As}
\end{figure}

Figure~\ref{fig:sec13_As} illustrates the effect for representative
rare-peak benchmarks motivated by the massive-galaxy discussion of
\citet{BoylanKolchin2023}. This is a sensitivity experiment rather than an
alternative fit to the primordial amplitude: it asks how a small upstream
perturbation propagates when the other ingredients of the model are held
fixed.

\subsection{Where the primordial sensitivity becomes large}
\label{subsec:sec13_response_map}

To localize the amplification in halo mass and redshift, define
\begin{equation}
\mathcal R_n(M_h,z)
=
\frac{
n_{\rm rare}(M_h,z;1.02A_{s,0})
}{
n_{\rm rare}(M_h,z;A_{s,0})
}.
\label{eq:sec13_Rn}
\end{equation}
In the ST high-$\nu$ approximation,
\begin{equation}
\mathcal R_n
\simeq
\exp\left[
\frac{a_{\rm ST}\nu_0^2(M_h,z)}{2}
\left(1-\frac{1}{1.02}\right)
\right].
\label{eq:sec13_Rn_ST}
\end{equation}

\begin{figure}
\centering
\includegraphics[width=\columnwidth]
{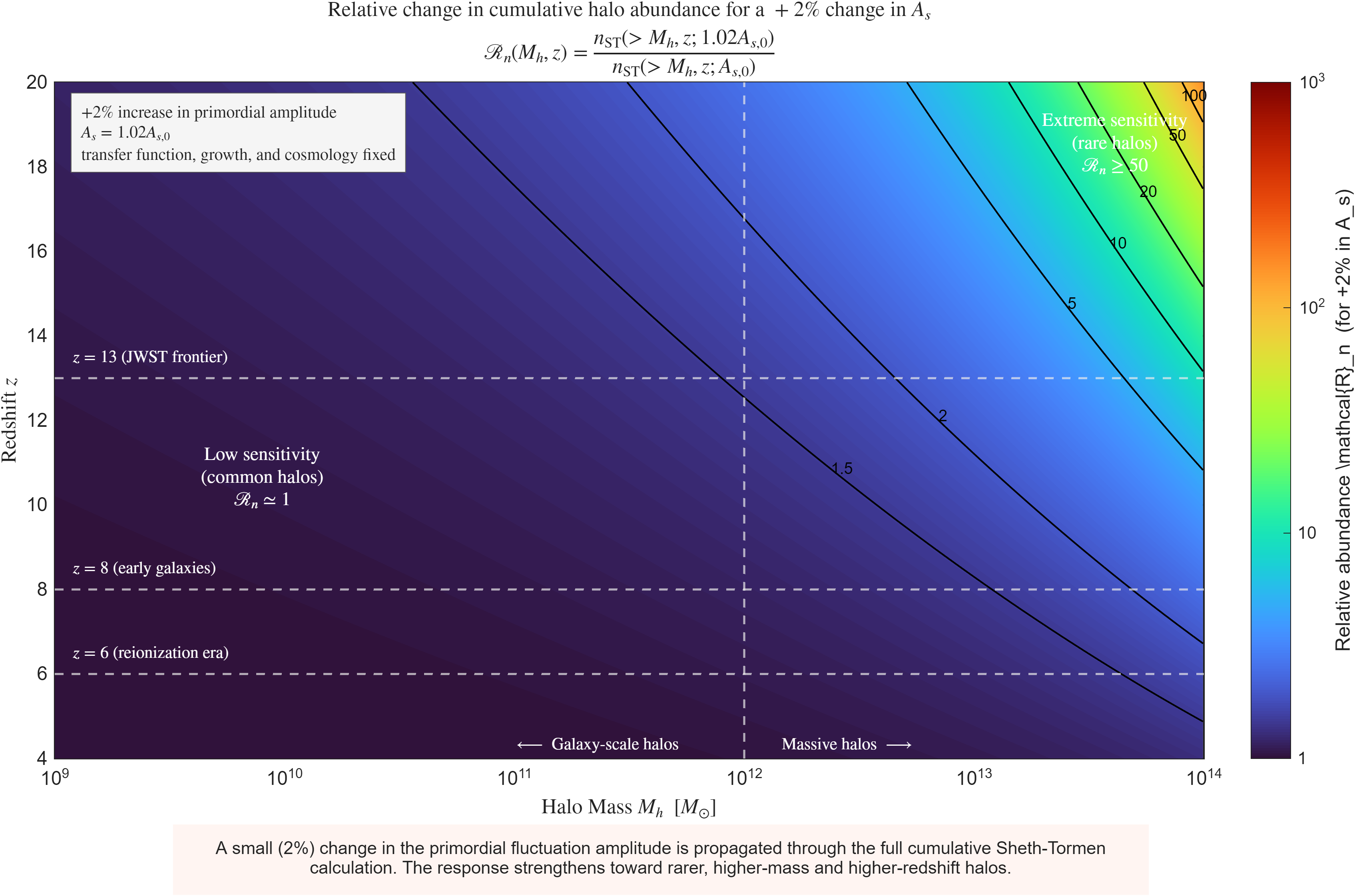}
\caption{Two-dimensional response of rare-halo abundance to a $+2$ per cent
change in the primordial amplitude. The response is weak for common halos
and increases toward high halo mass and high redshift, where collapse
requires increasingly rare peaks.}
\label{fig:sec13_response_map}
\end{figure}

Figure~\ref{fig:sec13_response_map} displays this response over the
$M_h$--$z$ plane. Common halos at relatively low redshift remain close to
$\mathcal R_n=1$, whereas the response rises toward the high-mass,
high-redshift region where $\nu$ is large. The map therefore identifies the
part of parameter space in which small upstream differences can acquire the
largest statistical leverage.

The same region also requires the greatest caution. Rare-halo calibration,
stellar-mass inference, halo-mass convention, survey volume, selection, and
stochastic halo-to-observable mappings all become increasingly important
toward the extreme tail.

\subsection{Separating primordial sensitivity from HMF calibration}
\label{subsec:sec13_st_gureft_results}

The second controlled experiment changes the halo-abundance prescription
while leaving the primordial and linear fluctuation fields unchanged. The
Sheth--Tormen and GUREFT branches therefore share
\begin{equation}
\begin{aligned}
\mathcal P_{\mathcal R}^{\rm ST}
&=
\mathcal P_{\mathcal R}^{\rm GUREFT},
\\
P_m^{\rm ST}
&=
P_m^{\rm GUREFT},
\\
\sigma_{\rm ST}
&=
\sigma_{\rm GUREFT}.
\end{aligned}
\label{eq:sec13_same_linear_field}
\end{equation}
The common peak height
\begin{equation}
\nu(M,z)=\frac{\delta_c}{\sigma(M,z)}
\label{eq:sec13_common_peak_height}
\end{equation}
may therefore be used as a diagnostic of the rarity of the underlying
linear fluctuation, even though the GUREFT multiplicity function is
parameterized directly in terms of $\sigma$.

The differential HMF response is
\begin{equation}
\mathcal R_{\rm HMF}(M_h,z)
=
\frac{
({\rm d}n/{\rm d}M_h)_{\rm GUREFT}
}{
({\rm d}n/{\rm d}M_h)_{\rm ST}
},
\label{eq:sec13_hmf_ratio}
\end{equation}
and the corresponding cumulative response is
\begin{equation}
\mathcal R_{n,{\rm HMF}}(>M_h,z)
=
\frac{
n_{\rm GUREFT}(>M_h,z)
}{
n_{\rm ST}(>M_h,z)
}.
\label{eq:sec13_cumulative_ratio}
\end{equation}
Neither quantity is assumed to be independent of mass. The ST--GUREFT
comparison is therefore fundamentally different from a global
normalization change: the two prescriptions can differ in both the shape
and normalization of the high-redshift halo population.

\begin{figure}
\centering
\includegraphics[width=\columnwidth]
{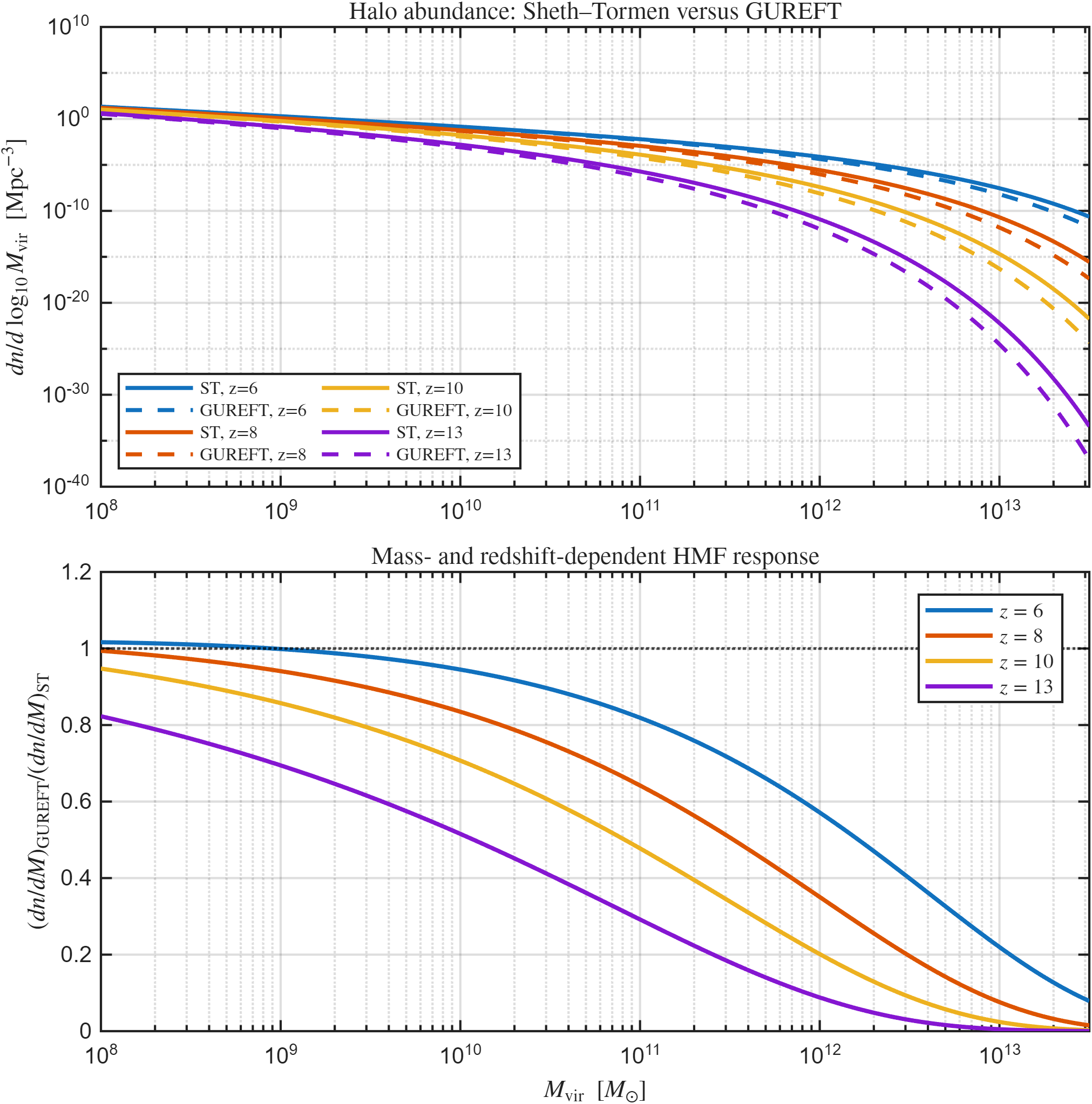}
\caption{Comparison of the Sheth--Tormen and GUREFT halo mass functions
at $z=6$, $8$, $10$, and $13$, evaluated using the same underlying matter
power spectrum and mass variance $\sigma(M,z)$. The upper panel shows the
differential halo abundance, with solid curves representing Sheth--Tormen
and dashed curves representing GUREFT. The lower panel shows the ratio
$({\rm d}n/{\rm d}M)_{\rm GUREFT}/({\rm d}n/{\rm d}M)_{\rm ST}$.
The departure from unity toward the rare-halo regime demonstrates that the
difference between the two prescriptions is mass- and redshift-dependent
rather than a global normalization shift.}
\label{fig:sec13_st_gureft_hmf}
\end{figure}

Figure~\ref{fig:sec13_st_gureft_hmf} shows this comparison at
$z=6$, $8$, $10$, and $13$. The same primordial spectrum, transfer
function, growth history, matter power spectrum, and self-consistently
calculated $\sigma(M,z)$ are used in both branches. The figure therefore
isolates the effect of changing the nonlinear halo-abundance prescription.

The GUREFT calibration covers the principal redshifts considered here,
including $z=6$, $8$, $10$, and $13$ \citep{Yung2024}. This avoids a
redshift extrapolation at those epochs, but it does not eliminate rare-tail
uncertainty. The most massive halos at the highest redshifts are
intrinsically sparsely sampled, so the extreme high-mass, high-redshift
corner is treated as a stress-test regime rather than as a precision
abundance determination.

\subsection{High-redshift AGN: an independent growth clock}
\label{subsec:sec13_agn_results}

The AGN branch introduces information not contained in the HMF alone. A
black hole observed at $z_{\rm obs}$ must satisfy both a host-supply
condition and a growth-time condition. For radiative efficiency
$\epsilon_{\rm rad}$ and Eddington ratio $\lambda$, the growth time is
\begin{equation}
\Delta t_{\rm grow}
=
t_{\rm ef}
\ln\left(\frac{M_{\rm BH}}{M_{\rm seed}}\right),
\label{eq:sec13_growth_time}
\end{equation}
with
\begin{equation}
t_{\rm ef}
\simeq
0.45\,{\rm Gyr}\,
\frac{\epsilon_{\rm rad}}
{(1-\epsilon_{\rm rad})\lambda}.
\label{eq:sec13_efold}
\end{equation}
The available interval is
\begin{equation}
\Delta t_{\rm avail}
=
t(z_{\rm obs})-t(z_{\rm seed}).
\label{eq:sec13_available_time}
\end{equation}

\begin{figure}
\centering
\includegraphics[width=\columnwidth]
{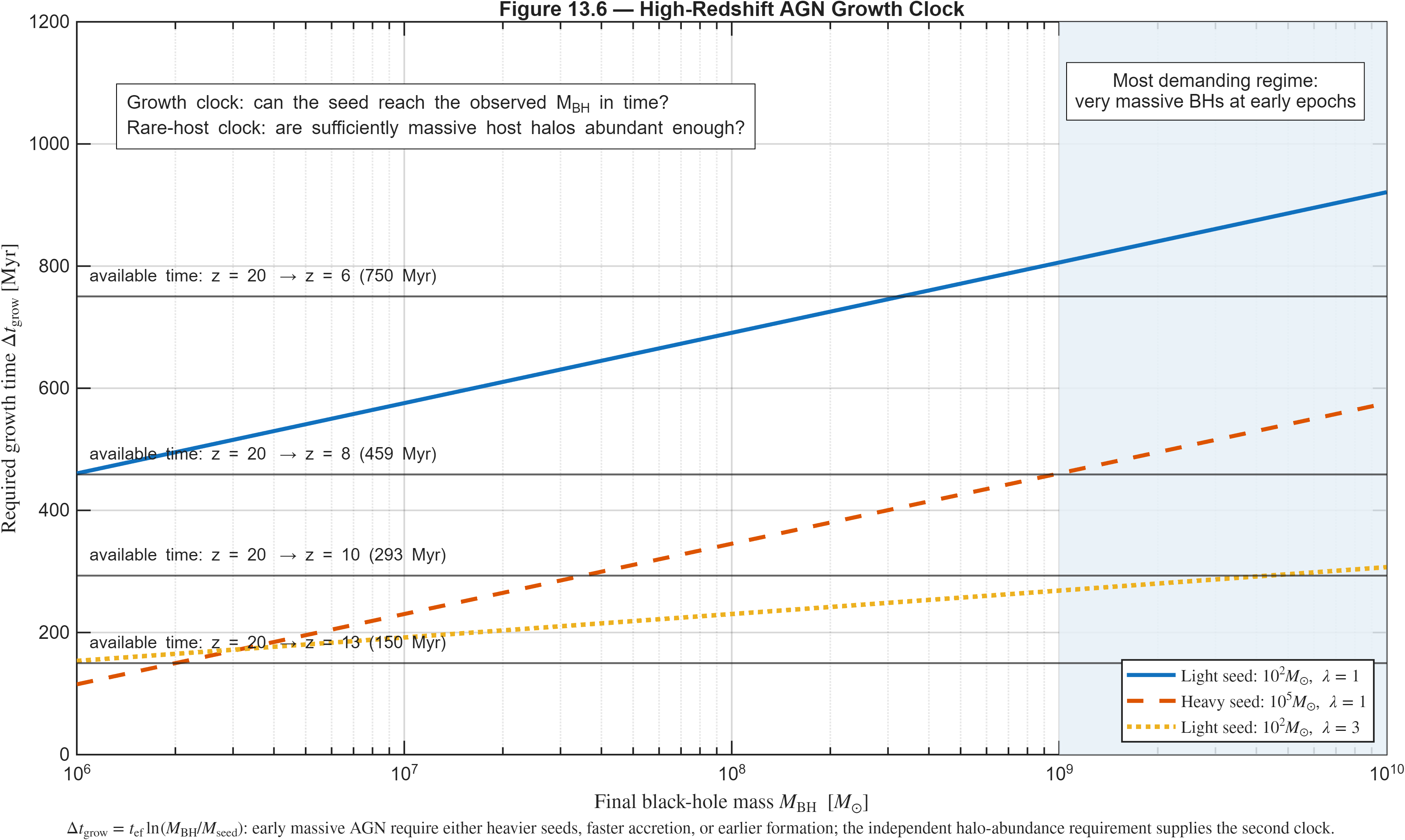}
\caption{Black-hole growth clock for representative seed and accretion
scenarios. Horizontal reference levels give the cosmic time available
between seed formation at $z=20$ and selected observation redshifts.
The temporal requirement is independent of the additional requirement
that sufficiently massive host halos exist in adequate numbers.}
\label{fig:sec13_agn_clock}
\end{figure}

Figure~\ref{fig:sec13_agn_clock} compares the required growth time with the
available cosmic time for representative light- and heavy-seed channels.
Heavy seeds reduce the required number of e-foldings, while larger effective
accretion rates reduce the growth time. The earliest and most massive black
holes occupy the most demanding part of this parameter space.

The corresponding host ceiling is
\begin{equation}
n_{{\rm AGN},X}
=
f_{\rm occ}f_{\rm duty}
n_X(>M_{h,\min},z),
\qquad
X\in\{{\rm ST,GUREFT}\},
\label{eq:sec13_agn_supply}
\end{equation}
under the constant occupation and duty-cycle approximation. High-redshift
AGN therefore impose the two independent consistency conditions
\begin{equation}
\begin{aligned}
\Delta t_{\rm grow}
&\leq
\Delta t_{\rm avail},
\\
n_{\rm AGN}^{\rm obs}
&\leq
f_{\rm occ}f_{\rm duty}
n_X(>M_{h,\min},z).
\end{aligned}
\label{eq:sec13_two_clocks}
\end{equation}
The first condition probes black-hole assembly, whereas the second probes
the cosmological supply of sufficiently massive hosts. Changing from ST to
GUREFT changes the host-supply calculation but does not alter the seed mass,
radiative efficiency, Eddington ratio, or black-hole growth law.

\subsection{Propagation into the UV population}
\label{subsec:sec13_uv_response}

The final numerical experiment propagates the two distinct sources of
sensitivity into luminosity space. The stellar UV luminosity function is
\begin{equation}
\Phi_X(M_{\rm UV},z)
=
\int
\frac{{\rm d}n_X}{{\rm d}M_h}
P(M_{\rm UV}\mid M_h,z)\,
{\rm d}M_h,
\qquad
X\in\{{\rm ST,GUREFT}\}.
\label{eq:sec13_uvlf_general}
\end{equation}
The deterministic halo-to-light mapping is recovered when the conditional
distribution collapses to a one-to-one relation.

For the controlled primordial-amplitude experiment, define
\begin{equation}
\mathcal R_{\Phi,A_s}(M_{\rm UV},z)
=
\frac{
\Phi_{\rm ST}(M_{\rm UV},z;1.02A_{s,0})
}{
\Phi_{\rm ST}(M_{\rm UV},z;A_{s,0})
}.
\label{eq:sec13_Rphi_As}
\end{equation}
The ST branch is used here because its high-$\nu$ form exposes the
primordial rare-tail amplification analytically.

The independent HMF response is
\begin{equation}
\mathcal R_{\Phi,{\rm HMF}}(M_{\rm UV},z)
=
\frac{
\Phi_{\rm GUREFT}(M_{\rm UV},z)
}{
\Phi_{\rm ST}(M_{\rm UV},z)
},
\label{eq:sec13_Rphi_HMF}
\end{equation}
with the primordial spectrum, linear matter field, mass variance, and
halo-to-light kernel held fixed. More generally,
\begin{equation}
\mathcal R_{\Phi,{\rm HMF}}
=
\frac{
\displaystyle
\int ({\rm d}n_{\rm GUREFT}/{\rm d}M_h)
P(M_{\rm UV}\mid M_h,z)\,{\rm d}M_h
}{
\displaystyle
\int ({\rm d}n_{\rm ST}/{\rm d}M_h)
P(M_{\rm UV}\mid M_h,z)\,{\rm d}M_h
}.
\label{eq:sec13_uvlf_hmf_integral_ratio}
\end{equation}
The HMF difference is therefore propagated through the same astrophysical
kernel rather than represented by a redshift-only normalization factor.

\begin{figure}
\centering
\includegraphics[width=\columnwidth]
{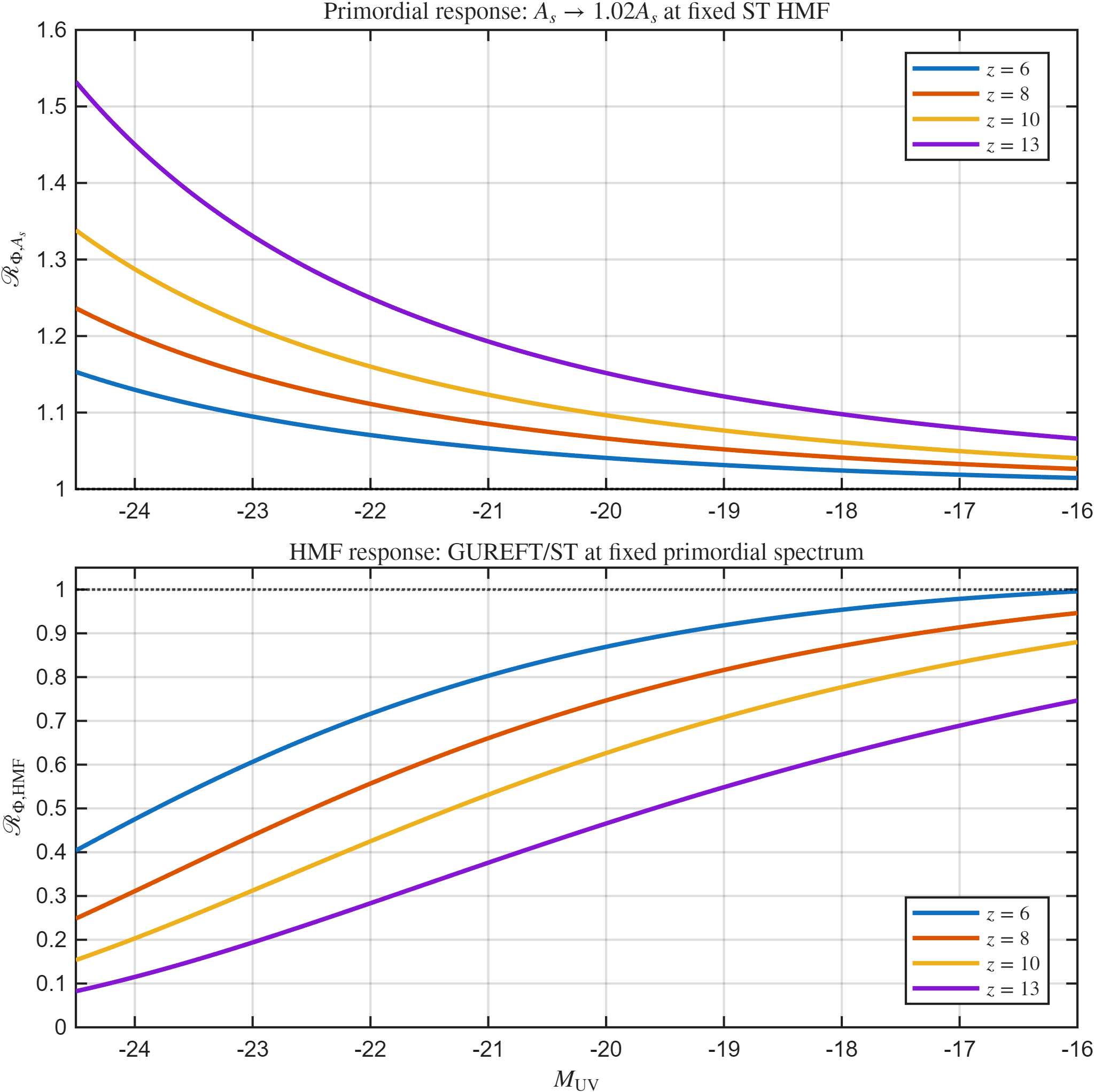}
\caption{Propagation of primordial-amplitude and halo-mass-function
sensitivity into the predicted UV galaxy population. The upper panel shows
$\mathcal R_{\Phi,A_s}
=\Phi_{\rm ST}(M_{\rm UV},z;1.02A_{s,0})/
\Phi_{\rm ST}(M_{\rm UV},z;A_{s,0})$
for a $+2$ per cent perturbation of the primordial amplitude with the
Sheth--Tormen prescription and halo-to-light mapping held fixed. The lower
panel shows
$\mathcal R_{\Phi,{\rm HMF}}
=\Phi_{\rm GUREFT}(M_{\rm UV},z)/
\Phi_{\rm ST}(M_{\rm UV},z)$
with the primordial spectrum, linear matter field, mass variance, and
halo-to-light mapping held fixed. The luminosity- and redshift-dependent
responses distinguish an upstream primordial perturbation from a change in
the nonlinear halo-abundance prescription.}
\label{fig:sec13_uv_response}
\end{figure}

Figure~\ref{fig:sec13_uv_response} compares the two response functions.
The primordial-amplitude response becomes stronger toward the bright end,
where the adopted mapping selects increasingly rare hosts. The ST--GUREFT
response is also luminosity dependent because different UV magnitudes probe
different parts of a mass- and redshift-dependent HMF ratio.

Increasing the primordial amplitude by $2$ per cent raises the predicted UV
abundance, with the response increasing toward brighter magnitudes and
higher redshift. This behaviour is the observable continuation of the
rare-peak amplification identified in
Figures~\ref{fig:sec13_As} and \ref{fig:sec13_response_map}.

The HMF response behaves differently. At fixed primordial spectrum,
$\Phi_{\rm GUREFT}/\Phi_{\rm ST}$ inherits the mass- and redshift-dependent
difference between the halo populations shown in
Figure~\ref{fig:sec13_st_gureft_hmf}. It therefore cannot be represented by
a constant renormalization of the luminosity function.

Figure~\ref{fig:sec13_uv_response} is a response diagnostic rather than a
formal fit to JWST data. Direct observational comparison additionally
requires completeness, redshift probability distributions, cosmic variance,
dust, bursty star formation, and scatter in the halo-to-light relation
\citep{Donnan2023,Donnan2024,Franco2025}. The numerical experiment instead
isolates how two changes introduced at different stages of the forward model
survive into the predicted UV population.

\subsection{Numerical conclusions}
\label{subsec:sec13_numerical_conclusions}

The numerical experiments establish four results.

First, the propagation from $A_s$ to $\sigma$ and $\nu$ is modest and
controlled. A percent-level primordial-amplitude perturbation produces a
smaller fractional change in the linear mass variance and peak height.

Second, the rare-tail mapping supplies a strong statistical lever arm. In
the ST high-$\nu$ limit, the exponential dependence on $\nu^2$ can amplify
that small upstream change into a much larger fractional abundance response.
The amplification is localized toward high halo mass and high redshift
rather than applying uniformly to the halo population.

Third, halo-abundance calibration constitutes a physically distinct source
of uncertainty. The ST--GUREFT experiment holds
$\mathcal P_{\mathcal R}$, $P_m$, and $\sigma$ fixed while changing the
nonlinear abundance assigned to the same underlying fluctuation field. The
resulting response depends on mass and redshift and propagates into
luminosity space through the halo-to-light mapping.

Fourth, high-redshift AGN add an independent temporal constraint. Their
interpretation requires both an adequate rare-host reservoir and sufficient
time for seed black holes to grow. The growth clock is unchanged by the
choice between ST and GUREFT, whereas the HMF prescription affects the
host-supply clock.

The primordial amplification pathway is
\begin{equation}
\delta\mathcal P_{\mathcal R}
\rightarrow
\delta\sigma
\rightarrow
\delta\nu
\rightarrow
\delta n_h
\rightarrow
\delta\Phi_{\rm UV},
\label{eq:sec13_amplification_path}
\end{equation}
while the complementary HMF control is
\begin{equation}
\begin{aligned}
\mathcal P_{\mathcal R},\,P_m,\,\sigma
&=\text{fixed},
\\
{\rm ST}\rightarrow{\rm GUREFT}
&\Longrightarrow
\delta n_h\rightarrow\delta\Phi_{\rm UV}.
\end{aligned}
\label{eq:sec13_hmf_control_path}
\end{equation}

These experiments do not imply that present JWST populations require a
modified primordial spectrum. They identify where primordial information
can be amplified and separate that sensitivity from uncertainty introduced
at the nonlinear halo-abundance stage. Section~\ref{sec:interpretation_quantum_structure}
uses this separation to determine what can, and cannot, be inferred about
primordial physics from high-redshift JWST populations.

%

\section{Interpreting JWST Through the Quantum-to-Structure Chain}
\label{sec:interpretation_quantum_structure}

The numerical experiments of Section~\ref{sec:numerical_information_flow}
provide a framework for interpreting high-redshift galaxies and active
galactic nuclei without assigning an apparent abundance discrepancy
prematurely to primordial physics. The calculation separates three distinct
operations: changing the primordial fluctuation field, changing the mapping
from that field to collapsed halos, and changing the astrophysical and
observational mappings that connect halos to detected sources. The relevant
forward sequence is
\begin{equation}
\begin{aligned}
\text{quantum vacuum}
&\rightarrow \mathcal R\rightarrow \mathcal P_{\mathcal R}(k)\\
&\rightarrow P_m(k,z)\rightarrow \sigma(M,z)\rightarrow \nu(M,z)\\
&\rightarrow \frac{{\rm d}n_X}{{\rm d}M_h}\rightarrow P(O\mid M_h,z)\\
&\rightarrow S(O,z)\rightarrow \text{JWST observables},
\end{aligned}
\label{eq:sec14_full_chain}
\end{equation}
where $X\in\{{\rm ST,GUREFT}\}$, $P(O\mid M_h,z)$ represents the
astrophysical mapping from halo mass to an observable $O$, and $S(O,z)$ is
the survey selection function. The reverse inference begins with observed
photons and attempts to move backward through these layers toward halo
rarity and ultimately the primordial fluctuation field. Each inversion
therefore introduces assumptions that must be tested independently.

\subsection{Rare-peak amplification of primordial sensitivity}
\label{subsec:sec14_localized_amplification}

The controlled $A_s$ experiment shows that the first stages of the response
are modest. At fixed transfer-function shape and growth history,
\begin{equation}
\sigma\propto A_s^{1/2},
\qquad
\nu\propto A_s^{-1/2}.
\label{eq:sec14_As_sigma_nu}
\end{equation}
A two per cent increase in $A_s$ therefore raises $\sigma$ by approximately
one per cent and lowers the corresponding peak height by approximately one
per cent. The stronger response arises only after this shift is propagated
through the exponentially suppressed rare-halo tail. For the Sheth--Tormen
high-$\nu$ limit,
\begin{equation}
n_{\rm rare}
\propto
\exp\left(-\frac{a\nu^2}{2}\right),
\label{eq:sec14_rare_tail}
\end{equation}
so that, to leading order,
\begin{equation}
{\rm d}\ln n_{\rm rare}
\simeq
\frac{a\nu^2}{2}\,
{\rm d}\ln A_s.
\label{eq:sec14_As_sensitivity}
\end{equation}
This leading-order derivative follows by differentiating the exponential $\exp[-a_{\rm ST}\nu^2/2]$ and using $\nu\propto A_s^{-1/2}$ at fixed spectral shape. The large response is therefore not an instability of the linear matter field; it arises from the statistics of sufficiently rare peaks. Multiplicity prefactors and finite-perturbation corrections are discussed in Appendix~\ref{app:rare_peak_sensitivity}.

Figure~\ref{fig:sec13_response_map} localizes this effect in mass and
redshift. The response increases toward the high-mass, high-redshift corner,
while much of the halo population remains comparatively insensitive. The
same region that provides the greatest leverage on primordial statistics is
also the region in which halo counts are smallest and uncertainties in halo
mass, abundance, and observable inference are most consequential. Primordial
sensitivity and inferential difficulty therefore increase together.

\subsection{Primordial sensitivity and halo calibration are distinct}
\label{subsec:sec14_primordial_vs_hmf}

The second controlled experiment changes the halo-abundance prescription
while holding the primordial and linear fluctuation field fixed. The
comparison is defined by
\begin{equation}
\begin{aligned}
\mathcal P_{\mathcal R}^{\rm ST}
&=\mathcal P_{\mathcal R}^{\rm GUREFT},\\
P_m^{\rm ST}&=P_m^{\rm GUREFT},\\
\sigma_{\rm ST}&=\sigma_{\rm GUREFT},
\end{aligned}
\label{eq:sec14_same_linear_field}
\end{equation}
with the same collapse threshold used to characterize peak rarity. The
change enters when the calculated fluctuation statistics are converted into
a collapsed halo population. A useful diagnostic is
\begin{equation}
\mathcal R_{\rm HMF}(M_h,z)
=
\frac{({\rm d}n/{\rm d}M_h)_{\rm GUREFT}}
     {({\rm d}n/{\rm d}M_h)_{\rm ST}}.
\label{eq:sec14_hmf_ratio}
\end{equation}
Unlike a redshift-only multiplicative correction,
$\mathcal R_{\rm HMF}$ depends on both halo mass and redshift. The
ST--GUREFT comparison therefore changes the shape as well as the
normalization of the predicted halo population.

Figure~\ref{fig:sec13_st_gureft_hmf} makes this separation explicit. The
same $\mathcal P_{\mathcal R}$, $P_m$, and $\sigma$ can lead to different
predicted halo abundances because Sheth--Tormen and GUREFT encode different
mappings from the fluctuation field to collapsed objects. The former is an
analytic rare-peak prescription, whereas the latter is calibrated to
high-redshift numerical simulations \citep{Yung2024}. A difference in halo
abundance therefore does not, by itself, identify a difference in the
primordial spectrum.

The GUREFT fit covers the cosmic-dawn redshifts used in the numerical
comparison, including $z=6$, $8$, $10$, and $13$. Its calibration is not,
however, equally constraining over the entire mass--redshift plane. The most
massive halos at the highest redshifts are intrinsically sparse and limited
by finite simulation volume. Results in this extreme corner are consequently
best interpreted as rare-peak stress tests rather than precision abundance
determinations. A primordial interpretation should therefore be robust to
reasonable uncertainty in the nonlinear halo-abundance prescription.

\subsection{Massive galaxies: from stellar mass to peak rarity}
\label{subsec:sec14_massive_galaxies}

For a massive high-redshift galaxy, the reverse inference begins with an
estimated stellar mass. The baryonic reservoir requires
\begin{equation}
M_\star
\leq
\epsilon_\star f_bM_h,
\label{eq:sec14_stellar_budget}
\end{equation}
and therefore
\begin{equation}
M_h
\geq
\frac{M_\star}{\epsilon_\star f_b}.
\label{eq:sec14_halo_floor}
\end{equation}
The inferred halo requirement then determines $\sigma(M_h,z)$ and
$\nu(M_h,z)$ and identifies the corresponding location in the rare-peak
response map.

Two related but distinct conditions follow. The first is an individual
mass-budget condition: whether a halo can contain enough baryons to support
the inferred stellar mass. The second is a population condition: whether
the cosmological model supplies enough halos above the required mass to
support the observed number density or stellar-mass density. These tests
should not be conflated \citep{BoylanKolchin2023}.

The rare-tail response also clarifies why uncertainties in the reverse
mapping become increasingly important for the most extreme systems. Changes
in stellar mass, redshift, lensing magnification, star-formation efficiency,
or halo-mass definition can shift the inferred host mass. In the steep
high-$\nu$ tail, even a moderate shift in $M_h$ can then produce a much
larger change in the predicted abundance. Rare galaxies are powerful probes
of the primordial-to-structure chain only when the downstream inference is
correspondingly well constrained.

\subsection{High-redshift AGN: growth clock and rare-host clock}
\label{subsec:sec14_agn_two_clocks}

High-redshift AGN add a temporal constraint to the halo-abundance problem. A
luminous AGN requires both a black hole that can reach the inferred mass
within the available cosmic time and a sufficiently abundant population of
host halos. The two requirements can be written
\begin{equation}
\begin{aligned}
\Delta t_{\rm grow}
&\leq
t(z_{\rm obs})-t(z_{\rm seed}),\\
n_{\rm AGN}^{\rm obs}
&\leq
f_{\rm occ}f_{\rm duty}
 n_X(>M_{h,\min},z),
\end{aligned}
\label{eq:sec14_two_clocks}
\end{equation}
where $X\in\{{\rm ST,GUREFT}\}$.

Figure~\ref{fig:sec13_agn_clock} isolates the growth-time requirement. For a
given final black-hole mass, the required number of accretion e-foldings
depends strongly on seed mass, radiative efficiency, Eddington ratio, and
accretion history. Heavier seeds reduce the required growth interval, while
sustained high or super-Eddington accretion can shorten the effective growth
time. The earliest massive black holes occupy the most demanding region of
this parameter space \citep{Inayoshi2020,Napolitano2025}.

The rare-host clock is logically separate. A black hole may satisfy the
growth-time requirement while its required host halo remains too rare under
a particular abundance prescription. Conversely, an adequate supply of
massive halos does not guarantee that their central black holes can reach the
observed mass by the required epoch. The two constraints must therefore be
satisfied simultaneously.

This structure makes high-redshift AGN complementary to massive galaxies and
the galaxy UV luminosity function. Galaxies primarily constrain baryonic
reservoirs and the conversion of halo mass into stellar light. AGN additionally
probe black-hole seeding, accretion, occupation, duty cycle, and the finite
age of the Universe.

\subsection{The UV luminosity function as a population-level transfer test}
\label{subsec:sec14_uvlf_interpretation}

The galaxy UV luminosity function tests the halo-to-light transformation at
the population level,
\begin{equation}
\Phi_X(M_{\rm UV},z)
=
\int
\frac{{\rm d}n_X}{{\rm d}M_h}
P(M_{\rm UV}\mid M_h,z)\,
{\rm d}M_h,
\label{eq:sec14_uvlf}
\end{equation}
where $X\in\{{\rm ST,GUREFT}\}$. Unlike an individual extreme object, the
UVLF samples a distribution of luminosities and therefore a corresponding
distribution of host-halo masses.

The upper panel of Figure~\ref{fig:sec13_uv_response} propagates the
controlled primordial-amplitude perturbation through the fixed
Sheth--Tormen HMF and the same halo-to-light mapping. It realizes the chain
\begin{equation}
\delta A_s
\rightarrow
\delta\sigma
\rightarrow
\delta\nu
\rightarrow
\delta n_h
\rightarrow
\delta\Phi_{\rm UV}.
\label{eq:sec14_uv_response_chain}
\end{equation}
The response strengthens toward the bright end because increasingly luminous
systems are associated, in the adopted mapping, with increasingly rare host
halos.

The lower panel performs a different experiment: the primordial spectrum and
halo-to-light mapping are fixed while the halo mass function is changed from
ST to GUREFT. The corresponding diagnostic is
\begin{equation}
\mathcal R_{\Phi,{\rm HMF}}(M_{\rm UV},z)
=
\frac{\Phi_{\rm GUREFT}(M_{\rm UV},z)}
     {\Phi_{\rm ST}(M_{\rm UV},z)}.
\label{eq:sec14_uv_hmf_ratio}
\end{equation}
Because the GUREFT-to-ST halo ratio depends on mass and redshift, the UVLF
response is generally luminosity dependent as well as redshift dependent.
It cannot be represented by a single global normalization factor.

A third class of changes acts still farther downstream. Altering
$P(M_{\rm UV}\mid M_h,z)$ through star-formation efficiency, burstiness,
dust attenuation, metallicity, or scatter can modify the predicted bright
end without changing either the primordial spectrum or the HMF. Comparisons
with JWST UV luminosity functions \citep{Donnan2023,Donnan2024,Franco2025}
therefore test a composite forward model. Primordial information can be
constrained only after the halo-abundance and halo-to-light transformations
are treated consistently.

\subsection{A hierarchy for localizing apparent tension}
\label{subsec:sec14_hierarchy}

The forward calculation naturally separates four layers of inference:
\begin{equation}
\begin{aligned}
{\rm I:}\;&
\mathcal P_{\mathcal R}(k)
&&\text{primordial/QFT},\\
{\rm II:}\;&
P_m,\sigma,\nu,\frac{{\rm d}n}{{\rm d}M_h}
&&\text{growth and halo statistics},\\
{\rm III:}\;&
P(O\mid M_h,z)
&&\text{astrophysical mapping},\\
{\rm IV:}\;&
S(O,z)
&&\text{observation and selection}.
\end{aligned}
\label{eq:sec14_layers}
\end{equation}
An apparent excess should first be established observationally, then tested
against uncertainties in the astrophysical mapping, and then against the
halo-abundance prescription. Only a discrepancy that remains after these
downstream layers are controlled can be localized credibly toward the
primordial sector.

The resulting evidence hierarchy is
\begin{equation}
\begin{aligned}
\text{candidate excess}
&\rightarrow
\text{secure population}\\
&\rightarrow
\text{robust abundance discrepancy}\\
&\rightarrow
\text{astrophysical robustness}\\
&\rightarrow
\text{HMF robustness}\\
&\rightarrow
\text{primordial interpretation}.
\end{aligned}
\label{eq:sec14_evidence_hierarchy}
\end{equation}
This ordering is particularly important in the high-mass, high-redshift
regime because it is simultaneously the region of greatest rare-peak
sensitivity and the region in which small shifts in inferred mass or
abundance have the largest consequences.

\subsection{What JWST measures and what must be inferred}
\label{subsec:sec14_what_jwst_measures}

JWST directly measures photons: fluxes, spectra, colors, line strengths,
morphologies, and their redshift dependence. Halo mass, peak height, and the
primordial curvature spectrum are inferred quantities. Their conceptual
separation may be summarized as
\begin{equation}
\text{photons}
\not\equiv
M_h
\not\equiv
\nu
\not\equiv
\mathcal P_{\mathcal R}.
\label{eq:sec14_not_equivalent}
\end{equation}
The connections between these quantities are physical models rather than
identities.

This separation specifies what is required for JWST to become an increasingly
precise cosmological probe. Spectroscopic confirmation strengthens the
observational layer; improved stellar-population modeling strengthens the
galaxy-to-halo mapping; larger survey volumes reduce statistical and
cosmic-variance uncertainties; high-redshift simulations improve the halo
abundance calibration; and joint constraints with the CMB restrict the
allowed primordial parameter space.

The quantum-to-structure framework therefore changes the interpretive
question. Rather than asking only whether an object appears ``too massive''
or ``too early,'' the relevant question is which link in the forward chain
must change to reproduce the observed population and whether that change is
consistent with independent constraints.

\subsection{Interpretive synthesis}
\label{subsec:sec14_synthesis}

The numerical results support a bounded interpretation: the rare-peak tail amplifies small changes in primordial amplitude, but the response depends strongly on halo mass and redshift. A change of halo mass function at fixed primordial spectrum and mass variance produces a distinct effect, while halo-to-light and selection assumptions introduce further uncertainty. Massive galaxies, the stellar UV luminosity function, and AGN growth-time constraints consequently test complementary links of the same framework. These findings motivate the submission-level conclusions below without claiming that existing JWST observations require nonstandard primordial physics.
\section{Conclusions and Outlook}
\label{sec:conclusions}

This work has developed a continuous framework connecting the quantum origin
of primordial fluctuations to the abundance and observable properties of
galaxies and active galactic nuclei (AGN) at cosmic dawn. The central
sequence is
\begin{equation}
\begin{aligned}
\text{quantum vacuum}
&\rightarrow {\cal R}
\rightarrow {\cal P}_{\cal R}(k)
\rightarrow P_m(k,z) \\
&\rightarrow \sigma(M,z)
\rightarrow \nu(M,z)
\rightarrow \frac{dn}{dM_h} \\
&\rightarrow \text{luminous populations}
\rightarrow \text{JWST}.
\end{aligned}
\label{eq:conclusion_chain}
\end{equation}
The sequence distinguishes primordial initial conditions, halo statistics, and observational inference.

The numerical experiments in Section~\ref{sec:numerical_information_flow}
provide the principal result. A controlled percent-level perturbation of the
primordial amplitude produces only a modest response in the linear mass
variance and peak height. The much larger effect arises when the shifted peak
height is propagated through the exponentially suppressed high-$\nu$ tail of
the halo distribution. This amplification is strongly localized in halo mass
and redshift: common halos remain comparatively insensitive, whereas
increasingly massive halos at increasingly early epochs acquire a much larger
abundance response. Rare cosmic-dawn populations therefore provide
potentially powerful leverage on primordial fluctuation statistics, but
precisely in the regime where halo statistics and observable inference are
most demanding.

This sensitivity does not provide a unique inference of primordial physics.
The controlled Sheth--Tormen (ST)--GUREFT comparison isolates the role of the
halo-abundance prescription. In this comparison,
${\cal P}_{\cal R}(k)$, $P_m(k,z)$, and $\sigma(M,z)$ are held fixed while
the mapping from the linear fluctuation field to the abundance of collapsed
halos is changed. The ST prescription provides a transparent analytic
rare-peak baseline, whereas the GUREFT halo mass function of
\citet{Yung2024} supplies a simulation-calibrated description constructed
for the high-redshift regime. Their difference is generally both
mass- and redshift-dependent and may be summarized by
\begin{equation}
{\cal R}_{\rm HMF}(M_h,z)
=
\frac{
\left(dn/dM_h\right)_{\rm GUREFT}
}{
\left(dn/dM_h\right)_{\rm ST}
}.
\label{eq:conclusion_hmf_ratio}
\end{equation}
A primordial perturbation and an HMF-calibration change can therefore alter
the same predicted high-redshift population while entering the causal chain
at physically distinct stages. The GUREFT calibration directly covers the
cosmic-dawn redshift interval considered here, although the most massive
halos at the highest redshifts remain intrinsically sparse. Predictions in
that extreme corner are consequently interpreted as rare-peak stress tests
rather than precision determinations of halo abundance.

The observable mapping introduces a further independent layer. For galaxies,
the HMF determines the cosmological supply of potential hosts, while
star-formation efficiency and history, stellar populations, metallicity,
dust, feedback, burstiness, and scatter determine how that supply is
transformed into stellar mass and ultraviolet luminosity. The UV-response
calculation shows that rare-tail sensitivity can survive this transformation
and become especially pronounced toward the bright end. At the same time,
the ST--GUREFT comparison demonstrates that changes in the halo-abundance
prescription can propagate through the same halo-to-light kernel and modify
the predicted luminosity function without changing the primordial spectrum.
Comparisons with JWST luminosity functions must therefore constrain
cosmological halo supply and astrophysical luminosity mapping jointly rather
than identify an abundance difference directly with primordial physics
\citep{Donnan2023,Donnan2024,Franco2025}.

Massive high-redshift galaxies and AGN provide complementary consistency
tests. For galaxies, an inferred stellar mass imposes a baryonic-reservoir
requirement and hence a minimum host-halo mass,
\begin{equation}
M_{h,\min}
=
\frac{M_\star}{\epsilon_\star f_b},
\label{eq:conclusion_galaxy_host}
\end{equation}
linking the observation to peak rarity and cumulative halo abundance
\citep{BoylanKolchin2023}. The limiting choice
$\epsilon_\star=1$ represents an absolute baryonic ceiling rather than a
realistic star-formation model. For AGN, the constraint is twofold: the
cosmological density field must provide a sufficient reservoir of rare host
halos, and the central black hole must be capable of reaching the inferred
mass within the finite cosmic time available. Seed mass, radiative
efficiency, Eddington ratio, occupation, and duty cycle therefore enter the
AGN problem in addition to halo abundance
\citep{Inayoshi2020,Napolitano2025}. The galaxy and AGN channels probe
overlapping halo statistics through different astrophysical kernels and
should therefore remain distinct.

The resulting interpretation is naturally hierarchical. JWST directly
measures photons rather than halo masses, peak heights, or the primordial
curvature spectrum. A candidate excess must first be established as a secure
population-level observational result. It must then be tested against the
halo-to-observable mapping and against uncertainty in the conversion from
linear peak statistics to nonlinear halo abundance. Only after those
downstream layers are controlled can a persistent discrepancy be propagated
credibly toward the primordial sector. Schematically,
\begin{equation}
\begin{aligned}
\text{JWST population}
&\rightarrow \text{observable inference}\\
&\rightarrow \text{halo requirement}\\
&\rightarrow \text{peak rarity}\\
&\rightarrow \text{primordial constraints}.
\end{aligned}
\label{eq:reverse_inference_chain}
\end{equation}
A convincing primordial interpretation would therefore require a discrepancy
to remain robust under observational, astrophysical, and HMF uncertainties
while remaining consistent with independent constraints on the primordial
spectrum.

The present framework does not require the conclusion that current JWST
observations demonstrate a failure of $\Lambda$CDM or of the standard
nearly scale-invariant primordial spectrum. Its immediate role is diagnostic:
it identifies where information is amplified, where degeneracies enter, and
which physical layer would have to change if a robust discrepancy persists.
This distinction is particularly important in the high-mass,
high-redshift regime, where leverage on primordial statistics is greatest but
uncertainties in halo calibration, stellar-mass inference, survey selection,
and stochastic halo-to-observable mapping can also have disproportionately
large effects.

Several extensions can sharpen this test. Larger-volume and
higher-resolution simulations can improve calibration of the extreme
high-redshift halo tail and reduce uncertainty in the conversion from the
linear density field to collapsed abundance. Probabilistic halo-to-light
models can propagate scatter, bursty star formation, dust, feedback, and
assembly history into predicted luminosity functions. The AGN branch can be
extended to a full luminosity-function calculation incorporating seed
distributions, occupation, duty cycle, accretion histories, obscuration, and
luminosity scatter. On the primordial side, the response formalism can be
generalized beyond an overall change in $A_s$ to scale-dependent features,
running of the spectral index, primordial non-Gaussianity, and alternative
inflationary spectra. Such extensions require particular care because a
simulation-calibrated HMF obtained for a standard CDM power spectrum need
not remain calibrated for arbitrary modifications of the primordial or
small-scale matter spectrum. Joint analysis with CMB constraints and
simulation suites constructed for non-standard spectra will therefore be
important for determining whether cosmic-dawn populations provide genuinely
complementary information about primordial physics.

The broader conclusion is that the path from quantum fields in the early
Universe to JWST observables is neither direct nor inaccessible. It is a
sequence of identifiable physical transformations whose sensitivities can be
calculated and whose uncertainties can be separated. In forward form,
\begin{equation}
\begin{aligned}
{\cal P}_{\cal R}(k)
&\rightarrow P_m(k,z)\rightarrow \sigma(M,z)\rightarrow \nu(M,z)\\
&\rightarrow \frac{dn}{dM_h}\rightarrow P(O|M_h,z)\\
&\rightarrow S(O,z)\rightarrow \text{observed population}.
\end{aligned}
\label{eq:conclusion_forward_inference}
\end{equation}
In reverse, population-level observations can constrain this chain only by
passing through the intervening halo, astrophysical, and selection models.
By treating high-redshift galaxies and AGN as complementary consistency
tests of the complete chain, rather than as isolated ``too early'' objects,
observations of cosmic dawn can progressively test how primordial quantum
fluctuations became the nonlinear structures observed in the Universe today.

\section*{Acknowledgements}

The authors gratefully acknowledge the support of the Johns Hopkins
University Whiting School EP Applied Physics Program. A.N. and R.S. also
acknowledge the support of the Johns Hopkins University Applied
Physics Laboratory.

\section*{Data Availability}

The numerical results presented in this work are derived from the analytic
models and observational datasets cited in the text. The numerical code used
to generate the theoretical calculations and figures is available from the
corresponding author upon reasonable request.

%


\appendix

\section{Numerical Implementation of the Quantum-to-Halo Pipeline}
\label{app:numerical_pipeline}

This appendix summarizes the numerical implementation underlying the results
of Section~\ref{sec:numerical_information_flow}. The purpose is
reproducibility rather than repetition of the physical derivations in the
main text. The calculation is organized so that each numerical operation
corresponds to one link in the chain
\begin{equation}
\begin{aligned}
\mathcal P_{\mathcal R}(k)
&\rightarrow P_m(k,z)
\rightarrow \sigma(M,z)
\rightarrow \nu(M,z) \\
&\rightarrow
\frac{{\rm d}n_X}{{\rm d}M_h}
\rightarrow n_X(>M_h,z),
\end{aligned}
\label{eq:appA_chain}
\end{equation}
where
\begin{equation}
X\in\{{\rm ST,GUREFT}\}
\label{eq:appA_hmf_branches}
\end{equation}
for the quantitative HMF comparison. Press--Schechter serves as a historical and numerical diagnostic.

\subsection{Baseline cosmology and numerical grids}
\label{app:baseline_grids}

The baseline calculation adopts
\begin{equation}
\Omega_m=0.3158,\qquad
\Omega_b=0.0493,\qquad
h=0.6732,
\end{equation}
with
\begin{equation}
n_s=0.96605,\qquad
\sigma_8=0.812,\qquad
\delta_c=1.686.
\end{equation}
The baryon fraction is
\begin{equation}
f_b=\frac{\Omega_b}{\Omega_m}\simeq0.156.
\label{eq:appA_fb}
\end{equation}

Both wavenumber and halo mass are sampled logarithmically. The $k$ range is
chosen sufficiently broad that the top-hat variance is insensitive to the
integration boundaries over the halo-mass interval used in the analysis.
Convergence is checked by increasing the number of grid points and extending
the $k$ and mass ranges until $\sigma(M)$, the differential HMF, and the
cumulative abundance are stable over the region displayed in the figures.

\subsection{Primordial and matter power spectra}
\label{app:power_implementation}

The primordial curvature spectrum is represented by
\begin{equation}
\mathcal P_{\mathcal R}(k)
=
A_s
\left(\frac{k}{k_*}\right)^{n_s-1}.
\label{eq:appA_PR}
\end{equation}
The corresponding linear matter spectrum is generated using the transfer
function and growth factor described in
Section~\ref{sec:transfer_growth}. For the numerical HMF calculation it is
convenient to write
\begin{equation}
P_m(k,z)
=
A\,k^{n_s}T^2(k)D^2(z),
\label{eq:appA_Pm}
\end{equation}
where $A$ is fixed by the adopted $\sigma_8$ normalization. The baseline
calculation uses the Eisenstein--Hu transfer prescription
\citep{EisensteinHu1998}. The linear growth factor is normalized to
\begin{equation}
D(0)=1.
\end{equation}

The controlled primordial-amplitude experiment changes the normalization
without changing the spectral shape, transfer function, or growth history:
\begin{equation}
A_s=(1+\epsilon)A_{s,0},
\qquad
P_m\rightarrow(1+\epsilon)P_m,
\label{eq:appA_As_perturbation}
\end{equation}
and hence
\begin{equation}
\sigma(M,z)
\rightarrow
(1+\epsilon)^{1/2}\sigma(M,z).
\label{eq:appA_sigma_perturbation}
\end{equation}
The perturbed spectra are not renormalized back to the fiducial
$\sigma_8=0.812$, because doing so would remove the amplitude perturbation
that the experiment is designed to propagate.

\subsection{Top-hat smoothing and mass variance}
\label{app:sigma_implementation}

A mass $M$ is associated with the comoving top-hat radius
\begin{equation}
R(M)
=
\left(
\frac{3M}{4\pi\bar\rho_{m,0}}
\right)^{1/3},
\label{eq:appA_RM}
\end{equation}
where $\bar\rho_{m,0}$ is the present mean matter density. The Fourier-space
top-hat window is
\begin{equation}
W(x)
=
3\frac{\sin x-x\cos x}{x^3},
\qquad
x=kR,
\label{eq:appA_window}
\end{equation}
and the variance is evaluated as
\begin{equation}
\sigma^2(M,z)
=
\frac{1}{2\pi^2}
\int_0^\infty
k^2P_m(k,z)W^2(kR)\,{\rm d}k.
\label{eq:appA_sigma}
\end{equation}
As an internal normalization check,
$R_8=8\,h^{-1}{\rm Mpc}$ reproduces the adopted $\sigma_8$ to the numerical
precision of the integration.

\subsection{Peak height and logarithmic derivative}
\label{app:peak_derivative}

The collapse peak height is
\begin{equation}
\nu(M,z)
=
\frac{\delta_c}{\sigma(M,z)}.
\label{eq:appA_nu}
\end{equation}
The HMF additionally requires
\begin{equation}
\left|
\frac{{\rm d}\ln\sigma^{-1}}{{\rm d}\ln M}
\right|.
\label{eq:appA_derivative}
\end{equation}
This derivative is evaluated on the logarithmic mass grid using a smooth
finite-difference gradient. The grid is refined until the resulting HMF is
stable against further changes in numerical resolution.

\subsection{Press--Schechter and Sheth--Tormen checks}
\label{app:hmf_implementation}

For multiplicity functions written in terms of peak height, the differential
HMF is
\begin{equation}
\frac{{\rm d}n}{{\rm d}M}
=
\frac{\bar\rho_{m,0}}{M^2}
f(\nu)
\left|
\frac{{\rm d}\ln\sigma^{-1}}{{\rm d}\ln M}
\right|.
\label{eq:appA_hmf_generic}
\end{equation}
For Press--Schechter,
\begin{equation}
f_{\rm PS}(\nu)
=
\sqrt{\frac{2}{\pi}}\,
\nu
\exp\left(-\frac{\nu^2}{2}\right),
\label{eq:appA_ps}
\end{equation}
while the adopted Sheth--Tormen multiplicity function is
\begin{equation}
\begin{aligned}
f_{\rm ST}(\nu)
={}&
A_{\rm ST}
\sqrt{\frac{2a_{\rm ST}}{\pi}}
\left[
1+(a_{\rm ST}\nu^2)^{-p_{\rm ST}}
\right] \\
&\times
\nu
\exp\left(-\frac{a_{\rm ST}\nu^2}{2}\right),
\end{aligned}
\label{eq:appA_st}
\end{equation}
with
\begin{equation}
A_{\rm ST}=0.3222,\qquad
a_{\rm ST}=0.707,\qquad
p_{\rm ST}=0.3
\end{equation}
\citep{PressSchechter1974,ShethTormen1999,ShethMoTormen2001}.
Press--Schechter is used primarily as a historical and diagnostic reference;
ST supplies the analytic rare-peak baseline used in the controlled
primordial-amplitude experiment.

\subsection{GUREFT high-redshift abundance branch}
\label{app:gureft_implementation}

The complementary high-redshift HMF is the simulation-calibrated GUREFT fit
of \citet{Yung2024}. In physical mass units it is implemented as
\begin{equation}
\frac{{\rm d}n_{\rm G}}{{\rm d}M_{\rm vir}}
=
f_{\rm G}(\sigma,z)
\frac{\bar\rho_{m,0}}{M_{\rm vir}^2}
\left|
\frac{{\rm d}\ln\sigma^{-1}}
{{\rm d}\ln M_{\rm vir}}
\right|,
\label{eq:appA_gureft_hmf}
\end{equation}
where
\begin{equation}
f_{\rm G}(\sigma,z)
=
A_{\rm G}(z)
\left[
\left(
\frac{\sigma}{b_{\rm G}(z)}
\right)^{-a_{\rm G}(z)}
+1
\right]
\exp\left[
-\frac{c_{\rm G}(z)}{\sigma^2}
\right].
\label{eq:appA_gureft_multiplicity}
\end{equation}
Each coefficient is represented by
\begin{equation}
\chi_i(z)
=
\chi_{0,i}
+\chi_{1,i}z
+\chi_{2,i}z^2,
\qquad
\chi_i\in
\{A_{\rm G},a_{\rm G},b_{\rm G},c_{\rm G}\}.
\label{eq:appA_gureft_redshift}
\end{equation}

For physical masses in $M_\odot$, the coefficients used in the numerical
implementation are
\begin{equation}
\begin{array}{c|rrr}
\chi_i & \chi_{0,i} & \chi_{1,i} & \chi_{2,i} \\
\hline
A_{\rm G} & 0.13765772 & -0.01003821 & 0.00102964 \\
a_{\rm G} & 1.06641384 &  0.02475576 & -0.00283342 \\
b_{\rm G} & 4.86693806 &  0.09212356 & -0.01426283 \\
c_{\rm G} & 1.19837952 & -0.00142967 & -0.00033074
\end{array}
\label{eq:appA_gureft_coefficients}
\end{equation}
from the physical-unit fit of \citet{Yung2024}.

The GUREFT branch uses the same self-consistently calculated
$\sigma(M,z)$ as the ST branch. The approximate $\sigma(M)$ fitting
relations supplied with the simulation fit are therefore not substituted
for the matter-power-spectrum calculation developed in the main text. This
choice preserves the common upstream chain
\begin{equation}
\mathcal P_{\mathcal R}
\rightarrow
P_m
\rightarrow
\sigma
\end{equation}
and isolates the HMF prescription in the ST--GUREFT comparison.

The GUREFT halo mass is $M_{\rm vir}$, based on the virial
spherical-overdensity convention used in the simulation analysis. When an
empirical halo-to-observable relation is expressed in another halo-mass
definition, a consistent mass conversion or an explicit systematic
allowance is required. The redshift-dependent fit is calibrated over
$6\leq z\leq19$; the redshifts $z=6$, $8$, $10$, and $13$ used in the
numerical comparison therefore lie within that interval. The extreme
high-mass tail remains sparsely sampled and is treated as a stress-test
regime rather than as a precision abundance prediction.

\subsection{Controlled ST--GUREFT comparison}
\label{app:st_gureft_control}

The HMF-calibration experiment is defined by
\begin{equation}
\begin{aligned}
\mathcal P_{\mathcal R}^{\rm ST}
&=
\mathcal P_{\mathcal R}^{\rm GUREFT},
\\
P_m^{\rm ST}
&=
P_m^{\rm GUREFT},
\\
\sigma_{\rm ST}
&=
\sigma_{\rm GUREFT},
\end{aligned}
\label{eq:appA_common_linear_field}
\end{equation}
while the mapping from $\sigma$ to collapsed halo abundance is changed. The
differential response is
\begin{equation}
\mathcal R_{\rm HMF}(M,z)
=
\frac{
({\rm d}n/{\rm d}M)_{\rm GUREFT}
}{
({\rm d}n/{\rm d}M)_{\rm ST}
}.
\label{eq:appA_hmf_ratio}
\end{equation}
Unlike a global normalization correction,
$\mathcal R_{\rm HMF}$ is generally both mass- and redshift-dependent.

\subsection{Cumulative number and mass densities}
\label{app:cumulative_integrals}

For either quantitative branch,
\begin{equation}
n_X(>M,z)
=
\int_M^\infty
\frac{{\rm d}n_X}{{\rm d}M'}\,{\rm d}M',
\qquad
X\in\{{\rm ST,GUREFT}\},
\label{eq:appA_cumulative_number}
\end{equation}
and
\begin{equation}
\rho_{h,X}(>M,z)
=
\int_M^\infty
M'
\frac{{\rm d}n_X}{{\rm d}M'}\,{\rm d}M'.
\label{eq:appA_cumulative_mass}
\end{equation}
These integrals are accumulated numerically from the high-mass end of the
grid downward. The upper mass boundary is extended until its contribution
to the mass interval of interest is negligible.

\subsection{Numerical validation}
\label{app:numerical_validation}

The numerical pipeline is examined through several internal consistency checks. First, the normalized matter power spectrum reproduces the adopted value of $\sigma_8$. Second, the mass variance $\sigma(M)$ decreases smoothly with increasing halo mass, while the peak height $\nu(M,z)$ increases toward the rare high-mass tail. Third, the differential halo mass functions are evaluated on logarithmically spaced mass grids, and their positivity and smoothness are examined over the plotted mass and redshift ranges. Fourth, the Sheth–Tormen and GUREFT branches use the same calculated $\sigma(M,z)$, ensuring that their comparison isolates the adopted halo-abundance prescription. Finally, the resulting ratio $R_\mathrm{HMF}(M,z)$ is evaluated as a function of both mass and redshift rather than being approximated by a single normalization factor.

These checks are kept separate from the observable mappings so that
numerical artifacts in the cosmological pipeline are not misidentified as
high-redshift astrophysical effects.

\section{Halo-to-Light Mapping and the UV Luminosity Function}
\label{app:halo_to_light}

This appendix specifies the mapping used to connect the theoretical halo
population to the rest-frame UV luminosity function discussed in
Sections~\ref{sec:uvlf_information_transfer} and
\ref{sec:numerical_information_flow}. The purpose is to keep the
cosmological supply of halos distinct from the baryonic mapping that makes
those halos observable.

\subsection{Halo mass, baryonic mass, and stellar conversion}
\label{app:halo_baryon_mapping}

For a halo of mass $M_h$, the associated baryonic reservoir is
\begin{equation}
M_b=f_bM_h,
\label{eq:appB_baryons}
\end{equation}
with $f_b=\Omega_b/\Omega_m$. A simple stellar-conversion prescription gives
\begin{equation}
M_\star
=
\epsilon_\star f_bM_h,
\label{eq:appB_stellar_mass}
\end{equation}
where $\epsilon_\star$ is an effective stellar-conversion efficiency. The
limiting condition
\begin{equation}
M_\star\leq f_bM_h
\label{eq:appB_absolute_ceiling}
\end{equation}
corresponds to $\epsilon_\star\leq1$ and provides an absolute baryonic
ceiling. It is not a realistic expectation that all halo baryons are
converted into stars.

For the GUREFT branch, $M_h$ is identified with the GUREFT virial mass
$M_{\rm vir}$. Empirical relations defined using a different halo-mass
convention require a consistent conversion before they are combined with
the GUREFT abundance.

\subsection{Star-formation timescale}
\label{app:sfr_timescale}

For the baseline deterministic mapping, the characteristic star-formation
timescale is
\begin{equation}
t_\star(z)
=
f_{\rm dyn}H^{-1}(z),
\label{eq:appB_tstar}
\end{equation}
where $f_{\rm dyn}$ parametrizes the fraction of a Hubble time over which
the stellar mass is assembled. The corresponding star-formation rate is
\begin{equation}
{\rm SFR}(M_h,z)
=
\frac{\epsilon_\star f_bM_h}{t_\star(z)}
=
\frac{\epsilon_\star f_b}{f_{\rm dyn}}
M_hH(z).
\label{eq:appB_sfr}
\end{equation}
This expression explains why the luminosity associated with a halo of fixed
mass need not decrease toward high redshift: a shorter characteristic
timescale can increase the instantaneous SFR even while halos of that mass
become substantially rarer.

\subsection{UV luminosity and absolute magnitude}
\label{app:uv_conversion}

The intrinsic rest-frame UV luminosity density is related to the SFR by
\begin{equation}
L_{\nu,{\rm UV}}
=
\kappa_{\rm UV}\,{\rm SFR},
\label{eq:appB_uv_sfr}
\end{equation}
where $\kappa_{\rm UV}$ depends on the stellar population, IMF, metallicity,
star-formation history, and UV calibration. The AB absolute magnitude is
\begin{equation}
M_{\rm UV}
=
51.60
-
2.5\log_{10}
\left[
\frac{L_{\nu,{\rm UV}}}
{{\rm erg\,s^{-1}\,Hz^{-1}}}
\right].
\label{eq:appB_Muv}
\end{equation}
Equations~(\ref{eq:appB_sfr})--(\ref{eq:appB_Muv}) define the deterministic
mapping
\begin{equation}
M_h
\longrightarrow
M_{\rm UV}
(M_h,z;\epsilon_\star,f_{\rm dyn},\kappa_{\rm UV}).
\label{eq:appB_deterministic_map}
\end{equation}

For the response calculation shown in
Figure~\ref{fig:sec13_uv_response}, the first-pass mapping uses
\begin{equation}
\epsilon_\star=0.10,\qquad
f_{\rm dyn}=0.10,
\label{eq:appB_response_parameters}
\end{equation}
and
\begin{equation}
\kappa_{\rm UV}
=
8.0\times10^{27}
\frac{{\rm erg\,s^{-1}\,Hz^{-1}}}
{M_\odot\,{\rm yr}^{-1}}.
\label{eq:appB_kappa_response}
\end{equation}
These values define the controlled response experiment rather than a
unique fit to the galaxy population.

\subsection{Deterministic UV luminosity function}
\label{app:deterministic_uvlf}

If the halo-to-light relation is monotonic and deterministic, number
conservation gives
\begin{equation}
\Phi_X(M_{\rm UV},z)\,{\rm d}M_{\rm UV}
=
\frac{{\rm d}n_X}{{\rm d}M_h}\,{\rm d}M_h.
\label{eq:appB_number_conservation}
\end{equation}
Therefore
\begin{equation}
\Phi_X(M_{\rm UV},z)
=
\left.
\frac{{\rm d}n_X}{{\rm d}M_h}
\left|
\frac{{\rm d}M_h}{{\rm d}M_{\rm UV}}
\right|
\right|_{M_h=M_h(M_{\rm UV},z)},
\label{eq:appB_uvlf_deterministic}
\end{equation}
where
\begin{equation}
X\in\{{\rm ST,GUREFT}\}.
\end{equation}
The Jacobian is essential because the HMF is a density per unit halo mass,
whereas the UVLF is a density per unit magnitude.

\subsection{General stochastic mapping}
\label{app:stochastic_uvlf}

A deterministic relation is useful for isolating the information flow, but
real galaxies exhibit scatter. The general forward model is
\begin{equation}
\Phi_X(M_{\rm UV},z)
=
\int_0^\infty
\frac{{\rm d}n_X}{{\rm d}M_h}
P(M_{\rm UV}\mid M_h,z)\,
{\rm d}M_h.
\label{eq:appB_uvlf_general}
\end{equation}
The conditional distribution can incorporate scatter in star-formation
efficiency, burstiness, dust attenuation, stellar-population properties,
metallicity, and assembly history. A survey selection function can then be
applied to the predicted observable population.

At the bright end, scatter can be especially important because the HMF is
steep. There are more moderate-mass halos available to scatter toward high
luminosity than rare massive halos available to scatter downward. This
asymmetry can increase the predicted abundance of bright galaxies without
changing either the primordial spectrum or the underlying HMF.

\subsection{Propagation of the ST--GUREFT difference}
\label{app:st_gureft_uvlf}

When both HMF branches use the same halo-to-light kernel,
\begin{equation}
\Phi_{\rm ST}(M_{\rm UV},z)
=
\int
\frac{{\rm d}n_{\rm ST}}{{\rm d}M_h}
P(M_{\rm UV}\mid M_h,z)\,
{\rm d}M_h,
\label{eq:appB_phi_st}
\end{equation}
and
\begin{equation}
\Phi_{\rm GUREFT}(M_{\rm UV},z)
=
\int
\frac{{\rm d}n_{\rm GUREFT}}{{\rm d}M_h}
P(M_{\rm UV}\mid M_h,z)\,
{\rm d}M_h.
\label{eq:appB_phi_gureft}
\end{equation}
The luminosity-space HMF response is therefore
\begin{equation}
\mathcal R_{\Phi,{\rm HMF}}
(M_{\rm UV},z)
=
\frac{
\Phi_{\rm GUREFT}(M_{\rm UV},z)
}{
\Phi_{\rm ST}(M_{\rm UV},z)
}.
\label{eq:appB_phi_hmf_ratio}
\end{equation}
Because the GUREFT/ST HMF ratio varies with halo mass and redshift,
$\mathcal R_{\Phi,{\rm HMF}}$ is generally luminosity- and redshift-dependent.

For a common deterministic monotonic mapping, the Jacobian cancels between
the two branches and
\begin{equation}
\mathcal R_{\Phi,{\rm HMF}}
(M_{\rm UV},z)
=
\left.
\mathcal R_{\rm HMF}(M_h,z)
\right|_{M_h=M_h(M_{\rm UV},z)}.
\label{eq:appB_deterministic_hmf_ratio}
\end{equation}
With scatter, this local relation no longer holds: the observable response
is a weighted convolution over a range of halo masses. This distinction is
why the HMF calibration cannot be represented by a single redshift-only
factor in luminosity space.

\subsection{Primordial-amplitude response in luminosity space}
\label{app:primordial_uv_response}

The controlled primordial experiment is propagated through the same
halo-to-light mapping. Its response is
\begin{equation}
\mathcal R_{\Phi,A_s}(M_{\rm UV},z)
=
\frac{
\Phi_{\rm ST}
(M_{\rm UV},z;1.02A_{s,0})
}{
\Phi_{\rm ST}
(M_{\rm UV},z;A_{s,0})
}.
\label{eq:appB_phi_As_ratio}
\end{equation}
The comparison between
$\mathcal R_{\Phi,A_s}$ and $\mathcal R_{\Phi,{\rm HMF}}$ therefore separates
an upstream perturbation of the primordial fluctuation amplitude from a
change introduced at the nonlinear halo-abundance stage. The two responses
shown in Figure~\ref{fig:sec13_uv_response} are calculated using the same
baseline halo-to-light prescription.

\subsection{Galaxy and AGN luminosity functions}
\label{app:galaxy_agn_separation}

The stellar galaxy UVLF and AGN luminosity function remain distinct
observable mappings. For stellar emission,
\begin{equation}
M_h
\rightarrow
M_b
\rightarrow
{\rm SFR}
\rightarrow
L_{\nu,{\rm UV}}
\rightarrow
M_{\rm UV}.
\label{eq:appB_galaxy_chain}
\end{equation}
For AGN, the general luminosity function is
\begin{equation}
\begin{aligned}
\Phi_{{\rm AGN},X}(L,z)
={}&
\int
\frac{{\rm d}n_X}{{\rm d}M_h}
f_{\rm occ}(M_h,z)
f_{\rm duty}(M_h,z)
\\
&\times
P(L\mid M_h,z,{\rm active})\,
{\rm d}M_h .
\end{aligned}
\label{eq:appB_agn_lf}
\end{equation}
Thus $f_{\rm occ}$ and $f_{\rm duty}$ are not inserted into the stellar
galaxy UVLF. Both galaxy and AGN populations draw from the same cosmological
halo reservoir, but they illuminate that reservoir through different
astrophysical kernels.

\section{Rare-Peak Sensitivity and Halo-Abundance Calibration}
\label{app:rare_peak_sensitivity}

This appendix develops the sensitivity relations underlying the rare-peak
discussion in Sections~\ref{sec:numerical_information_flow} and
\ref{sec:interpretation_quantum_structure}. The purpose is not to infer the
primordial amplitude from an individual JWST source, but to quantify why
rare high-redshift populations can respond strongly to small changes in the
linear variance and to distinguish that response from uncertainty in the
halo-abundance prescription.

\subsection{Rare-tail approximation}
\label{app:rare_tail_approx}

For the Gaussian Press--Schechter form, the high-$\nu$ dependence is
dominated by
\begin{equation}
n_{\rm rare}
\propto
\exp\left(-\frac{\nu^2}{2}\right).
\label{eq:appC_ps_tail}
\end{equation}
For Sheth--Tormen the corresponding exponential dependence is
\begin{equation}
n_{\rm rare}
\propto
\exp\left(-\frac{a_{\rm ST}\nu^2}{2}\right),
\label{eq:appC_st_tail}
\end{equation}
with $a_{\rm ST}=0.707$. Since
\begin{equation}
\nu=\frac{\delta_c}{\sigma},
\end{equation}
a small change in the variance gives
\begin{equation}
{\rm d}\ln\nu
=
-{\rm d}\ln\sigma.
\label{eq:appC_dnu}
\end{equation}
Taking the logarithmic differential of
Equation~(\ref{eq:appC_st_tail}) gives, in the rare-tail approximation,
\begin{equation}
{\rm d}\ln n_{\rm rare}
\simeq
a_{\rm ST}\nu^2\,
{\rm d}\ln\sigma.
\label{eq:appC_sensitivity_sigma}
\end{equation}
Thus
\begin{equation}
\frac{\Delta n_{\rm rare}}{n_{\rm rare}}
\sim
a_{\rm ST}\nu^2
\frac{\Delta\sigma}{\sigma}
\label{eq:appC_fractional_sigma}
\end{equation}
for sufficiently small perturbations and large $\nu$, with the
understanding that multiplicity prefactors and the mass derivative of
$\sigma$ have been neglected in this leading sensitivity estimate.

\subsection{Sensitivity to the primordial amplitude}
\label{app:As_sensitivity}

If the shape of the primordial spectrum, transfer function, and growth
history are held fixed,
\begin{equation}
P_m\propto A_s,
\qquad
\sigma\propto A_s^{1/2}.
\label{eq:appC_sigma_As}
\end{equation}
Consequently,
\begin{equation}
{\rm d}\ln\sigma
=
\frac12{\rm d}\ln A_s,
\end{equation}
and Equation~(\ref{eq:appC_sensitivity_sigma}) becomes
\begin{equation}
{\rm d}\ln n_{\rm rare}
\simeq
\frac{a_{\rm ST}\nu^2}{2}
{\rm d}\ln A_s.
\label{eq:appC_sensitivity_As}
\end{equation}
This relation captures the rare-tail lever arm: a high-$\nu$ population can
exhibit a substantially larger fractional abundance response than the
fractional perturbation imposed on $A_s$.

For a finite change relative to $A_{s,0}$,
\begin{equation}
\nu(A_s)
=
\nu_0
\left(
\frac{A_s}{A_{s,0}}
\right)^{-1/2},
\label{eq:appC_nu_finite}
\end{equation}
and the leading ST exponential gives
\begin{equation}
\frac{n_{\rm rare}(A_s)}
{n_{\rm rare}(A_{s,0})}
\simeq
\exp
\left\{
-\frac{a_{\rm ST}}{2}
\left[
\nu^2(A_s)-\nu_0^2
\right]
\right\}.
\label{eq:appC_finite_ratio}
\end{equation}
For the $+2$ per cent experiment,
\begin{equation}
\nu^2(1.02A_{s,0})
=
\frac{\nu_0^2}{1.02},
\end{equation}
so that
\begin{equation}
\mathcal R_{n,A_s}
\simeq
\exp\left[
\frac{a_{\rm ST}\nu_0^2}{2}
\left(
1-\frac{1}{1.02}
\right)
\right].
\label{eq:appC_plus2_response}
\end{equation}
The experiment measures sensitivity; it does not imply freedom to vary
$A_s$ independently of CMB and other primordial constraints.

\subsection{Why sensitivity is not unique inference}
\label{app:sensitivity_not_inference}

A large derivative in Equation~(\ref{eq:appC_sensitivity_As}) does not imply
that a measured excess of rare objects uniquely determines $A_s$. The
observable abundance depends schematically on
\begin{equation}
n_{\rm obs}
=
\mathcal F
\left[
\mathcal P_{\mathcal R},
T,
D,
f_{\rm HMF},
P(O\mid M_h,z),
S(O,z)
\right],
\label{eq:appC_degeneracy}
\end{equation}
where $f_{\rm HMF}$ denotes the collapse-abundance prescription,
$P(O\mid M_h,z)$ the astrophysical mapping, and $S(O,z)$ the survey
selection function. High sensitivity to one upstream parameter therefore
coexists with degeneracy among several downstream quantities.

\subsection{ST--GUREFT halo-abundance response}
\label{app:st_gureft_scaling}

The controlled HMF experiment holds the primordial and linear fields fixed:
\begin{equation}
\mathcal P_{\mathcal R},\quad
P_m,\quad
\sigma
\quad\text{fixed},
\qquad
{\rm ST}\rightarrow{\rm GUREFT}.
\label{eq:appC_st_gureft_control}
\end{equation}
The differential response is
\begin{equation}
\mathcal R_{\rm HMF}(M,z)
=
\frac{
({\rm d}n/{\rm d}M)_{\rm GUREFT}
}{
({\rm d}n/{\rm d}M)_{\rm ST}
},
\label{eq:appC_hmf_response}
\end{equation}
while the cumulative-number response is
\begin{equation}
\mathcal R_{n,{\rm HMF}}(>M,z)
=
\frac{
n_{\rm GUREFT}(>M,z)
}{
n_{\rm ST}(>M,z)
},
\label{eq:appC_cumulative_response}
\end{equation}
and the cumulative collapsed-mass response is
\begin{equation}
\mathcal R_{\rho,{\rm HMF}}(>M,z)
=
\frac{
\rho_{h,{\rm GUREFT}}(>M,z)
}{
\rho_{h,{\rm ST}}(>M,z)
}.
\label{eq:appC_mass_response}
\end{equation}

These ratios are not reducible to a universal redshift-only factor. Their
mass dependence is precisely what allows the HMF choice to change the shape
of a predicted high-redshift population rather than only its normalization.

For a common deterministic halo-to-light mapping,
\begin{equation}
\mathcal R_{\Phi,{\rm HMF}}(M_{\rm UV},z)
=
\left.
\mathcal R_{\rm HMF}(M_h,z)
\right|_{M_h=M_h(M_{\rm UV},z)},
\label{eq:appC_uv_deterministic_response}
\end{equation}
whereas a stochastic mapping produces the weighted ratio
\begin{equation}
\mathcal R_{\Phi,{\rm HMF}}
=
\frac{
\displaystyle
\int
({\rm d}n_{\rm GUREFT}/{\rm d}M_h)
P(M_{\rm UV}\mid M_h,z)\,{\rm d}M_h
}{
\displaystyle
\int
({\rm d}n_{\rm ST}/{\rm d}M_h)
P(M_{\rm UV}\mid M_h,z)\,{\rm d}M_h
}.
\label{eq:appC_uv_stochastic_response}
\end{equation}
Likewise, for identical AGN occupation and duty-cycle prescriptions,
\begin{equation}
\mathcal R_{\rm AGN}
=
\frac{
\displaystyle
\int_{M_{h,\min}}^\infty
f_{\rm occ}f_{\rm duty}
({\rm d}n_{\rm GUREFT}/{\rm d}M_h)\,{\rm d}M_h
}{
\displaystyle
\int_{M_{h,\min}}^\infty
f_{\rm occ}f_{\rm duty}
({\rm d}n_{\rm ST}/{\rm d}M_h)\,{\rm d}M_h
}.
\label{eq:appC_agn_response}
\end{equation}
Thus the ST--GUREFT comparison changes the supply of collapsed hosts without
changing the upstream primordial spectrum or the linear variance.

\subsection{Domain of the HMF comparison}
\label{app:gureft_domain}

The GUREFT fit provides a high-redshift calibration appropriate to the
cosmic-dawn regime considered in this work \citep{Yung2024}. The adopted
redshifts $z=6$, $8$, $10$, and $13$ lie within its fitted redshift interval.
Nevertheless, the most massive halos at the highest redshifts remain
statistically sparse. The resulting ST--GUREFT comparison should therefore
be interpreted as a controlled assessment of HMF sensitivity, with the
extreme rare tail treated as a stress-test regime.

A further limitation applies if the primordial or matter power-spectrum
shape is changed substantially. The GUREFT calibration was obtained for a
standard CDM cosmology and is not automatically guaranteed for arbitrary
localized features, cutoffs, non-standard dark-matter transfer functions,
or other large departures from the calibrated spectrum. The present
ST--GUREFT experiment therefore keeps the upstream linear field fixed.
Extensions to non-standard spectra require an HMF calibration demonstrated
to remain valid for those spectra or a dedicated simulation suite.

\subsection{Interpretive boundary}
\label{app:interpretive_boundary}

The two numerical experiments probe different physical sensitivities:
\begin{equation}
\begin{aligned}
\delta\mathcal P_{\mathcal R}
&\rightarrow
\delta\sigma
\rightarrow
\delta\nu
\rightarrow
\delta n_h,
\\
\mathcal P_{\mathcal R},P_m,\sigma
&=\text{fixed},
\qquad
{\rm ST}\rightarrow{\rm GUREFT}
\rightarrow
\delta n_h.
\end{aligned}
\label{eq:appC_two_sensitivities}
\end{equation}
Both changes can subsequently propagate through the same astrophysical and
selection kernels into observable populations, but they enter the
calculation at different physical stages. Keeping them separate is essential
when interpreting high-redshift JWST populations.

The appendices therefore close the technical loop of the manuscript.
Appendix~\ref{app:numerical_pipeline} specifies how primordial information
is propagated numerically into halo statistics;
Appendix~\ref{app:halo_to_light} specifies how the halo population is mapped
into observable light; and the present appendix quantifies why the rare tail
can amplify upstream differences while remaining subject to HMF,
astrophysical, and observational degeneracies.

\end{document}